\documentclass[a4paper,12pt]{article}

\usepackage{cite}
\usepackage{amsmath}
\usepackage{amssymb}
\usepackage{bm}
\usepackage{color}
\usepackage{colortbl}

\usepackage{xcolor}
\usepackage{graphicx}
\usepackage{tikz}

\numberwithin{equation}{section}

\allowdisplaybreaks

\usepackage{geometry}
\makeatletter
\newsavebox{\@brx}
\newcommand{\llangle}[1][]{\savebox{\@brx}{\(\m@th{#1\langle}\)}%
  \mathopen{\copy\@brx\kern-0.5\wd\@brx\usebox{\@brx}}}
\newcommand{\rrangle}[1][]{\savebox{\@brx}{\(\m@th{#1\rangle}\)}%
  \mathclose{\copy\@brx\kern-0.5\wd\@brx\usebox{\@brx}}}
\makeatother

\newcounter{aff}

\begin{document}
\begin{titlepage}
\begin{flushright}
{\footnotesize NITEP 191, OCU-PHYS 587}
\end{flushright}
\begin{center}
{\Large\bf
Affine Symmetries for ABJM Partition Function\\[6pt]
and its Generalization
}\\
\bigskip\bigskip
{\large
Sanefumi Moriyama\footnote{\tt moriyama@omu.ac.jp}
and
Tomoki Nosaka\footnote{\tt nosaka@yukawa.kyoto-u.ac.jp}
}\\
\bigskip
${}^\dagger$\,{\it Kavli Institute for Theoretical Sciences, University of Chinese Academy of Sciences,}\\
{\it No. 3, Nanyitiao, Zhongguancun, Haidian District, Beijing, China 100190}\\[3pt]
${}^*$\,{\it Department of Physics, Graduate School of Science,\\
Osaka Metropolitan University, Sumiyoshi-ku, Osaka, Japan 558-8585}\\[3pt]
${}^*$\,{\it Nambu Yoichiro Institute of Theoretical and Experimental Physics (NITEP),}\\
{\it Osaka Metropolitan University, Sumiyoshi-ku, Osaka, Japan 558-8585}\\[3pt]
${}^*$\,{\it Osaka Central Advanced Mathematical Institute (OCAMI),}\\
{\it Osaka Metropolitan University, Sumiyoshi-ku, Osaka, Japan 558-8585}
\end{center}

\begin{abstract}
Partially motivated by the fact that the grand partition function of the ABJM theory or its generalization is expressed by a spectral operator enjoying symmetries of the Weyl group, it was found that the grand partition function satisfies the $q$-Painlev\'e equation, which is constructed from the affine Weyl group. 
In this paper we clarify the affine symmetries of the grand partition function.
With the affine symmetries, we find that the grand partition function extends naturally outside the fundamental domain of duality cascades and once the Painlev\'e equation holds in the fundamental domain, so does it outside.
\end{abstract}

\end{titlepage}

\tableofcontents

\section{Introduction}
\label{sec_intro}

Symmetry serves a crucial role in understanding physics.
Especially, infinite-dimensional symmetries control physical systems so strongly that only finite quantities remain independent.
It was known \cite{MP} that the grand canonical partition function of the Aharony-Bergman-Jafferis-Maldacena (ABJM) theory \cite{ABJM} constructed with one NS5-brane and one $(1,k)$5-brane, after normalized by the lowest order, is given by the Fredholm determinant of a spectral operator enjoying symmetries of the $A_1$ Weyl group.
Later, it was found \cite{BGT3} that the grand partition function satisfies the $q$-Painlev\'e equation $q\text{P}_{\text{III}_3}$, which associates to symmetries of the {\it affine} $A_1$ Weyl group.
Also, all of these relations are generalized to the theory with two NS5-branes and two $(1,k)$5-branes \cite{MN3}, where the spectral operator enjoys symmetries of the $D_5$ Weyl group \cite{KMN} and the grand partition function satisfies the $q$-Painlev\'e equation $q\text{P}_\text{VI}$ associated to the {\it affine} $D_5$ Weyl group \cite{BGKNT,MN9}.
In this paper we would like to point out a big difference of symmetries between non-affine cases and affine ones.
Namely, we study how the affine symmetries work for grand partition functions where only the non-affine symmetries of spectral operators were clarified so far.
These affine symmetries allow us to determine the grand partition functions in the whole infinite parameter space of relative ranks from a finite domain.
Before that, however, we shall explain the similarities between the non-affine symmetries of spectral operators and the affine symmetries of Painlev\'e equations first.

The ABJM theory \cite{ABJM,HLLLP2,ABJ} is the ${\cal N}=6$ super Chern-Simons theory with gauge group $\text{U}(N_1)\times\text{U}(N_2)$, Chern-Simons levels $(k,-k)$ and two pairs of bifundamental matters.
This theory describes the worldvolume theory of M2-branes with $\min(N_1,N_2)$ M2-branes and $|N_2-N_1|$ fractional M2-branes on the background ${\mathbb C}^4/{\mathbb Z}_k$.
This is obtained from the brane configuration of D3-branes on a circle in IIB string theory with an NS5-brane and a $(1,k)$5-brane placed perpendicularly and tilted relatively to preserve supersymmetries.
Using the localization technique \cite{Kapustin:2009kz}, the partition function defined with the infinite-dimensional path integral reduces to a finite-dimensional matrix model.

For the case of equal ranks $N_1=N_2=N$, after obtaining the behavior of $N^{\frac{3}{2}}$ for the free energy \cite{DMP1}, the perturbative corrections in $N^{-1}$ for the partition function are summed up to an Airy function \cite{FHM}.
The Airy function suggests us to move to the grand canonical ensemble by introducing a fugacity dual to the rank.
The grand canonical partition function normalized by the lowest order of the partition function is found to be expressed by the Fredholm determinant of a spectral operator \cite{MP}.
The expression further leads us to the studies of full non-perturbative corrections \cite{HMO2,CM,HMO3}.
Finally, it was found that the grand potential is given by the free energy of topological strings on a background local ${\mathbb P}^1\times{\mathbb P}^1$ \cite{HMMO}.

All these studies extend to the case of non-equal ranks with two formalisms associated respectively to open strings and close strings.
The open string formalism \cite{MM} corrects the Fredholm determinant by multiplying contributions similar to Wilson loops and easily applies to numerical studies and various integrable identities \cite{G,JT,2DTL}, while the closed string formalism is more deep.
The closed string formalism \cite{MS2,MN5,closed} keeps the Fredholm determinant and incorporates the relative rank by changing parameters of the spectral operator \cite{KMZ}.
The spectral operator is regarded as a quantized algebraic curve ${\mathbb P}^1\times{\mathbb P}^1$ and enjoys symmetries of the $A_1$ Weyl group.

The arguments extend to more complicated backgrounds for M2-branes, where the quantized algebraic curve is also generalized.
In this paper we consider the four-node circular-quiver super Chern-Simons theory with gauge group $\text{U}(N_1)_0\times \text{U}(N_2)_k\times \text{U}(N_3)_0\times \text{U}(N_4)_{-k}$
(subscripts denoting the Chern-Simons levels) and bifundamental matters connecting adjacent unitary groups, which is constructed with two NS5-branes and two $(1,k)$5-branes \cite{MNN,MNY,KMN,KM}.
For this theory, the corresponding curve has symmetries of the $D_5$ Weyl group.
Generally, matrix models for brane configurations with $p$ NS5-branes and $q$ $(1,k)$5-branes were also studied \cite{MP,MN1,MN2,Hatsuda:2015lpa} and named as $(p,q)$ models.

Correspondingly, Painlev\'e equations also relate to similar symmetries.
Originally Painlev\'e equations were studied aiming at special functions for non-linear second-order differential equations whose solution does not have movable singularities except poles.
From a modern viewpoint, Painlev\'e equations are classified by algebraic curves \cite{Sakai}.
Explicitly, various $q$-Painlev\'e equations are constructed from exceptional affine Weyl groups.
See \cite{ME8,MY} for quantum representations of the affine $E_8$ Weyl group constructed recently.

With the similarity in mind, it is natural that the grand partition functions of the ABJM theory and the above-mentioned four-node $(2,2)$ model were found to satisfy respectively the $q$-Painlev\'e equations $q\text{P}_{\text{III}_3}$ \cite{BGT3} and $q\text{P}_{\text{VI}}$ \cite{BGKNT,MN9}, which enjoy respectively symmetries of the affine $A_1$ and $D_5$ Weyl groups.
Here we would like to stress, however, that there is a big difference between the Weyl group and the affine Weyl group.
Among other, the affine Weyl group contains elements of discrete translations which are absent in the Weyl group.
Hence, the main project to be studied in this paper is how symmetries of the discrete translations are realized in the grand partition functions.

Related to this project, duality cascades \cite{Elitzur:1997fh,Klebanov:2000hb} in the ABJM theory and its generalizations \cite{Aharony:2009fc,Evslin:2009pk} induced by the Hanany-Witten brane transitions \cite{Hanany:1996ie} were studied in \cite{HK,FMMN,FMS}.
In \cite{FMMN,FMS}, the working hypothesis of duality cascades was proposed as the following steps;
(1) we first select one of the lowest ranks in the brane configuration as a reference and fix it temporally;
(2) we change the reference when we encounter lower ranks in applying the Hanany-Witten transitions arbitrarily without crossing the reference;
(3) we repeat the above step until no lower ranks appear.
Note that here we are allowed to add the overall rank uniformly when we encounter negative ranks since we are considering the supersymmetric grand partition function with a source of the overall rank.
It was then asked, starting from an arbitrary brane configuration, whether duality cascades always end and, if yes, whether the endpoint is uniquely determined.
From the charge conservations, the change of references in duality cascades is realized as discrete translations in the parameter space of relative ranks.
By defining the fundamental domain to be the parameter domain where no more duality cascades are applicable, this question was translated into the question whether the fundamental domain forms a parallelotope (that is, a polytope which can tile the whole parameter space by discrete translations) \cite{FMMN}.
This question was answered positively in \cite{FMS} for general supersymmetric configurations of D3-branes on a circle stretching between perpendicular 5-branes.
Especially, for those with symmetries of Weyl groups, the parallelotopes form the affine Weyl chambers \cite{FMMN}.
Thus, the translations in the affine Weyl group are interpreted physically as duality cascades.
So, our original project on understanding how the grand partition function is transformed under the translations is recast physically into the transformation of the grand partition function under duality cascades.

For the ABJM theory, the transformation rule of the grand partition function was derived directly from the matrix model \cite{HK}.
Interestingly, as we explain in the next section, the transformation rule is compatible with the $q$-Painlev\'e equation $q\text{P}_{\text{III}_3}$ found in \cite{BGT3}.
Namely, although $q\text{P}_{\text{III}_3}$ was originally confirmed only inside the fundamental domain \cite{BGT3}, it also holds exactly on the boundary where the equation relates the grand partition functions in the fundamental domain and those outside obtained by the transformation rule under the translation.
With the transformation rule established, we can further show that $q\text{P}_{\text{III}_3}$ continues to hold outside everywhere in the parameter space.

It is preferable to derive the transformation rule for the $(2,2)$ model also directly from the matrix model as in the ABJM theory and discuss the $q$-Painlev\'e equation $q\text{P}_\text{VI}$ with it.
However, such an approach is difficult for the $(2,2)$ model due to the convergence condition for the integrations in the matrix model.
Indeed, as discussed in \cite{MN9} and some earlier works \cite{Yaakov:2013fza,Nosaka:2017ohr}, the convergence of the matrix model is guaranteed only in a range of relative ranks where the theory is ``good'' in the sense of \cite{Gaiotto:2008ak}.
For this reason, in this paper we explore the project of studying the transformation under duality cascades by overlooking these ambiguities and providing a few studies from slightly different approaches.

First, we trust the validity of analysis of the matrix model in the fundamental domain \cite{MN9} and try to extend it outside using $q\text{P}_{\text{VI}}$.
With this method, we obtain the exact expressions of the matrix model slightly beyond the fundamental domain.
Using these expressions, we can read off the transformation rules under the translations of duality cascades.
Here each transformation rule is supplemented by the change of the overall rank $N$ which is explained from the physical argument of duality cascades.
The rules thus found around the fundamental domain are surprisingly simple (see \cite{MN9} or \eqref{22cascades} later), which we propose to remain valid even far away from the fundamental domain.
Second, alternatively, we can ``derive'' the transformation rules directly from the matrix model.
Despite the ambiguities in the convergence condition, a prescription to rewrite the matrix model into residue sums gives the transformation rules for exchanging 5-branes which duality cascades are composed of.
Interestingly, these two approaches reach the same conclusion.
Namely, the prescription with residue sums precisely reproduces the transformation rules obtained from $q\text{P}_{\text{VI}}$.

Namely, the main claim in this paper is that we can extend the grand partition function outside the fundamental domain of duality cascades naturally by translations in the relative ranks and lifts in the overall rank.
As a result, the grand partition function defined on the whole parameter space of the relative ranks enjoys the $q$-Painlev\'e equations.
The result is, of course, not just a mathematical extension of the solution under the consistency with the $q$-Painlev\'e equations.
From the consistency with the physical argument of duality cascades and the formal derivation from the matrix model, we strongly expect that this solution is a natural extension of the physical partition function of the super Chern-Simons theory.

Let us stress that, while the grand partition function of the ABJM theory was found to satisfy $q\text{P}_{\text{III}_3}$ \cite{BGT3} and its generalization was found to satisfy $q\text{P}_{\text{VI}}$ \cite{MN9}, symmetries of the affine Weyl group was not fully clarified so far.
Also, while duality cascades were studied by proposing a working hypothesis and the fundamental domain after duality cascades was found to be the affine Weyl chamber in \cite{FMMN}, the studies with the grand partition function were missing.
By fully relating them, we find that the grand partition function extends outside the fundamental domain and is compatible with the $q$-Painlev\'e equations.

This paper is organized as follows.
In the next section, we first combine the studies of \cite{BGT3} and \cite{HK} consistently and show that, with the transformation rule of the grand partition function of the ABJM theory under the translation, the $q$-Painlev\'e equation $q\text{P}_{\text{III}_3}$ continues to hold outside the fundamental domain.
Then, in section \ref{22} we apply these viewpoints to the $(2,2)$ model.
We propose the transformation rules of the grand partition function under duality cascades and show that, with the transformation rules, the $q$-Painlev\'e equation $q\text{P}_{\text{VI}}$ continues to hold outside the fundamental domain.
In section \ref{sec_cascadefrommatrixmodel} we propose a prescription to evaluate the partition functions of the $(2,2)$ model outside the fundamental domain, which allows us to derive the transformation rules proposed in section \ref{22}.
In section \ref{translation} we provide further supporting (but not necessarily fully independent) evidences for the cascade relations.
Finally we conclude with some future directions.
In the appendix we list exact expressions of the $(2,2)$ model for $k=1$ and $k=2$.

\section{Affine symmetries for ABJM theory}
\label{sec_ABJM}

In this section we shall explain that the grand partition function of the ABJM theory defined in the fundamental domain extends naturally to the whole parameter space of the relative rank with shifts in the total rank, and show that the resulting transformation rule of the grand partition function under the translation is compatible with the $q\text{P}_{\text{III}_3}$ Painlev\'e equation.

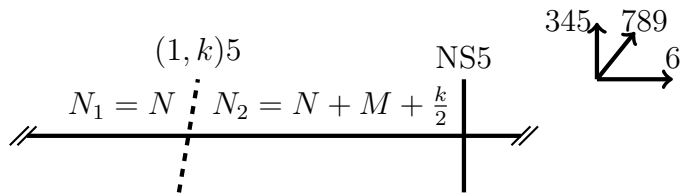
\begin{figure}[!t]
\centering
\begin{tikzpicture}[scale=0.25]
\draw [ultra thick] (0,3) -- (26,3);
\draw [thick] (-0.5,2.5) -- (0.5,3.5);
\draw [thick] (-0.9,2.5) -- (0.1,3.5);
\draw [thick] (25.5,2.5) -- (26.5,3.5);
\draw [thick] (25.9,2.5) -- (26.9,3.5);
\draw [ultra thick, dashed] (8,0) -- (9,6);
\node [above] at (9,6) {$(1,k)5$};
\draw [ultra thick] (23,0) -- (23,6);
\node [above] at (23,6) {$\text{NS}5$};
\node [above] at (5,3.3) {$N_1=N$};
\node [above] at (16,3) {$N_2=N+M+\frac{k}{2}$};
\draw [ultra thick, ->] (30,6) -- (30,9);
\draw [ultra thick, ->] (30,6) -- (34,6);
\draw [ultra thick, ->] (30,6) -- (32,8.5);
\node [above] at (28.5,8) {$345$};
\node [above] at (34,6) {$6$};
\node [above] at (32.5,8) {$789$};
\end{tikzpicture}
\caption{The brane configuration for the ABJM theory.
The solid horizontal line denotes D3-branes, the solid vertical line denotes the NS5-brane, while the dashed tilted line denotes the $(1,k)$5-brane.
Aside from the common directions $012$ for all the branes, D3-branes extend to the direction $6$, with two sets of perpendicular directions denoted respectively by $345$ and $789$.
The NS5-brane extends to the directions of $345$, while the $(1,k)$5-brane extends to the directions of $345$ tilted to those of $789$ by a common angle $\arctan k$.
}
\label{ABJMbraneconfig}
\end{figure}

Before discussing the grand partition function, let us concentrate on the brane configuration first.
The brane configuration of the ABJM theory \cite{ABJM,HLLLP2,ABJ} with gauge group $\text{U}(N_1)_k\times\text{U}(N_2)_{-k}$ consists of D3-branes on a circle with an NS5-brane and a $(1,k)$5-brane placed perpendicularly and tilted relatively by an angle, where the numbers of D3-branes $(N_1,N_2)$ in each interval separated by 5-branes can be different.
See figure \ref{ABJMbraneconfig}.
In \cite{FMMN,FMS} it was proposed to choose the lowest rank as a reference and cut the circle into a segment.
If we denote the cut by a bracket, the NS5-brane by $\bullet$ and the $(1,k)$5-brane by $\circ$ with the number of D3-branes in each interval specified, the brane configuration can be denoted as
\begin{align}
{\langle} N_1\circ N_2\bullet{\rangle}={\langle} N\circ N+M+{\textstyle\frac{k}{2}}\bullet{\rangle},
\label{ABJMbrane}
\end{align}
where we label the overall rank by $N$ and the relative rank by $M$.
Note that the parameterization of the ranks is slightly different from the conventional one ${\langle} N\circ N+L\bullet{\rangle}$ used in literatures.
We shift the relative rank deliberately by $\frac{k}{2}$ so that the Weyl reflection can be denoted simply by $M\to -M$.
Since ranks $N$ and $N+M+\frac{k}{2}$ are integers, $M$ has to be either integers or half-integers depending on whether $k$ is even or odd.
It is known that physics for the brane configurations related by the Hanany-Witten transitions \cite{Hanany:1996ie}
\begin{align}
\langle\cdots K
\begin{array}{c}
\bullet\vspace{-0.4cm}\\
\circ
\end{array}
L
\begin{array}{c}
\circ\vspace{-0.4cm}\\
\bullet
\end{array}
M\cdots\rangle\Rightarrow\langle\cdots K
\begin{array}{c}
\circ\vspace{-0.4cm}\\
\bullet
\end{array}
K-L+M+k
\begin{array}{c}
\bullet\vspace{-0.4cm}\\
\circ
\end{array}
M\cdots\rangle,
\label{HW}
\end{align}
are the same, up to an overall phase factor for the partition function.
We define the fundamental domain of duality cascades by requiring that no lower ranks appear in the Hanany-Witten transitions, which is an interval
\begin{align}
|M|\le\frac{k}{2}.
\label{ABJMFD}
\end{align}
This is a one-dimensional parallelotope tiling the one-dimensional space by translations.
Outside the fundamental domain \eqref{ABJMFD}, $M<-\frac{k}{2}$ or $\frac{k}{2}<M$, the overall rank $N$ and the relative rank $M$ are transformed as $(N,M)\to(N',M')$ \cite{HK} in duality cascades with $(N',M')=(N+M+\frac{k}{2},M+k)$ or $(N',M')=(N-M+\frac{k}{2},M-k)$ respectively.
Forgetting about the overall rank $N$, the transformation is simply the translation of the relative rank $M$, $M\to M\pm k$, which alleviates the breakdown of the inequalities \eqref{ABJMFD}.
The transformation rule for the overall rank $N$ will also play an important role later.

Next, let us turn to the grand partition function.
We introduce the notation corresponding to the brane configuration \eqref{ABJMbrane}
\begin{align}
Z_{k,M}(N)=Z_k(N_1=N,N_2=N+M+{\textstyle\frac{k}{2}}),
\label{ABJMPF}
\end{align}
for the partition function\footnote{
Compared with the standard notation in \cite{DMP1}, we multiply an extra phase factor so that the resulting partition function is real and positive \cite{MM}.
}
\begin{align}
Z_k(N_1,N_2)&=e^{\frac{\pi i}{6k}((N_2-N_1)^3+(N_2-N_1))}
i^{-\frac{1}{2}(N_1^2-N_2^2)}\int\frac{d^{N_1}\mu}{(2\pi)^{N_1}N_1!}\frac{d^{N_2}\nu}{(2\pi)^{N_2}N_2!}
e^{\frac{ik}{4\pi}(\sum_m\mu_m^2-\sum_n\nu_n^2)}\nonumber\\
&\times\frac{\prod_{m<m'}(2\sinh\frac{\mu_m-\mu_{m'}}{2})^2
\prod_{n<n'}(2\sinh\frac{\nu_n-\nu_{n'}}{2})^2}
{\prod_{m=1}^{N_1}\prod_{n=1}^{N_2}(2\cosh\frac{\mu_m-\nu_n}{2})^2}.
\label{ABJMMM}
\end{align}
Note that the integration contour is taken to be the real axis ${\mathbb R}$ for all the variables here and hereafter unless specified.
We also define the grand partition function by introducing a fugacity dual to the overall rank $N$,
\begin{align}
\Xi_{k,M}(\kappa)=\sum_{N=0}^\infty Z_{k,M}(N)\kappa^N.
\label{ABJMGPF}
\end{align}

\begin{figure}[!t]
\centering\includegraphics[scale=0.6]{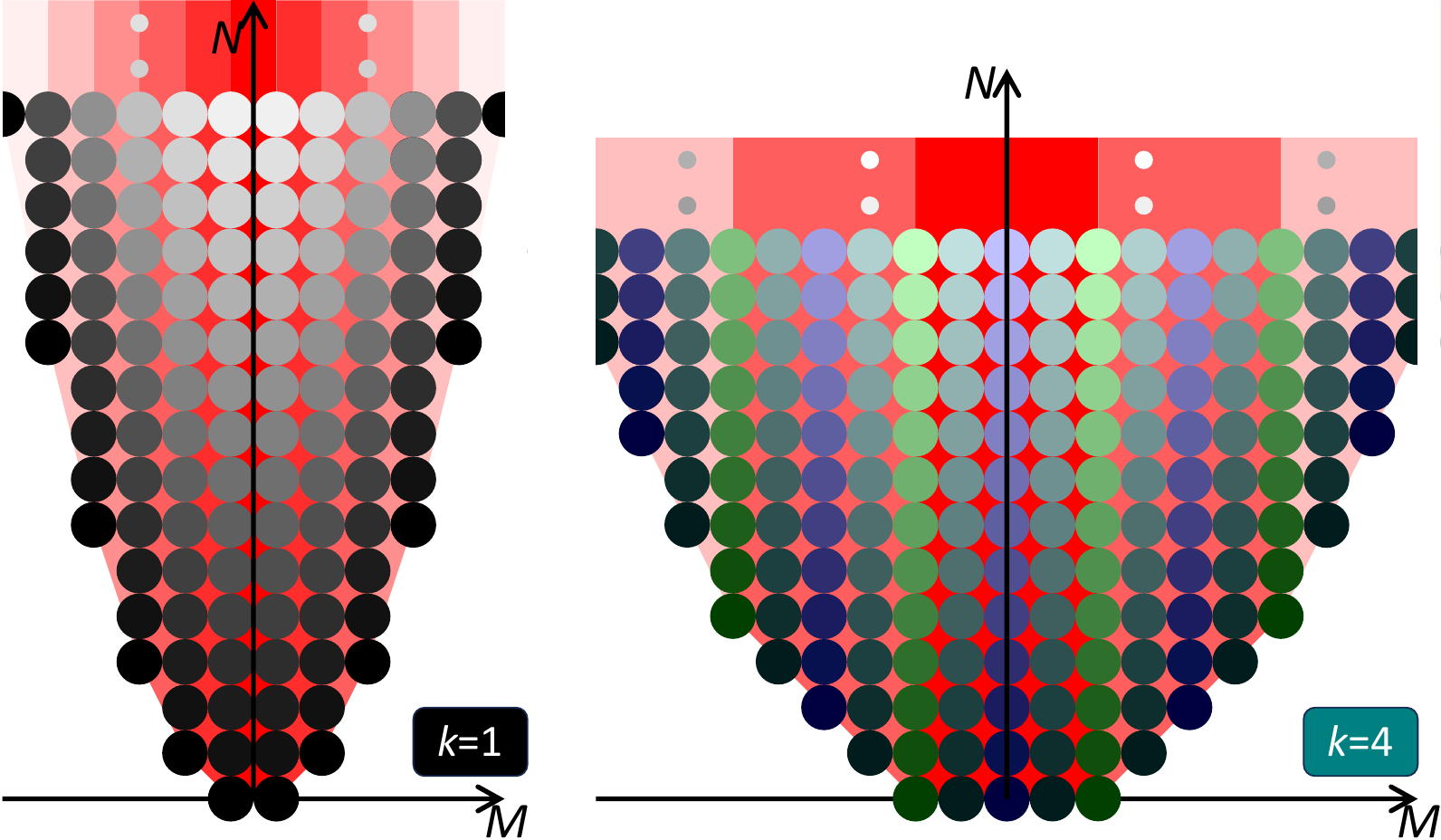}
\caption{Non-vanishing partition functions $Z_{k,M}(N)$ in the grand partition function $\Xi_{k,M}(\kappa)$ of the ABJM theory for $k=1$ (left) and $k=4$ (right).
We plot with a colored circle when the partition function $Z_{k,M}(N)$ is non-vanishing, with different colors schematically representing different values.
The fundamental domain and its copies in the parameter space of the relative rank $M$ are distinguished by red regions with different shades.
All partition functions reduce to those in the fundamental domain of duality cascades uniquely, which implies conversely that the same pattern of colored circles is repeated in different copies of the fundamental domain.
}
\label{ABJMGC}
\end{figure}

From the matrix model \eqref{ABJMMM}, we can study how the partition function transforms under duality cascades and derive \cite{HK}
\begin{align}
Z_{k,M+k}\Bigl(N+M+\frac{k}{2}\Bigr)
=Z_{k,M}(N),
\label{HondaKubo}
\end{align}
which is zero, $Z_{k,M}(N)=0$, if $N<0$ when reducing to the fundamental domain \eqref{ABJMFD} with duality cascades.
This is expected from the transformation of $(N+M+\frac{k}{2},M+k)\to(N,M)$ under duality cascades, since duality cascades do not change physics.
Furthermore, in terms of the grand partition function, \eqref{HondaKubo} implies 
\begin{align}
\Xi_{k,M+k}(\kappa)
=\kappa^{M+\frac{k}{2}}
\Xi_{k,M}(\kappa).
\label{HondaKuboXi}
\end{align}
This transformation rule shows that the partition function $Z_{k,M}(N)$ or the grand partition function $\Xi_{k,M}(\kappa)$ extends naturally outside the fundamental domain \eqref{ABJMFD} by repetition of the same values if the starting point of $N$ for non-vanishing partition functions $Z_{k,M}(N)$ is taken care of.
See figure \ref{ABJMGC} for the lowest value of $N$ for non-vanishing $Z_{k,M}(N)$ and the repetitive pattern of the identical exact values of $Z_{k,M}(N)$.
Since the relations \eqref{HondaKubo}, \eqref{HondaKuboXi} fix the behavior of the partition functions under duality cascades, we refer to them as cascade relations.

On the other hand, the grand partition function of the ABJM theory is known to satisfy \cite{BGT3}
\begin{align}
e^{-\frac{\pi i}{k}M}\Xi_{k,M}(\kappa)^2
+e^{\frac{\pi i}{k}M}\Xi_{k,M}(-\kappa)^2
-\Xi_{k,M+1}(-i\kappa)\Xi_{k,M-1}(i\kappa)=0,
\label{abjmbilinear}
\end{align}
which is nothing but the Hirota bilinear form of the third $q$-Painlev\'e equation $q\text{P}_{\text{III}_3}$.
Note that in \cite{BGT3} the bilinear equation was checked for the grand partition function only when all the grand partition functions are in the fundamental domain \eqref{ABJMFD}, namely, $|M|\le \frac{k}{2}-1$ for \eqref{abjmbilinear} (or at least the extension outside was not mentioned explicitly).

In this section we shall see that the transformation rule \eqref{HondaKuboXi} extends the domain of validity of the bilinear equation \eqref{abjmbilinear} from $|M|\le\frac{k}{2}-1$ to $M=\pm\frac{k}{2}$ first and then to the entire parameter space of $M$ in a consistent way with the same set of exact values of the partition function \eqref{HondaKubo}.
Conversely, one would say that the consistency with the bilinear equation \eqref{abjmbilinear} as well as the physical argument of duality cascades constrains the grand partition function non-trivially so that we can extend the grand partition function outside the fundamental domain uniquely as in \eqref{HondaKuboXi}.
This gives us an insight useful in generalizing the cascade relation \eqref{HondaKuboXi} for the $(2,2)$ model where the direct derivation from the matrix model is difficult.

We stress that, with the cascade relation \eqref{HondaKuboXi}, both the exact values of the partition function and the bilinear equations for the whole infinite parameter space of the relative rank $M$ are determined from those in the finite fundamental domain.
Especially for $k=1$, as we see explicitly below, there is only one independent grand partition function and one independent bilinear equation.
Others are simply generated by \eqref{HondaKuboXi}.

\subsection{$M=\pm\frac{k}{2}$}
First let us consider the $q\text{P}_{\text{III}_3}$ bilinear equation \eqref{abjmbilinear} with $M$ formally set to $M=\pm\frac{k}{2}$.
Although the equation contains the grand partition function outside the fundamental domain $\Xi_{k,\pm(\frac{k}{2}+1)}(\kappa)$, by using the cascade relation \eqref{HondaKuboXi} we can rewrite it in terms of $\Xi_{k,\mp(\frac{k}{2}-1)}(\kappa)$ in the fundamental domain.
Hence, we obtain
\begin{align}
\Xi_{k,-\frac{k}{2}}(\kappa)^2-\Xi_{k,-\frac{k}{2}}(-\kappa)^2
-\kappa\Xi_{k,-\frac{k}{2}+1}(-i\kappa)\Xi_{k,\frac{k}{2}-1}(i\kappa)=0,
\label{ABJMbilinearatM-k/2}
\end{align}
for both the cases of $M=\pm\frac{k}{2}$.
Interestingly, by using the known exact values of $Z_{k,M}(N)$ for $|M|\le \frac{k}{2}$ \cite{Honda:2014npa}, we find that this equation is also satisfied.
Namely, from the viewpoint of the bilinear equation \eqref{abjmbilinear}, the cascade relation \eqref{HondaKubo} at $M=-\frac{k}{2}-1$ and $M=\frac{k}{2}$ can be interpreted as the rule to extend the bilinear equation from $|M|\le \frac{k}{2}-1$ to $|M|\le \frac{k}{2}$ in a consistent way with the actual values of $Z_{k,M}(N)$.

For example, for $k=1$, if we further use the relation $\Xi_{1,\frac{1}{2}}(\kappa)=\Xi_{1,-\frac{1}{2}}(\kappa)$ obtained from \eqref{HondaKuboXi}, the grand partition functions reduce to a single function $\Xi_1(\kappa)=\Xi_{1,\frac{1}{2}}(\kappa)=\Xi_{1,-\frac{1}{2}}(\kappa)$ and we can write \eqref{ABJMbilinearatM-k/2} only in terms of $\Xi_{1}(\kappa)$ as
\begin{align}
\Xi_{1}(\kappa)^2-\Xi_{1}(-\kappa)^2-\kappa\Xi_{1}(-i\kappa)\Xi_{1}(i\kappa)=0,
\label{abjmk1n0}
\end{align}
which is indeed satisfied by
\begin{align}
\Xi_{1}(\kappa)=\sum_{N=0}^\infty Z_{1,\pm\frac{1}{2}}(N)\kappa^N=1+\frac{1}{4}\kappa+\frac{1}{16\pi}\kappa^2+\frac{\pi-3}{64\pi}\kappa^3+\frac{10-\pi^2}{1024\pi^2}\kappa^4+{\cal O}(\kappa^5),
\end{align}
with the exact values of the partition function $Z_{1,\pm\frac{1}{2}}(N)$ \cite{Hatsuda:2012hm} substituted.

It may be interesting to ask whether the bilinear equation \eqref{abjmk1n0} determines the grand partition function at all orders of $\kappa$.
We find that this is only partially possible in the sense that while the next odd order of $\kappa$ is fixed if the lower orders are given, this does not work for the next even order.
Indeed the irrational number $\pi$ in $Z_{1,\pm\frac{1}{2}}(2)=1/(16\pi)$ never appears from the equation \eqref{abjmk1n0} and the values $Z_{1,\pm\frac{1}{2}}(0)=1$ and $Z_{1,\pm\frac{1}{2}}(1)=1/4$.
This might be related to certain degeneracy of the equation.
Interestingly, after the mass deformation in \cite{HLLLP2,Gomis:2008vc}, we obtain a recurrence relation for $Z_{k,M}(N)$ which is non-degenerate at all orders \cite{Nosaka:2020tyv,Nosaka:2024gle}.

\subsection{$|M|>\frac{k}{2}$}
Next let us consider the bilinear equation \eqref{abjmbilinear} for $|M|>\frac{k}{2}$.
When $|M|>\frac{k}{2}$, the relation \eqref{HondaKubo} implies that the lowest value of $N$ for non-vanishing $Z_{k,M}(N)$ is greater than zero.
As a result, the small $\kappa$ expansion of the grand partition function $\Xi_{k,M}(\kappa)$ starts with a non-zero power of $\kappa$ \eqref{HondaKuboXi}.
To understand the general structure, it is useful to rewrite \eqref{HondaKuboXi} as $\Xi_{k,M+k}(\kappa)=\kappa^{\frac{1}{2k}((M+k)^2-M^2)}\Xi_{k,M}(\kappa)$, which further implies
\begin{align}
\Xi_{k,M+nk}(\kappa)=\kappa^{\frac{1}{2k}((M+nk)^2-M^2)}\Xi_{k,M}(\kappa),
\label{ABJMXitransformation}
\end{align}
for $n\in\mathbb{Z}$.

As already discussed in \cite{HK}, the transformation rule for the partition function \eqref{HondaKubo} is motivated by the behavior of the brane configuration under duality cascades induced by the Hanany-Witten transitions of 5-brane exchanges \eqref{HW}, such as
\begin{align}
&\cdots\Rightarrow{\langle}15\bullet 10\circ{\rangle}
\to{\langle} 10\circ 15\bullet{\rangle}
\Rightarrow{\langle} 10\bullet 6\circ{\rangle}
\to{\langle} 6\circ 10\bullet{\rangle}
\Rightarrow{\langle} 6\bullet 3\circ{\rangle}\nonumber\\
&\to{\langle} 3\circ 6\bullet{\rangle}
\Rightarrow{\langle} 3\bullet 1\circ{\rangle}
\to{\langle} 1\circ 3\bullet{\rangle}
\Rightarrow{\langle} 1\bullet 0\circ{\rangle}
\to{\langle} 0\circ 1\bullet{\rangle}
\Rightarrow{\langle} 0\bullet 0\circ{\rangle},
\end{align}
for $k=1$.
Here we have denoted the Hanany-Witten transitions \eqref{HW} by $\Rightarrow$ and the change of references by $\rightarrow$.
Note that $\rightarrow$ preserves the partition function since the partition function \eqref{ABJMMM} is independent of the choice of the reference, while $\Rightarrow$ may change the partition function by an $N$-independent overall factor in principle (see section \ref{sec_cascadefrommatrixmodel} for more details) though this does not happen in the current ABJM case \eqref{HondaKubo}.
From this computation we find out that the general expression for the lowest power of $\kappa$ in the grand partition function $\Xi_{1,M}(\kappa)$ for $k=1$ where non-vanishing partition functions start from is given by
\begin{align}
\Xi_{1,M}(\kappa)=\kappa^{\frac{1}{2}(M^2-\frac{1}{4})}\Xi_{1}(\kappa).
\label{xi1}
\end{align}
It is interesting to observe that, with the extension \eqref{xi1} all of the bilinear equations \eqref{abjmbilinear} with arbitrary $M$ reduce to the single relation \eqref{abjmk1n0}.
Later we will see this property works for general $k$.
For larger $k$, although the general expression for the lowest power of $\kappa$ is more intricate (which involves the residue class, $M$ mod $k$), non-vanishing partition functions continue to start from $N=0$ in the fundamental domain.
From this viewpoint, we may rewrite \eqref{ABJMXitransformation} again as
\begin{align}
\Xi_{k,M}(\kappa)=\kappa^{\frac{1}{2k}(M^2-M_{\text{FD}}^2)}\Xi_{k,M_{\text{FD}}}(\kappa),
\label{ABJMXitransformationwithMFD}
\end{align}
where $M_{\text{FD}}$ is the corresponding relative rank in the fundamental domain which $M$ reduces to through duality cascades.

We would like to claim for general $k$ that, with the cascade relation \eqref{HondaKuboXi}, once the bilinear equation \eqref{abjmbilinear} is satisfied for the relative rank $M$ in the fundamental domain, the equation also works for $M$ outside.
For this purpose, we study how each term of this equation transforms under the translation $M\to M+k$.
Since we have the transformation rule
\begin{align}
\frac{\Xi_{k,M+k}(\kappa)}{\Xi_{k,M}(\kappa)}
=\kappa^{M+\frac{k}{2}},\quad
\frac{e^{\pm\frac{\pi i}{k}(M+k)}}{e^{\pm\frac{\pi i}{k}M}}=-1,
\end{align}
under the translation, each term transforms by the factor
\begin{align}
&-(\kappa^{M+\frac{k}{2}})^2,
\quad-((-\kappa)^{M+\frac{k}{2}})^2,
\quad(-i\kappa)^{M+1+\frac{k}{2}}(i\kappa)^{M-1+\frac{k}{2}},
\end{align}
all of which are identical.
This shows that the bilinear equation is invariant under the translation, of which general duality cascades are composed.

To summarize, we have combined the works on the cascade relation \cite{HK,FMMN,FMS} and the $q\text{P}_{\text{III}_3}$ bilinear equation \cite{BGT3} consistently.
Although the fundamental domain of the ABJM theory in duality cascades was discussed in \cite{HK,FMMN,FMS} and the fundamental domain was found to be a parallelotope, which is a segment in one dimension, the relation to the $q\text{P}_{\text{III}_3}$ bilinear equation was not discussed.
Also, although the grand partition function of the ABJM theory was found to satisfy $q\text{P}_{\text{III}_3}$ in the fundamental domain in \cite{BGT3}, studies on the affine Weyl group were missing.
Here we combine these two aspects of the ABJM theory.
We extend the grand partition function to the whole parameter space of the relative rank $M$ with the cascade relation \eqref{ABJMXitransformationwithMFD} by multiplying suitable powers of the fugacity $\kappa$ determined from duality cascades and claim that it satisfies $q\text{P}_{\text{III}_3}$ for the whole space of the relative rank $M$.

All we have found in this section is not confined to the ABJM theory.
Indeed, we shall see how the results are generalized to the four-node (2,2) model in the next section.

\section{Affine symmetries for four-node $(2,2)$ model}
\label{22}

In the previous section, in extending the domain where the grand partition function of the ABJM theory satisfies the bilinear form of the $q$-Painlev\'e equation $q\text{P}_{\text{III}_3}$ \eqref{abjmbilinear} \cite{BGT3}, we have explained explicitly how the grand partition function transforms under the translation of duality cascades \eqref{HondaKuboXi} and called it the cascade relation.
In this section we shall turn to the study of the cascade relations for the super Chern-Simons theory with gauge group $\text{U}(N_1)_0\times\text{U}(N_2)_{k}\times\text{U}(N_3)_0\times\text{U}(N_4)_{-k}$ (subscripts denoting the Chern-Simons levels) and bifundamental matters, known as the four-node $(2,2)$ model.
Namely, based on the physical argument of duality cascades for the brane configuration and the analysis of the partition function in \cite{MN9} as well as an analogy from the ABJM theory, we propose the cascade relations for the $(2,2)$ model and show the consistency with $q\text{P}_\text{VI}$ with them.

The theory we consider is a generalization of the ABJM theory in the following two senses.
First, the brane configuration for the ABJM theory consists of one NS5-brane and one $(1,k)$5-brane, while that for the above super Chern-Simons theory consists of two of them.
Second, the ABJM theory satisfies the $q\text{P}_{\text{III}_3}$ bilinear equation \cite{BGT3}, while the above super Chern-Simons theory is still integrable and satisfies the 40 $q\text{P}_{\text{VI}}$ bilinear equations \cite{MN9}.
This theory is called the four-node $(2,2)$ model after the brane configuration.

Let us denote the relative ranks by ${\bm M}=(M_0,M_1,M_3)$, introduce the FI parameters ${\bm Z}=(Z_1,Z_3)$ and often further combine the parameters as $\widetilde{\bm M}=(M_0,M_1,M_3,Z_1,Z_3)$.
With these parameters, we consider the brane configuration
\begin{align}
&\langle N_1\mathop{\bullet}^4_0
N_2\mathop{\circ}^2_{-Z_1}
N_3\mathop{\circ}^1_0
N_4\mathop{\bullet}^3_{Z_3}\rangle
=
\langle N\mathop{\bullet}^4_0
N-M_0-M_3+k\mathop{\circ}^2_{-Z_1}
N-M_1-M_3+k\mathop{\circ}^1_0
N+M_0-M_3+k\mathop{\bullet}^3_{Z_3}\rangle\nonumber \\
&=\langle{}\overset{4}{\bullet}
\overset{2}{\circ}
\overset{1}{\circ}
\overset{3}{\bullet}{}\rangle,
\label{nosakaconfig}
\end{align}
with 5-branes labeled and the FI parameters included following the notation of \cite{MN9} (the last abbreviation is used in section \ref{sec_cascadefrommatrixmodel}).
From the localization technique, we define the partition function $Z_{k,\widetilde{\bm M}}(N)=Z_{k,({\bm M},{\bm Z})}(N)$ (or, when the dependence on the FI parameters ${\bm Z}$ is understood tacitly without confusions, simply $Z_{k,{\bm M}}(N)$) for this brane configuration as
\begin{align}
&Z_{k,\widetilde{\bm M}}(N)=\frac{e^{iP_{k,{\widetilde{\bm M}}}(N)}}{N_1!N_2!N_3!N_4!}\int
\frac{d^{N_1}\lambda^{(1)}}{(2\pi)^{N_1}}
\frac{d^{N_2}\lambda^{(2)}}{(2\pi)^{N_2}}
\frac{d^{N_3}\lambda^{(3)}}{(2\pi)^{N_3}}
\frac{d^{N_4}\lambda^{(4)}}{(2\pi)^{N_4}}
e^{\frac{ik}{4\pi}(
\sum_{i=1}^{N_2}(\lambda_i^{(2)})^2
-\sum_{i=1}^{N_4}(\lambda_i^{(4)})^2
)}\nonumber\\
&\quad\times e^{Z_1(\sum_{i=1}^{N_2}\lambda_i^{(2)}-\sum_{i=1}^{N_3}\lambda_i^{(3)})}
e^{-Z_3(\sum_{i=1}^{N_2}\lambda_i^{(4)}-\sum_{i=1}^{N_3}\lambda_i^{(1)})}\nonumber\\
&\quad\times\frac{
\prod_{i<j}^{N_1}(2\sinh\frac{\lambda_i^{(1)}-\lambda_j^{(1)}}{2})^2
\prod_{i<j}^{N_2}(2\sinh\frac{\lambda_i^{(2)}-\lambda_j^{(2)}}{2})^2
}{
\prod_{i=1}^{N_1}\prod_{j=1}^{N_2}2\cosh\frac{\lambda_i^{(1)}-\lambda_j^{(2)}}{2}
\prod_{i=1}^{N_2}\prod_{j=1}^{N_3}2\cosh\frac{\lambda_i^{(2)}-\lambda_j^{(3)}}{2}
}
\nonumber\\
&\quad\times
\frac{
\prod_{i<j}^{N_3}(2\sinh\frac{\lambda_i^{(3)}-\lambda_j^{(3)}}{2})^2
\prod_{i<j}^{N_4}(2\sinh\frac{\lambda_i^{(4)}-\lambda_j^{(4)}}{2})^2
}{
\prod_{i=1}^{N_3}\prod_{j=1}^{N_4}2\cosh\frac{\lambda_i^{(3)}-\lambda_j^{(4)}}{2}
\prod_{i=1}^{N_4}\prod_{j=1}^{N_1}2\cosh\frac{\lambda_i^{(4)}-\lambda_j^{(1)}}{2}
},
\label{ZkM}
\end{align}
where $(N_1,N_2,N_3,N_4)$ are labeled by $N_1=N$, $N_2=N-M_0-M_3+k$, $N_3=N-M_1-M_3+k$, $N_4=N+M_0-M_3+k$ as in \eqref{nosakaconfig} and $e^{iP_{k,{\widetilde {\bm M}}}(N)}$ is \cite{MN9}
\begin{align}
e^{iP_{k,{\widetilde {\bm M}}}(N)}&
=\exp\biggl[\pi i\Bigl(M_0N+\frac{M_0^3-M_0}{3k}+\frac{M_0(M_1^2+M_3^2)}{2k}-2M_0M_3+\frac{3kM_0}{2}-2M_1(Z_1+Z_3)\nonumber \\
&\qquad +\frac{(M_0-M_1)Z_1^2}{k}+\frac{(M_0-M_3+k)Z_3^2}{k}+\frac{(M_0-M_1-M_3-k)Z_1Z_3}{k}\Bigr)\biggr].
\label{PkMN}
\end{align}
Here we have chosen the overall phase $e^{iP_{k,{\widetilde {\bm M}}}(N)}$ in the same way as adopted in \cite{MN9}\footnote{
This phase $e^{iP_{k,{\widetilde {\bm M}}}(N)}$ is defined as $e^{iP_{k,{\widetilde {\bm M}}}(N)}=e^{i(\Theta_{k,{\widetilde {\bm M}}}-\Theta'_{k,{\widetilde {\bm M}}})}
i^{-\frac{N_2^2}{2}+\frac{N_4^2}{2}}$, with $N_2=N-M_0-M_3+k$ and $N_4=N+M_0-M_3+k$.
The phases $\Theta_{k,{\widetilde {\bm M}}}$ and $\Theta_{k,{\widetilde {\bm M}}}'$ are given by (2.16) in \cite{MN9}.
Here the power of $i$ comes from the phase introduced by (2.3) in \cite{MN9} associated with the non-zero Chern-Simons levels.
}
so that the coefficients of the $q\text{P}_{\text{VI}}$ bilinear equations simplify as \eqref{D5bilinear} below.
We shall propose the cascade relations for the $(2,2)$ model from the physical argument of duality cascades for the brane configuration \eqref{nosakaconfig} and the analysis of the partition function \eqref{ZkM}.

\begin{figure}[!t]
\centering\includegraphics[scale=0.6]{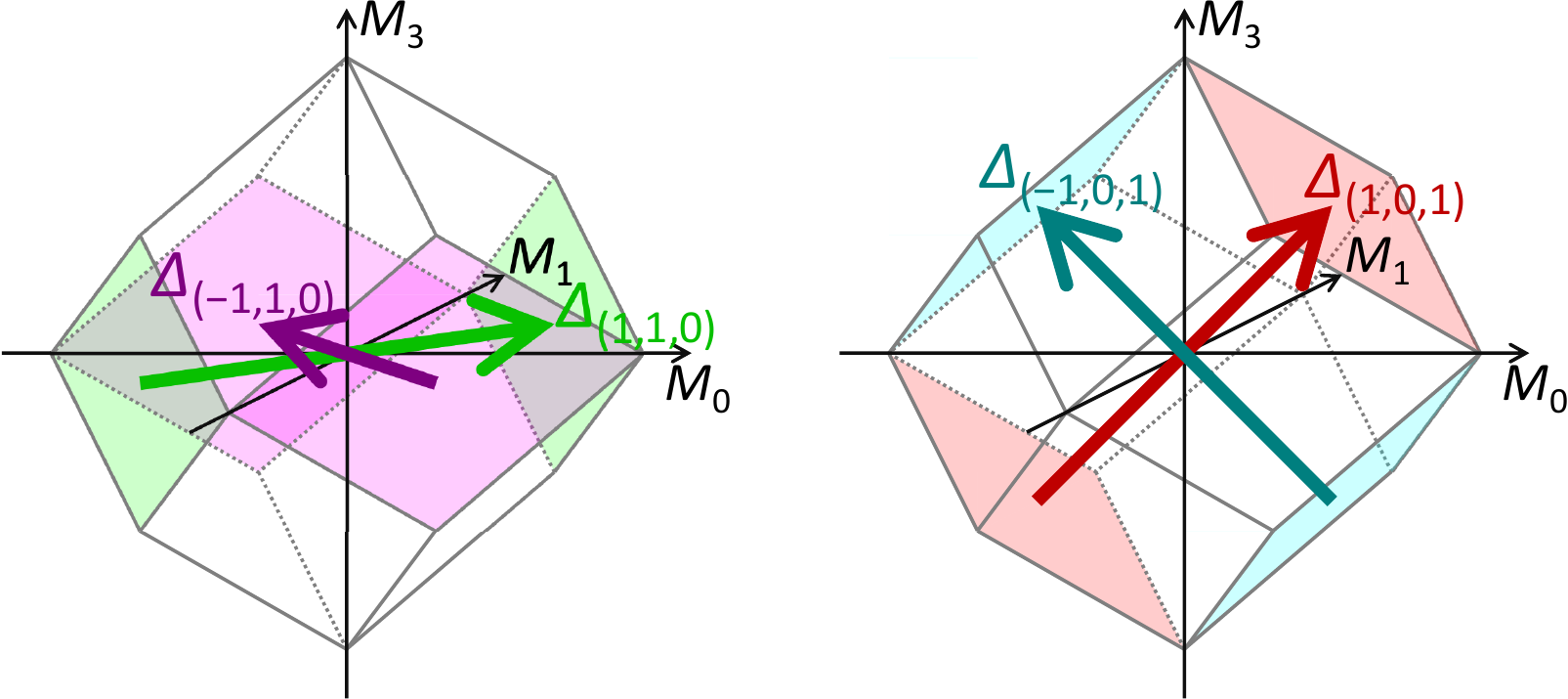}
\caption{The directions of the translations $\Delta_{(\pm 1,1,0)}$ and $\Delta_{(\pm 1,0,1)}$ are given respectively in the left and right figures.
The rhombic dodecahedron is a parallelotope, which tiles the entire three-dimensional space of the relative ranks ${\bm M}$ by the discrete translations and the directions of the translations are those which map one face into the opposite one.
The directions of $\Delta_{(1,1,0)}$, $\Delta_{(-1,1,0)}$, $\Delta_{(1,0,1)}$ and $\Delta_{(-1,0,1)}$ are represented by the green, magenta, red and cyan arrows respectively.
}
\label{22translationdirection}
\end{figure}

As discussed in \cite{KMN}, the grand partition function normalized by the partition function at $N=0$, or more generally that at the lowest value of $N$ for non-vanishing partition functions, enjoys symmetries of the $D_5$ Weyl group generated by
\begin{align}
s_1: M_3\leftrightarrow Z_3,\quad
s_2: M_3\leftrightarrow -Z_3,\quad
s_3: M_0\leftrightarrow M_3,\quad
s_4: M_0\leftrightarrow -M_1,\quad
s_5: M_1\leftrightarrow Z_1,
\end{align}
part of which is associated with the 5-brane exchange in the Hanany-Witten transition.
Note that here we adopt the same terminology of the Hanany-Witten transition even for the exchange of 5-branes of the same kind with no extra D3-branes generated,
\begin{align}
\langle\cdots K
\begin{array}{c}
\bullet\vspace{-0.4cm}\\
\circ
\end{array}
L
\begin{array}{c}
\bullet\vspace{-0.4cm}\\
\circ
\end{array}
M\cdots\rangle\Rightarrow\langle\cdots K
\begin{array}{c}
\bullet\vspace{-0.4cm}\\
\circ
\end{array}
K-L+M
\begin{array}{c}
\bullet\vspace{-0.4cm}\\
\circ
\end{array}
M\cdots\rangle.
\end{align}
Also, as in the case for the ABJM theory, in \cite{FMMN} duality cascades for this theory were discussed and the fundamental domain was found to be
\begin{align}
\pm M_0\pm M_1\le k,\quad
\pm M_0\pm M_3\le k,\quad
\pm M_1\pm M_3\le k.
\label{22FD}
\end{align}
Namely, after duality cascades, the relative rank ${\bm M}$ reduces to a point in the fundamental domain \eqref{22FD} uniquely. 
In duality cascades, references are changed when lower ranks compared with the reference appear.
These processes are realized by discrete translations in the parameter space of the relative ranks.
Especially, when lower ranks appear, one of the inequalities for the fundamental domain \eqref{22FD} breaks and the breakdown is alleviated by the discrete translation of changing references.
Depending on various intervals of lower ranks, there are various directions for the translations.
Nevertheless, all directions are compatible in the parameter space of the relative ranks \cite{FMS}, which guarantees that, after duality cascades, the relative rank reduces to a point in the fundamental domain \eqref{22FD} uniquely.
Since the parameter space is three-dimensional, only three of the translations are independent and the following four \cite{FMMN,MN9}
\begin{align}
\Delta_{(\pm 1,1,0)}&:(N,M_0,M_1,M_3)\mapsto(N\pm M_0+M_1+k,M_0\pm k,M_1+k,M_3),\nonumber\\
\Delta_{(\pm 1,0,1)}&:(N,M_0,M_1,M_3)\mapsto(N\pm M_0+M_3+k,M_0\pm k,M_1,M_3+k),
\label{Deltas}
\end{align}
are enough for our current study of how the grand partition function transforms under the translations.
The directions of the translations are depicted in figure \ref{22translationdirection}.

In \cite{MN9,BGKNT} it was found that the grand partition functions in the fundamental domain \eqref{22FD} satisfy the 40 bilinear equations
\begin{align}
&e^{-\frac{\pi i}{2k}(\sigma_cc+\sigma_dd+\sigma_ee)}S_{{\widetilde {\bm M}}}^{(1)}\Xi^\text{B}_{a_\pm b_\pm c_\varnothing d_\varnothing e_\varnothing}(\kappa)
+e^{\frac{\pi i}{2k}(\sigma_cc+\sigma_dd+\sigma_ee)}S_{{\widetilde {\bm M}}}^{(2)}\Xi^\text{B}_{a_\pm b_\mp c_\varnothing d_\varnothing e_\varnothing}(-\kappa)\nonumber\\
&\quad+S_{{\widetilde {\bm M}}}^{(3)}\Xi^\text{B}_{a_\varnothing b_\varnothing c_{\pm\sigma_c} d_{\pm\sigma_d} e_{\pm\sigma_e}}(\mp i\kappa)=0.
\label{D5bilinear}
\end{align}
Each of the 40 bilinear equations is characterized by two variables out of five including the relative ranks and the FI parameters $\widetilde{\bm M}=(M_0,M_1,M_3,Z_1,Z_3)$ along with patterns of shifts in the remaining three variables, $(a,b;\sigma_c,\sigma_d,\sigma_e)$.
Here $(a,b,c,d,e)$ denotes a permutation of $\{M_0,M_1,M_3,Z_1,Z_3\}$ and $(\sigma_c,\sigma_d,\sigma_e)$ is a pattern of shifts $(1,1,1)$, $(1,-1,-1)$, $(-1,1,-1)$, $(-1,-1,1)$.
In \eqref{D5bilinear} various bilinears $\Xi^\text{B}$ are defined as
\begin{align}
&\Xi^\text{B}_{a_\pm b_\pm c_\varnothing d_\varnothing e_\varnothing}(\kappa)=\prod_{\pm}\Xi_{k,{\widetilde {\bm M}}_\alpha\pm\frac{1}{2}(\delta^a_\alpha+\delta^b_\alpha)}(\kappa),\quad
\Xi^\text{B}_{a_\pm b_\mp c_\varnothing d_\varnothing e_\varnothing}(-\kappa)=\prod_{\pm}\Xi_{k,{\widetilde {\bm M}}_\alpha\pm\frac{1}{2}(\delta^a_\alpha-\delta^b_\alpha)}(-\kappa),\nonumber\\
&\Xi^\text{B}_{a_\varnothing b_\varnothing c_{\pm\sigma_c} d_{\pm\sigma_d} e_{\pm\sigma_e}}(\mp i\kappa)=\prod_{\pm}\Xi_{k,{\widetilde {\bm M}}_\alpha\pm\frac{1}{2}(\sigma_c\delta^c_\alpha+\sigma_d\delta^d_\alpha+\sigma_e\delta^e_\alpha)}(\mp i\kappa),
\label{22bilinear}
\end{align}
and the prefactors are ($\bar 1=3,\bar 3=1$)
\begin{align}
&(S^{(1)}_{\widetilde {\bm M}},S^{(2)}_{\widetilde {\bm M}},S^{(3)}_{\widetilde {\bm M}})=(S_i^+,S_i^-,S^{\sigma_{M_0}}_{\bar i}),\quad\text{for}\;(a,b)=(M_i,Z_i),\nonumber\\
&(S^{(1)}_{\widetilde {\bm M}},S^{(2)}_{\widetilde {\bm M}},S^{(3)}_{\widetilde {\bm M}})=(1,1,S^{\sigma_{Z_i}}_{\bar i}),\quad\text{for}\;(a,b)=(M_0,M_i),\nonumber\\
&(S^{(1)}_{\widetilde {\bm M}},S^{(2)}_{\widetilde {\bm M}},S^{(3)}_{\widetilde {\bm M}})=(1,1,S^{\sigma_{M_i}}_{\bar i}),\quad\text{for}\;(a,b)=(M_0,Z_i),\nonumber\\
&(S^{(1)}_{\widetilde {\bm M}},S^{(2)}_{\widetilde {\bm M}},S^{(3)}_{\widetilde {\bm M}})=(1,1,1),\quad\text{otherwise},
\label{prefactor}
\end{align}
with
\begin{align}
S_i^\pm=2\sin\frac{\pi(M_i\pm Z_i)}{k}.
\end{align}
More explicitly when necessary, the readers can always consult tables 2 and 3 in \cite{MN9} where the 40 bilinear equations are given line-by-line (in terms of the lowest partition functions).
These bilinear equations are associated to the $q$-Painlev\'e equations $q\text{P}_\text{VI}$.

By tracking the change of $N$ under duality cascades \eqref{Deltas}, we propose the general expression of the cascade relation
\begin{align}
\frac{\Xi_{k,{\bm M}}(\kappa)}{Z_{k,{\bm M}}}
=\kappa^{\frac{1}{2k}({\bm M}^2-{\bm M}_\text{FD}^2)}
\frac{\Xi_{k,{\bm M}_\text{FD}}(\kappa)}{Z_{k,{\bm M}_{\text{FD}}}}.
\label{XioverZcascaderule22model}
\end{align}
Hereafter we suppress the dependences on the FI parameters ${\bm Z}$ for the (grand) partition functions as they do not change under duality cascades.
\eqref{XioverZcascaderule22model} relates the grand partition function for ${\bm M}=(M_0,M_1,M_3)$ outside the fundamental domain to that for ${\bm M}_{\text{FD}}$ inside which is the corresponding relative rank of ${\bm M}$ obtained after applying duality cascades.
Here $Z_{k,{\bm M}}$ is the non-vanishing partition function $Z_{k,{\bm M}}(N)$ of the lowest rank $N$ for given ${\bm M}$, while $Z_{k,{\bm M}_{\text{FD}}}$ is the partition function at $N=0$ in the fundamental domain.

The expression \eqref{XioverZcascaderule22model} is analogous to the cascade relation for the ABJM theory \eqref{ABJMXitransformationwithMFD}.
Note, however, that although no extra factors appear in the cascade relation for the grand partition function of the ABJM theory \eqref{HondaKubo}, this does not work for the current $(2,2)$ model in general.
Unfortunately, the argument of duality cascades provides only the power of $\kappa$ in the cascade relation \eqref{XioverZcascaderule22model}.
Nevertheless, we propose that \eqref{XioverZcascaderule22model}
is valid if we normalize by the lowest non-vanishing partition function.
Still we need to determine the concrete expression for the ratio between $Z_{k,{\bm M}}$ and $Z_{k,{\bm M}_{\text{FD}}}$.
In \cite{BGKNT,MN9} an explicit determinant expression for $Z_{k,{\bm M}}(0)$ without integration was found which is valid at least in a subdomain of the fundamental domain \eqref{22FD}.
Assuming that the expression applies to the whole fundamental domain, we can evaluate the ratio when both of ${\bm M}$ before and after the translation are on the boundary of the fundamental domain.
Furthermore, by assuming that the $q\text{P}_{\text{VI}}$ bilinear equations are valid on the boundary where the equations involve both $\Xi_{k,\bm M}(\kappa)$ in the fundamental domain and those outside, we can also determine $Z_{k,{\bm M}}$ when ${\bm M}$ is slightly outside.
These analyses were done in \cite{MN9} for $k=1$ (and also for $k=2$ only on the boundaries of the fundamental domain).
From \eqref{XioverZcascaderule22model} with these results substituted, we finally propose that duality cascades cause the changes of the grand partition function as
\begin{align}
&\Delta_{(1,1,0)}\Xi=\frac{\Xi_{k,(M_0+k,M_1+k,M_3)}(\kappa)}{\Xi_{k,(M_0,M_1,M_3)}(\kappa)}=e^{-2\pi ik(Z_1+Z_3)}f_{M_1}(Z_1)\kappa^{M_0+M_1+k},\nonumber \\
&\Delta_{(-1,1,0)}\Xi=\frac{\Xi_{k,(M_0-k,M_1+k,M_3)}(\kappa)}{\Xi_{k,(M_0,M_1,M_3)}(\kappa)}=e^{-2\pi ik(Z_1+Z_3)}f_{M_1}(Z_1)\kappa^{-M_0+M_1+k},\nonumber \\
&\Delta_{(1,0,1)}\Xi=\frac{\Xi_{k,(M_0+k,M_1,M_3+k)}(\kappa)}{\Xi_{k,(M_0,M_1,M_3)}(\kappa)}=f_{M_3}(Z_3)\kappa^{M_0+M_3+k},\nonumber \\
&\Delta_{(-1,0,1)}\Xi=\frac{\Xi_{k,(M_0-k,M_1,M_3+k)}(\kappa)}{\Xi_{k,(M_0,M_1,M_3)}(\kappa)}=f_{M_3}(Z_3)\kappa^{-M_0+M_3+k},
\label{22cascades}
\end{align}
where
\begin{align}
f_M(Z)=(e^{\pi iZ}-(-1)^{2M}e^{-\pi iZ})^{-2M}
=\begin{cases}
(2\cos\pi Z)^{-2M},&\text{ for }M\in{\mathbb Z}+\frac{1}{2},\\
(-4\sin^2\pi Z)^{-M},&\text{ for }M\in{\mathbb Z}.
\end{cases}
\label{fM}
\end{align}
We shall postpone to section \ref{sec_cascadefrommatrixmodel} a formal derivation from the matrix model for the cascade relations \eqref{22cascades} with \eqref{fM}.

In the next subsection we explain how the cascade relations guarantee that the grand partition function satisfying the $q\text{P}_\text{VI}$ bilinear equations \eqref{D5bilinear} in the fundamental domain extends to that satisfying the same equations in the whole parameter space.

\subsection{Bilinear equations outside fundamental domain}
\label{inv}

In this subsection, we would like to show that, with the cascade relations \eqref{22cascades} given, once it was found that the grand partition function satisfies the 40 bilinear equations \eqref{D5bilinear} in the fundamental domain, it satisfies the same bilinear equations in the whole parameter space of the relative ranks.

We shall explain with an example of the bilinear equation labeled by $(a,b;\sigma_c,\sigma_d,\sigma_e)=(M_0,M_1;+,+,+)$,
\begin{align}
&e^{-\frac{\pi i}{2k}(M_3+Z_1+Z_3)}\Xi_{k,(M_0+\frac{1}{2},M_1+\frac{1}{2},M_3,Z_1,Z_3)}(\kappa)
\Xi_{k,(M_0-\frac{1}{2},M_1-\frac{1}{2},M_3,Z_1,Z_3)}(\kappa)\nonumber\\
&\quad
+e^{\frac{\pi i}{2k}(M_3+Z_1+Z_3)}
\Xi_{k,(M_0+\frac{1}{2},M_1-\frac{1}{2},M_3,Z_1,Z_3)}(-\kappa)
\Xi_{k,(M_0-\frac{1}{2},M_1+\frac{1}{2},M_3,Z_1,Z_3)}(-\kappa)\nonumber\\
&\quad
+2\sin\frac{\pi(M_3+Z_3)}{k}
\Xi_{k,(M_0,M_1,M_3+\frac{1}{2},Z_1+\frac{1}{2},Z_3+\frac{1}{2})}(-i\kappa)
\Xi_{k,(M_0,M_1,M_3-\frac{1}{2},Z_1-\frac{1}{2},Z_3-\frac{1}{2})}(i\kappa)=0.
\label{01+++}
\end{align}
By transforming with $\Delta_{(\pm 1,1,0)}$, each of the three terms in \eqref{01+++} transforms as
\begin{align}
f_{M_1+\frac{1}{2}}(Z_1)f_{M_1-\frac{1}{2}}(Z_1),\quad
f_{M_1+\frac{1}{2}}(Z_1)f_{M_1-\frac{1}{2}}(Z_1),\quad
f_{M_1}({\textstyle Z_1+\frac{1}{2}})f_{M_1}({\textstyle Z_1-\frac{1}{2}}),
\end{align}
while with $\Delta_{(\pm 1,0,1)}$, each term transforms as
\begin{align}
(-i)f_{M_3}(Z_3)^2,\quad
(i)f_{M_3}(Z_3)^2(-1),\quad
(-1)f_{M_3+\frac{1}{2}}({\textstyle Z_3+\frac{1}{2}})f_{M_3-\frac{1}{2}}({\textstyle Z_3-\frac{1}{2}})(-i).
\end{align}
Note that the phase factors (omitted if it is $1$) prior to the function $f_M(Z)$ come from those in the bilinear equations, while those behind $f_M(Z)$ are from the powers of $\kappa$.
Thus, the validity of the bilinear equations after the transformations relies on the relations
\begin{align}
\psi^\pm_i=\frac{f_{M_i+\frac{1}{2}}(Z_i\pm\frac{1}{2})f_{M_i-\frac{1}{2}}(Z_i\mp\frac{1}{2})}{f_{M_i}(Z_i)^2}=-1,\quad
\phi_i=\frac{f_{M_i}(Z_i+\frac{1}{2})f_{M_i}(Z_i-\frac{1}{2})}{f_{M_i+\frac{1}{2}}(Z_i)f_{M_i-\frac{1}{2}}(Z_i)}=1,
\label{phipsi}
\end{align}
which can be shown explicitly with \eqref{fM}.
Although we have picked up only an example of $(M_0,M_1;+,+,+)$, the same computations work for all other examples with different choices of variables $(a,b)$ and shifts $(\sigma_c,\sigma_d,\sigma_e)$.
In table \ref{bilinearphipsi}, we list the relations on which the validity of each bilinear equation under each transformation relies.
Since the relations do not depend on the shift patterns $(\sigma_c,\sigma_d,\sigma_e)$, we omit them in the table.
It is interesting to find that the validity of the 40 bilinear equations after applying the four translations relies only on a few relations \eqref{phipsi}.
To summarize, all of the 40 bilinear equations continue to be valid under the discrete translations in all of the directions.
For this reason, the cascade relations \eqref{22cascades} guarantee that, if the 40 $q\text{P}_\text{VI}$ bilinear equations hold in the fundamental domain, they are also satisfied outside.

\begin{table}[!t]
\begin{center}
\begin{tabular}{c|cc}
$q\text{P}_\text{VI}\;\backslash\;\text{transl.}$&$\Delta_{(\pm 1,1,0)}$&$\Delta_{(\pm 1,0,1)}$\\\hline
$(M_1,M_3)$&$\phi_1=1$&$\phi_3=1$\\
$(M_1,Z_1)$&$\psi^\pm_1=-1$&$\psi^\pm_3=-1$\\
$(M_1,Z_3)$&$\phi_1=1$&$\phi_3=1$\\
$(M_3,Z_1)$&$\phi_1=1$&$\phi_3=1$\\
$(M_3,Z_3)$&$\psi^\pm_1=-1$&$\psi^\pm_3=-1$\\
$(Z_1,Z_3)$&$\phi_1=1$&$\phi_3=1$
\end{tabular}\qquad\begin{tabular}{c|cc}
$q\text{P}_\text{VI}\;\backslash\;\text{transl.}$&$\Delta_{(\pm 1,1,0)}$&$\Delta_{(\pm 1,0,1)}$\\\hline
$(M_0,M_1)$&$\phi_1=1$&$\psi^\pm_3=-1$\\
$(M_0,M_3)$&$\psi^\pm_1=-1$&$\phi_3=1$\\
$(M_0,Z_1)$&$\phi_1=1$&$\psi^\pm_3=-1$\\
$(M_0,Z_3)$&$\psi^\pm_1=-1$&$\phi_3=1$
\end{tabular}
\end{center}
\caption{Relations \eqref{phipsi} on which the validity of each bilinear equation \eqref{22bilinear} labeled by $(a,b;\sigma_c,\sigma_d,\sigma_e)$ under each transformation \eqref{22cascades} relies.
Since the computations do not depend on the shifts $(\sigma_c,\sigma_d,\sigma_e)$, we label the bilinear equations only by the combinations of the variables $(a,b)$.}
\label{bilinearphipsi}
\end{table}

\section{Cascade relations from matrix model}
\label{sec_cascadefrommatrixmodel}

In this section, we shall provide a formal derivation for the cascade relations of the grand partition function \eqref{22cascades}.
In the case of the ABJM theory, the cascade relation under the translation $M\rightarrow M+k$ \eqref{HondaKuboXi} follows directly from the matrix model $Z_{k,M}(N)$ \eqref{ABJMMM} as shown in \cite{HK}.
It would be natural to ask whether we can understand the cascade relations for the four-node $(2,2)$ model \eqref{22cascades} from the original matrix model $Z_{k,{\bm M}}(N)$ \eqref{ZkM}.
As in the case of the ABJM theory and as we see explicitly below, since the translations of ${\bm M}$ are realized by moving one of the 5-branes in the IIB brane setup along the compactified direction cyclically \cite{FMMN}, we naturally expect that the cascade relations \eqref{22cascades} can be derived from the transformation rules of the partition function under 5-brane exchanges.

Note that in contrast with the case of the ABJM theory it is a subtle issue whether the partition function of the $(2,2)$ model is even well-defined or not outside the fundamental domain.
The exchange of 5-branes of the same kind may change the rank into that of ``bad'' theories \cite{Gaiotto:2008ak}, where the partition function becomes divergent.
The divergence is due to the integrand without an exponential decay or a Fresnel oscillation at infinity, and is not resolved by the FI parameters.
Nevertheless, in the following we observe that, if we adopt a prescription by formally replacing the integrations with residue sums for the exchange of 5-branes of the same kind, we can precisely reproduce from the matrix model the cascade relations \eqref{22cascades} required for the consistency of the $q\text{P}_{\text{VI}}$ bilinear equations \eqref{D5bilinear}.
In this sense, rather than a derivation of \eqref{22cascades}, our results in this section may be viewed as a proposal for an extension of the matrix model with the prescription of residue sums which reproduces \eqref{22cascades}.
As a result, this extension is consistent with the physical argument of duality cascades and the validity of the $q\text{P}_{\text{VI}}$ bilinear equations.

To make the argument precise, let us first introduce a new notation for the matrix model which is directly related to the brane setup.
To avoid a confusion with the notation in the previous sections, in this section we shall denote by $W$ the unintegrated partition function with all of the extra phases dropped for a subset of the brane configuration.
For the brane configuration ${}_\mu\langle N_1\begin{matrix}
\phantom{\mathop{}_{Z}}\vspace{-0.4cm}\\
\bullet\vspace{-0.4cm}\\
\circ\vspace{-0.4cm}\\
\mathop{}_{Z}
\end{matrix}
N_2\rangle_\nu$ where only an NS5-brane $\bullet$ or a $(1,k)$5-brane $\circ$ with the FI parameter $Z$ separates $N_1$ and $N_2$ D3-branes, we associate the building blocks
\begin{align}
& W\Big({}_\mu\langle N_1\mathop{\bullet}_{Z} N_2\rangle_\nu\Big)
=e^{-Z(\sum_{i=1}^{N_1}\mu_i-\sum_{i=1}^{N_2}\nu_i)}\frac{
\prod_{i<j}^{N_1}2\sinh\frac{\mu_i-\mu_j}{2}
\prod_{i<j}^{N_2}2\sinh\frac{\nu_i-\nu_j}{2}
}{
\prod_{i=1}^{N_1}
\prod_{j=1}^{N_2}
2\cosh\frac{\mu_i-\nu_j}{2}},\nonumber \\
&
W\Big({}_\mu\langle N_1\mathop{\circ}_{Z} N_2\rangle_\nu\Big)
=e^{-Z(\sum_{i=1}^{N_1}\mu_i-\sum_{i=1}^{N_2}\nu_i)}e^{\frac{ik}{4\pi}(\sum_{i=1}^{N_1}\mu_i^2-\sum_{i=1}^{N_2}\nu_i^2)}\frac{
\prod_{i<j}^{N_1}2\sinh\frac{\mu_i-\mu_j}{2}
\prod_{i<j}^{N_2}2\sinh\frac{\nu_i-\nu_j}{2}
}{
\prod_{i=1}^{N_1}
\prod_{j=1}^{N_2}
2\cosh\frac{\mu_i-\nu_j}{2}}.
\label{bracketW2}
\end{align}
The brane configurations corresponding to the above building blocks are summarized in figure \ref{braketvsbranes}.
From them we can define the unintegrated partition function for a general brane configuration as
\begin{align}
&W\Big({}_\mu
\langle
N_1
\begin{matrix}
\phantom{\mathop{}_{Z}}\vspace{-0.4cm}\\
\bullet\vspace{-0.4cm}\\
\circ\vspace{-0.4cm}\\
\mathop{}_{Z_1}
\end{matrix}
N_2
\cdots
N_{\ell-1}
\begin{matrix}
\phantom{\mathop{}_{Z}}\vspace{-0.4cm}\\
\bullet\vspace{-0.4cm}\\
\circ\vspace{-0.4cm}\\
\mathop{}_{Z_{\ell-1}}
\end{matrix}
N_{\ell}
\rangle_\nu\Big)=
\frac{1}{N_2!\cdots N_{\ell-1}!}
\int
\frac{d^{N_2}\lambda^{(2)}_i}{(2\pi)^{N_2}}
\cdots
\frac{d^{N_{\ell-1}}\lambda^{(\ell-1)}_i}{(2\pi)^{N_{\ell-1}}}\nonumber\\
&\quad\quad\times
W\Big({}_\mu
\langle
N_1
\begin{matrix}
\phantom{\mathop{}_{Z}}\vspace{-0.4cm}\\
\bullet\vspace{-0.4cm}\\
\circ\vspace{-0.4cm}\\
\mathop{}_{Z_1}
\end{matrix}
N_2
\rangle_{\lambda^{(2)}}\Big)
\cdots
W\Big({}_{\lambda^{(\ell-1)}}
\langle
N_{\ell-1}
\begin{matrix}
\phantom{\mathop{}_{Z}}\vspace{-0.4cm}\\
\bullet\vspace{-0.4cm}\\
\circ\vspace{-0.4cm}\\
\mathop{}_{Z_{\ell-1}}
\end{matrix}
N_\ell
\rangle_{\nu}\Big).
\label{Wunint}
\end{align}

\begin{figure}[!t]
\begin{center}
\begin{align*}
&{}_\mu\langle N_1\mathop{\bullet}_{Z} N_2\rangle_\nu
=\mbox{\raisebox{-1.35cm}{
\begin{tikzpicture}[scale=0.25]
\draw [ultra thick] (0,3) -- (4,3);
\draw [ultra thick] (2,0) -- (2,6);
\node [above] at (0.5,3) {$N_1$};
\node [below] at (0.5,3) {$\mu_i$};
\node [above] at (3.5,3) {$N_2$};
\node [below] at (3.5,3) {$\nu_i$};
\node [above] at (2,6) {NS5};
\node [below] at (2,0) {$-iZ$};
\end{tikzpicture}
}
}\qquad\qquad
{}_\mu\langle N_1\mathop{\circ}_{Z} N_2\rangle_\nu
=\mbox{\raisebox{-1.35cm}{
\begin{tikzpicture}[scale=0.25]
\draw [ultra thick] (0,3) -- (4,3);
\draw [ultra thick,dashed] (1.5,0) -- (2.5,6);
\node [above] at (0.5,3) {$N_1$};
\node [below] at (0.5,3) {$\mu_i$};
\node [above] at (3.5,3) {$N_2$};
\node [below] at (3.5,3) {$\nu_i$};
\node [above] at (3,5.6) {$(1,k)$5};
\node [below] at (1,0) {$-iZ$};
\end{tikzpicture}
}
}
\end{align*}\vspace{-1cm}
\caption{The brane configurations corresponding to ${}_\mu\langle N_1\begin{matrix}
\phantom{\mathop{}_{Z}}\vspace{-0.4cm}\\
\bullet\vspace{-0.4cm}\\
\circ\vspace{-0.4cm}\\
\mathop{}_{Z}
\end{matrix}
N_2\rangle_\nu$ whose unintegrated partition functions are given in \eqref{bracketW2}.
}
\label{braketvsbranes}
\end{center}
\end{figure}

Similarly, let us introduce $W_{k,{\bm M}}(N)$ for the configuration \eqref{nosakaconfig} with phases dropped from the partition function $Z_{k,{\bm M}}(N)$ \eqref{ZkM} which is used for the bilinear equations \eqref{D5bilinear},
\begin{align}
&Z_{k,{\bm M}}(N)=e^{iP_{k,{\bm M}}(N)}W_{k,{\bm M}}(N),
\label{convert}
\end{align}
where $e^{iP_{k,{\bm M}}(N)}$ is given by \eqref{PkMN}.
We have referred to \eqref{bracketW2} as building blocks, since $W_{k,{\bm M}}(N)$ is constructed from \eqref{bracketW2}.
Note that the cascade relations \eqref{22cascades} are obtained as long as the local transformation rules for the relevant parts of \eqref{Wunint} are known.
In the following we first discuss the effects of various 5-brane exchanges in terms of $W$ \eqref{Wunint}, and then convert the obtained transformation rules into those for $Z_{k,{\bm M}}(N)$ via \eqref{convert} and discuss the cascade relations \eqref{22cascades}.

\subsection{5-brane exchanges}

Now let us consider the effects of 5-brane exchanges on the matrix model $W$.
We first consider the exchange of 5-branes of different kinds, $\text{NS5}\leftrightarrow (1,k)5$ or $(1,k)5\leftrightarrow\text{NS}5$.
This case was studied rigorously in \cite{HK}, where the local transformation rules of the unintegrated partition functions involving only the 5-branes exchanged were found to be
\begin{align}
&W\Big({}_\mu\langle N_L\mathop{\bullet}_{Z_L}N_C\mathop{\circ}_{Z_R}N_R\rangle_\nu\Big)
=\Omega_+^{(Z_L,Z_R)}\Big[\begin{matrix}N_C\\{\widetilde N}_C\end{matrix}\Big]
W\Big({}_\mu\langle N_L\mathop{\circ}_{Z_R}{\widetilde N}_C\mathop{\bullet}_{Z_L}N_R\rangle_\nu\Big),\nonumber\\
&W\Big({}_\mu\langle N_L\mathop{\circ}_{Z_L}N_C\mathop{\bullet}_{Z_R}N_R\rangle_\nu\Big)
=\Omega_-^{(Z_L,Z_R)}\Big[\begin{matrix}N_C\\{\widetilde N}_C\end{matrix}\Big]
W\Big({}_\mu\langle N_L\mathop{\bullet}_{Z_R}{\widetilde N}_C\mathop{\circ}_{Z_L}N_R\rangle_\nu\Big),
\label{differentkinds}
\end{align}
with ${\widetilde N}_C=N_L+N_R-N_C+k$ and
\begin{align}
\Omega_+^{(Z_L,Z_R)}\Big[\begin{matrix}N_C\\{\widetilde N}_C\end{matrix}\Big]
=\bigg(\Omega_-^{(Z_L,Z_R)}\Big[\begin{matrix}N_C\\{\widetilde N}_C\end{matrix}\Big]\bigg)^{-1}
=e^{\pi i(-\frac{1}{6}-\frac{k^2}{12}+\frac{k(N_C+{\widetilde N}_C)}{4}+\frac{(N_C-{\widetilde N}_C)^2}{4}+(Z_L-Z_R)^2)}.
\label{OmegaOmegaprime}
\end{align}
Here for later purpose we have expressed the rank-dependence of the transformation factors $\Omega_\pm$ by using only $N_C,{\widetilde N}_C$.
Note that, as is expected, $\Omega_+$ and $\Omega_-$ are mutually reciprocal, since they are factors for inverse transformations.

Next let us consider the exchange of 5-branes of the same kind, $\text{NS}5\leftrightarrow\text{NS}5$ or $(1,k)5\leftrightarrow (1,k)5$.
Since the two cases result in the same transformation factor, let us concentrate on the exchange of two $\text{NS}5$-branes associated with the FI parameters $Z_L,Z_R$ and separating $N_L,N_C,N_R$ D3-branes.
Similarly to the above case, let us consider again the unintegrated partition functions involving only the 5-branes exchanged.
The unintegrated partition function before the exchange is given by
\begin{align}
&W\Big({}_\mu\langle N_L\mathop{\bullet}_{Z_L}N_C\mathop{\bullet}_{Z_R}N_R\rangle_\nu\Big)
=\frac{1}{N_C!}\int\frac{d^{N_C}\lambda_i}{(2\pi)^{N_C}}
W\Big({}_\mu\langle N_L\mathop{\bullet}_{Z_L} N_C\rangle_\lambda\Big)
W\Big({}_\lambda\langle N_C\mathop{\bullet}_{Z_R} N_R\rangle_\nu\Big),
\label{NSNSbefore}
\end{align}
while that after the exchange is given by the same expression $W\Big({}_\mu\langle N_L\underset{Z_R}{\bullet}{\widetilde N}_C\underset{Z_L}{\bullet}N_R\rangle_\nu\Big)$ with $N_C$ replaced by ${\widetilde N}_C=N_L+N_R-N_C$ and the two FI parameters $Z_L$ and $Z_R$ swapped.
We can evaluate the integration in \eqref{NSNSbefore} by using the Vandermonde determinant formula
\begin{align}
\prod_{i<j}^N2\sinh\frac{x_i-x_j}{2}
=\det_{i,s}^N\Bigl[e^{(\frac{N+1}{2}-s)x_i}\Bigr],
\end{align}
and the Cauchy-Binet formula
\begin{align}
\frac{1}{N!}\int d^Nx \det_{i,j}^N\Bigl[f_i(x_j)\Bigr]\det_{i,j}^N\Bigl[g_i(x_j)\Bigr]=\det_{i,j}^N\Bigl[\int dx f_i(x)g_j(x)\Bigr],
\end{align}
as
\begin{align}
&W\Big({}_\mu\langle N_L\mathop{\bullet}_{Z_L}N_C\mathop{\bullet}_{Z_R}N_R\rangle_\nu\Big)
=\Bigl(\prod_{i<j}^{N_L}2\sinh\frac{\mu_i-\mu_j}{2}\Bigr)
\Bigl(\prod_{i<j}^{N_R}2\sinh\frac{\nu_i-\nu_j}{2}\Bigr)\nonumber \\
&\quad \times e^{-Z_L\sum_{i=1}^{N_L}\mu_i+Z_R\sum_{i=1}^{N_R}\nu_i}\det_{r,s}^{N_C}\Big[I_{N_L+N_R}(N_C+1-r-s+Z_L-Z_R,\{\mu_i\}_{i=1}^{N_L}\cup \{\nu_i\}_{i=1}^{N_R})\Big],
\label{Wblock}
\end{align}
with
\begin{align}
I_{n}(\alpha,\{\beta_a\})=\int\frac{dx}{2\pi}\frac{e^{\alpha x}}{\prod_{a=1}^n2\cosh\frac{x-\beta_a}{2}}.
\label{formulaI}
\end{align}
If we assume that the integration is convergent, $I_{n}(\alpha,\{\beta_a\})$ is given by a residue sum,
\begin{align}
I_{n}(\alpha,\{\beta_a\})=I_{n}^{\text{(res)}}(\alpha,\{\beta_a\})=\frac{1}{e^{-\pi i\alpha}-(-1)^n e^{\pi i\alpha}}\sum_{a=1}^n\frac{e^{\alpha\beta_a}}{\prod_{a'(\neq a)}2i\sinh\frac{\beta_a-\beta_{a'}}{2}}.
\label{formulaIresiduesum}
\end{align}
Similarly, $W\Big({}_\mu\langle N_L\underset{Z_R}{\bullet}{\widetilde N}_C\underset{Z_L}{\bullet}N_R\rangle_\nu\Big)$ is obtained simply by replacing $N_C$ with $\widetilde N_C$ and swapping $Z_L$ and $Z_R$ in \eqref{Wblock}.

\begin{figure}[!t]
\begin{center}
\includegraphics[width=5cm]{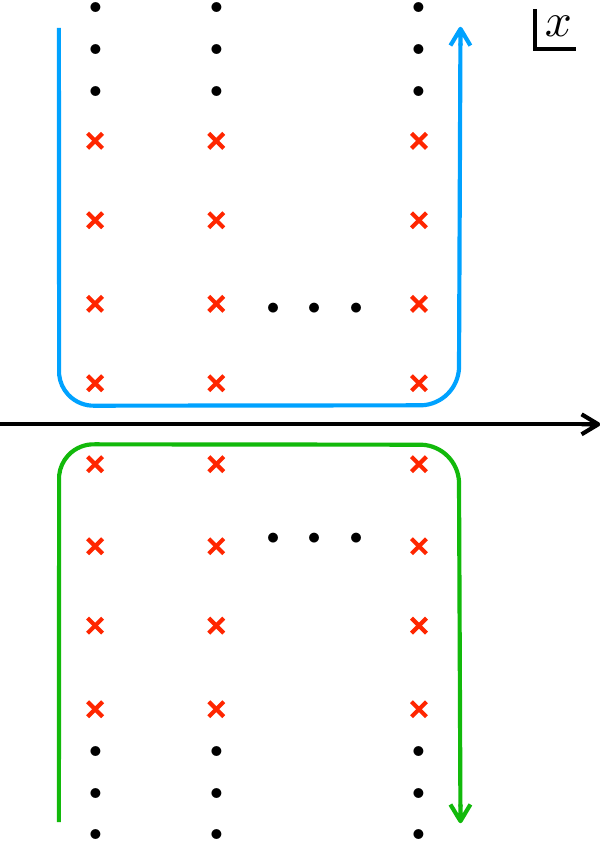}
\caption{
Contours which justify the prescription of replacing the integrations with residue sums, $I_n=I_n^{\text{(res)}}$, in the evaluation of $W({}_\mu\langle N_L\mathop{\bullet}_{Z_L}N_C\mathop{\bullet}_{Z_R}N_R\rangle_\nu)$ \eqref{Wblock}.
The blue line (and the green one) denotes the contour for $\text{Im}[Z_L-Z_R]>0$ (and that for $\text{Im}[Z_L-Z_R]<0$ respectively), while the red crosses denote the poles of $\prod_{i=1}^{N_L}\frac{1}{2\cosh\frac{x-\mu_i}{2}}\prod_{i=1}^{N_R}\frac{1}{2\cosh\frac{x-\nu_i}{2}}$.
}
\label{Yaakovcontour}
\end{center}
\end{figure}

As we have commented at the beginning of this section, convergence of the integration after the 5-brane exchange is not guaranteed even when the integration for the original configuration \eqref{NSNSbefore} is convergent.
The convergence condition for the integration \eqref{NSNSbefore} is
\begin{align}
\frac{N_L+N_R}{2}-N_C+1-|\text{Re}[Z_L-Z_R]|>0,
\label{convergenceNSNSbefore}
\end{align}
which is obtained by considering the asymptotic behavior of the
integrand as the absolute value of one of the integration variables,
say $|\lambda_1|$, grows,
\begin{align}
W\Big({}_\mu\langle N_L\mathop{\bullet}_{Z_L} N_C\rangle_\lambda\Big)
W\Big({}_\lambda\langle N_C\mathop{\bullet}_{Z_R} N_R\rangle_\nu\Big)
\sim e^{-(\frac{N_L+N_R}{2}-N_C+1\mp(Z_L-Z_R))|\lambda_1|},\quad \lambda_1\rightarrow \pm\infty.
\end{align}
On the other hand, the convergence condition for the integration after the 5-brane exchange is \eqref{convergenceNSNSbefore} with $N_C$ replaced with ${\widetilde N}_C$, which is different from \eqref{convergenceNSNSbefore}.
Nevertheless, let us tentatively ignore the issue of divergences and assume that both of the unintegrated partition functions, \eqref{NSNSbefore} and its replacement, are evaluated by $I^{\text{(res)}}_{n}(\alpha,\{\beta_a\})$ \eqref{formulaIresiduesum}.
This is also justified if we choose the integration contour for $I_{N_L+N_R}$ in \eqref{Wblock} to
surround only the poles of $\frac{1}{2\cosh\frac{x-\mu_i}{2}}\frac{1}{2\cosh\frac{x-\nu_i}{2}}$ in the upper-half (or lower-half) plane if $\text{Im}[Z_L-Z_R]>0$ (or $\text{Im}[Z_L-Z_R]<0$ respectively), as depicted in figure \ref{Yaakovcontour}.
Then we obtain\footnote{
This factor depending on the FI parameters was also derived by \cite{Kim:2012uz,Yaakov:2013fza} in a slightly different formulation (see \cite{Giacomelli:2023zkk} for further argument).
A similar factor was also found in the relation of partition functions between a pair of linear-quiver Chern-Simons theories in \cite{Nosaka:2017ohr} where the brane configurations are related through an exchange of 5-branes of the same kind.
}
\begin{align}
&W\Big({}_\mu\langle N_L\mathop{\bullet}_{Z_L}N_C\mathop{\bullet}_{Z_R}N_R\rangle_\nu\Big)
=\Gamma^{(Z_L,Z_R)}\Big[\begin{matrix}N_C\\{\widetilde N}_C\end{matrix}\Big]
W\Big({}_\mu\langle N_L\mathop{\bullet}_{Z_R}{\widetilde N}_C\mathop{\bullet}_{Z_L}N_R\rangle_\nu\Big),\nonumber \\
&W\Big({}_\mu\langle N_L\mathop{\circ}_{Z_L}{\widetilde N}_C\mathop{\circ}_{Z_R}N_R\rangle_\nu\Big)
=\Gamma^{(Z_L,Z_R)}\Big[\begin{matrix}N_C\\{\widetilde N}_C\end{matrix}\Big]
W\Big({}_\mu\langle N_L\mathop{\circ}_{Z_R}N_C\mathop{\circ}_{Z_L}N_R\rangle_\nu\Big),
\label{samekinds}
\end{align}
with ${\widetilde N}_C=N_L+N_R-N_C$ and
\begin{align}
\Gamma^{(Z_L,Z_R)}\Big[\begin{matrix}N_C\\{\widetilde N}_C\end{matrix}\Big]=(e^{-\pi i(Z_L-Z_R)}-(-1)^{{\widetilde N}_C-N_C}e^{\pi i(Z_L-Z_R)})^{{\widetilde N}_C-N_C}.
\label{Gamma}
\end{align}
Here we have used the determinant formula
\begin{align}
e^{-Z\sum_{i=1}^\ell x_i}
\frac{
\det_{r,s}^{m}[I_{\ell}^{\text{(res)}}(m+1-r-s+Z,\{x_i\}_{i=1}^\ell)]
}{
\det_{r,s}^{n}[I_{\ell}^{\text{(res)}}(n+1-r-s-Z,\{x_i\}_{i=1}^\ell)]
}=(e^{-\pi iZ}-(-1)^{n-m} e^{\pi iZ})^{n-m},
\end{align}
with $\ell=m+n$, which we have checked for $\ell\le 7$.
Note that in \eqref{samekinds} the equality holds only after ignoring the issue of divergences.
Surprisingly the transformation rule \eqref{samekinds} with \eqref{Gamma} is not obtained by taking the naive limit $k\to 0$ from \eqref{differentkinds} with \eqref{OmegaOmegaprime}.

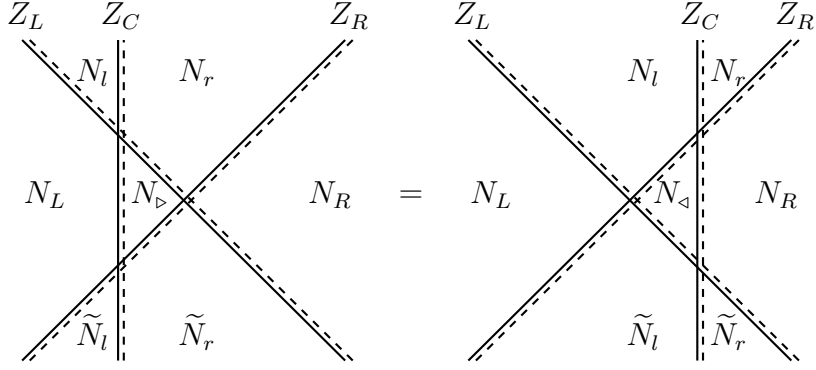
\begin{figure}[!t]
\begin{align*}
\mbox{\raisebox{-2.11cm}{
\begin{tikzpicture}[scale=0.25]
\draw [thick] (-0.2,0) -- (16.8,17);
\draw [thick,dashed] (0.2,0) -- (17.2,17);
\draw [thick] (4.85,0) -- (4.85,17);
\draw [thick,dashed] (5.15,0) -- (5.15,17);
\draw [thick] (-0.2,17) -- (16.8,0);
\draw [thick,dashed] (0.2,17) -- (17.2,0);
\node [above] at (0,17) {$Z_L$};
\node [above] at (5,17) {$Z_C$};
\node [above] at (17,17) {$Z_R$};
\node [above] at (3.5,14) {$N_l$};
\node [above] at (9,14) {$N_r$};
\node [above] at (1,7.5) {$N_L$};
\node [above] at (6.5,7.5) {$N_\triangleright$};
\node [above] at (16,7.5) {$N_R$};
\node [above] at (3.5,0) {${\widetilde N}_l$};
\node [above] at (9,0) {${\widetilde N}_r$};
\end{tikzpicture}
}}
=
\mbox{\raisebox{-2.11cm}{
\begin{tikzpicture}[scale=0.25]
\draw [thick] (-0.2,0) -- (16.8,17);
\draw [thick,dashed] (0.2,0) -- (17.2,17);
\draw [thick] (11.85,0) -- (11.85,17);
\draw [thick,dashed] (12.15,0) -- (12.15,17);
\draw [thick] (-0.2,17) -- (16.8,0);
\draw [thick,dashed] (0.2,17) -- (17.2,0);
\node [above] at (0,17) {$Z_L$};
\node [above] at (12,17) {$Z_C$};
\node [above] at (17,17) {$Z_R$};
\node [above] at (9,14) {$N_l$};
\node [above] at (13.5,14) {$N_r$};
\node [above] at (1,7.5) {$N_L$};
\node [above] at (10.5,7.5) {$N_\triangleleft$};
\node [above] at (16,7.5) {$N_R$};
\node [above] at (9,0) {${\widetilde N}_l$};
\node [above] at (13.5,0) {${\widetilde N}_r$};
\end{tikzpicture}
}}
\end{align*}
\caption{A schematic picture of the exchanges of 5-branes.
Here the ranks after the exchanges of 5-branes may be shifted by $k$ depending on the types of 5-branes (see \eqref{YBothers}).
We regard 5-branes as ``scattering particles'' and denote the numbers of D3-branes in each interval.
Although two processes are possible to move from 
$\langle\cdots
N_L
\begin{matrix}
\mathop{}^1\vspace{-0.45cm}\\
\bullet\vspace{-0.4cm}\\
\circ\vspace{-0.4cm}\\
\mathop{}_{Z_L}
\end{matrix}
N_l
\begin{matrix}
\mathop{}^2\vspace{-0.45cm}\\
\bullet\vspace{-0.4cm}\\
\circ\vspace{-0.4cm}\\
\mathop{}_{Z_C}
\end{matrix}
N_r
\begin{matrix}
\mathop{}^3\vspace{-0.45cm}\\
\bullet\vspace{-0.4cm}\\
\circ\vspace{-0.4cm}\\
\mathop{}_{Z_R}
\end{matrix}
N_R\cdots\rangle$ to $\langle\cdots
N_L
\begin{matrix}
\mathop{}^3\vspace{-0.45cm}\\
\bullet\vspace{-0.4cm}\\
\circ\vspace{-0.4cm}\\
\mathop{}_{Z_R}
\end{matrix}
{\widetilde N}_l
\begin{matrix}
\mathop{}^2\vspace{-0.45cm}\\
\bullet\vspace{-0.4cm}\\
\circ\vspace{-0.4cm}\\
\mathop{}_{Z_C}
\end{matrix}
{\widetilde N}_r
\begin{matrix}
\mathop{}^1\vspace{-0.45cm}\\
\bullet\vspace{-0.4cm}\\
\circ\vspace{-0.4cm}\\
\mathop{}_{Z_L}
\end{matrix}
N_R\cdots\rangle$,
the values of partition functions should not depend on the processes.}
\label{YB}
\end{figure}

Here let us make a detour to discuss interesting properties relating to the factors $\Omega$ \eqref{OmegaOmegaprime} and $\Gamma$ \eqref{Gamma}.
As is clear from physical intuition that the exchange does not rely on the intermediate processes, these factors should satisfy the braiding relations or the Yang-Baxter relations.
Let us consider a brane configuration
${}_\mu
\langle
N_L
\begin{matrix}
\mathop{}^1\vspace{-0.45cm}\\
\bullet\vspace{-0.4cm}\\
\circ\vspace{-0.4cm}\\
\mathop{}_{Z_L}
\end{matrix}
N_l
\begin{matrix}
\mathop{}^2\vspace{-0.45cm}\\
\bullet\vspace{-0.4cm}\\
\circ\vspace{-0.4cm}\\
\mathop{}_{Z_C}
\end{matrix}
N_r
\begin{matrix}
\mathop{}^3\vspace{-0.45cm}\\
\bullet\vspace{-0.4cm}\\
\circ\vspace{-0.4cm}\\
\mathop{}_{Z_R}
\end{matrix}
N_R\rangle_\nu$.
When bringing it into
${}_\mu
\langle
N_L
\begin{matrix}
\mathop{}^3\vspace{-0.45cm}\\
\bullet\vspace{-0.4cm}\\
\circ\vspace{-0.4cm}\\
\mathop{}_{Z_R}
\end{matrix}
{\widetilde N}_l
\begin{matrix}
\mathop{}^2\vspace{-0.45cm}\\
\bullet\vspace{-0.4cm}\\
\circ\vspace{-0.4cm}\\
\mathop{}_{Z_C}
\end{matrix}
{\widetilde N}_r
\begin{matrix}
\mathop{}^1\vspace{-0.45cm}\\
\bullet\vspace{-0.4cm}\\
\circ\vspace{-0.4cm}\\
\mathop{}_{Z_L}
\end{matrix}
N_R\rangle_\nu$
with $\widetilde N_l=N_L-N_r+N_R$ and $\widetilde N_r=N_L-N_l+N_R$ by exchanging 5-branes, we have two processes.
Namely, we can either follow the exchanges of
$\begin{matrix}
\mathop{}^1\vspace{-0.5cm}\\
\bullet\vspace{-0.4cm}\\
\circ
\end{matrix}
\leftrightarrow
\begin{matrix}
\mathop{}^2\vspace{-0.5cm}\\
\bullet\vspace{-0.4cm}\\
\circ
\end{matrix}$,
$\begin{matrix}
\mathop{}^1\vspace{-0.5cm}\\
\bullet\vspace{-0.4cm}\\
\circ
\end{matrix}
\leftrightarrow
\begin{matrix}
\mathop{}^3\vspace{-0.5cm}\\
\bullet\vspace{-0.4cm}\\
\circ
\end{matrix}$,
$\begin{matrix}
\mathop{}^2\vspace{-0.5cm}\\
\bullet\vspace{-0.4cm}\\
\circ
\end{matrix}
\leftrightarrow
\begin{matrix}
\mathop{}^3\vspace{-0.5cm}\\
\bullet\vspace{-0.4cm}\\
\circ
\end{matrix}$, or the exchanges of
$\begin{matrix}
\mathop{}^2\vspace{-0.5cm}\\
\bullet\vspace{-0.4cm}\\
\circ
\end{matrix}
\leftrightarrow
\begin{matrix}
\mathop{}^3\vspace{-0.5cm}\\
\bullet\vspace{-0.4cm}\\
\circ
\end{matrix}$,
$\begin{matrix}
\mathop{}^1\vspace{-0.5cm}\\
\bullet\vspace{-0.4cm}\\
\circ
\end{matrix}
\leftrightarrow
\begin{matrix}
\mathop{}^3\vspace{-0.5cm}\\
\bullet\vspace{-0.4cm}\\
\circ
\end{matrix}$,
$\begin{matrix}
\mathop{}^1\vspace{-0.5cm}\\
\bullet\vspace{-0.4cm}\\
\circ
\end{matrix}
\leftrightarrow
\begin{matrix}
\mathop{}^2\vspace{-0.5cm}\\
\bullet\vspace{-0.4cm}\\
\circ
\end{matrix}$.
See figure \ref{YB} for a schematical picture of the exchanges.
Since the final results of the partition function should not depend on the processes of exchanges, we have the relation
\begin{align}
\Gamma^{(Z_L,Z_C)}\Big[\begin{matrix}N_l\\N_\triangleright\end{matrix}\Big]
\Gamma^{(Z_L,Z_R)}\Big[\begin{matrix}N_r\\\widetilde N_r\end{matrix}\Big]
\Gamma^{(Z_C,Z_R)}\Big[\begin{matrix}N_\triangleright\\\widetilde N_l\end{matrix}\Big]
=\Gamma^{(Z_C,Z_R)}\Big[\begin{matrix}N_r\\N_\triangleleft\end{matrix}\Big]
\Gamma^{(Z_L,Z_R)}\Big[\begin{matrix}N_l\\\widetilde N_l\end{matrix}\Big]
\Gamma^{(Z_L,Z_C)}\Big[\begin{matrix}N_\triangleleft\\\widetilde N_r\end{matrix}\Big],
\label{YBGamma}
\end{align}
with $N_\triangleright=N_L-N_l+N_r$ and $N_\triangleleft=N_l-N_r+N_R$.
Similarly, we can replace some of NS5-branes by $(1,k)$5-branes.
Then, the Yang-Baxter relations
\begin{align}
\Gamma\Big[\begin{matrix}N_l\\N_\triangleright\end{matrix}\Big]
\Omega_{\pm}\Big[\begin{matrix}N_r\\\widetilde N_r+k\end{matrix}\Big]
\Omega_{\pm}\Big[\begin{matrix}N_\triangleright\\\widetilde N_l+2k\end{matrix}\Big]
&=\Omega_{\pm}\Big[\begin{matrix}N_r\\N_\triangleleft+k\end{matrix}\Big]
\Omega_{\pm}\Big[\begin{matrix}N_l\\\widetilde N_l+2k\end{matrix}\Big]
\Gamma\Big[\begin{matrix}N_\triangleleft+k\\\widetilde N_r+k\end{matrix}\Big],\nonumber\\
\Omega_{\pm}\Big[\begin{matrix}N_l\\N_\triangleright+k\end{matrix}\Big]
\Gamma\Big[\begin{matrix}N_r\\\widetilde N_r+k\end{matrix}\Big]
\Omega_{\mp}\Big[\begin{matrix}N_\triangleright+k\\\widetilde N_l+k\end{matrix}\Big]
&=\Omega_{\mp}\Big[\begin{matrix}N_r\\N_\triangleleft+k\end{matrix}\Big]
\Gamma\Big[\begin{matrix}N_l\\\widetilde N_l+k\end{matrix}\Big]
\Omega_{\pm}\Big[\begin{matrix}N_\triangleleft+k\\\widetilde N_r+k\end{matrix}\Big],\nonumber\\
\Omega_{\mp}\Big[\begin{matrix}N_l\\N_\triangleright+k\end{matrix}\Big]
\Omega_{\mp}\Big[\begin{matrix}N_r\\\widetilde N_r+2k\end{matrix}\Big]
\Gamma\Big[\begin{matrix}N_\triangleright+k\\\widetilde N_l+k\end{matrix}\Big]
&=\Gamma\Big[\begin{matrix}N_r\\N_\triangleleft\end{matrix}\Big]
\Omega_{\mp}\Big[\begin{matrix}N_l\\\widetilde N_r+k\end{matrix}\Big]
\Omega_{\mp}\Big[\begin{matrix}N_\triangleleft\\\widetilde N_r+2k\end{matrix}\Big],
\label{YBothers}
\end{align}
should hold as well with $\widetilde N_l$, $\widetilde N_r$, $N_\triangleright$ and $N_\triangleleft$ shifted by $k$ due to the Hanany-Witten transitions \eqref{HW}.
We have omitted the FI parameters $(Z_L,Z_C,Z_R)$ in \eqref{YBothers} since the dependence is the same as in \eqref{YBGamma}.

\subsection{Cascade relations}

Now that we have the local transformation rules \eqref{differentkinds} and \eqref{samekinds} of the unintegrated partition function $W$ under 5-brane exchanges, let us study the behavior of $Z_{k,{\bm M}}(N)$ under the cyclic 5-brane exchanges which induce the translations of $(M_0,M_1,M_3)$ in \eqref{22cascades}.
For example, let us consider $\Delta_{(1,1,0)}: (M_0,M_1,M_3)\rightarrow (M_0+k,M_1+k,M_3)$.
The translation $\Delta_{(1,1,0)}$ is realized by moving the $(1,k)5$-brane $\overset{1}{\underset{0}{\circ}}$ leftward cyclically along the compactified direction by exchanging with $\overset{2}{\underset{-Z_1}{\circ}}$, $\overset{4}{\underset{0}{\bullet}}$ and $\overset{3}{\underset{Z_3}{\bullet}}$ subsequently.
Then, the configuration in \eqref{nosakaconfig} denoted by $\langle{}\overset{4}{\bullet}
\overset{2}{\circ}
\overset{1}{\circ}
\overset{3}{\bullet}
{}\rangle$ changes as
\begin{align}
&\langle{}\overset{4}{\bullet}
\overset{2}{\circ}
\overset{1}{\circ}
\overset{3}{\bullet}
{}\rangle
=\langle N\mathop{\bullet}^4_0
N-M_0-M_3+k\mathop{\circ}^2_{-Z_1}
N-M_1-M_3+k\mathop{\circ}^1_0
N+M_0-M_3+k\mathop{\bullet}^3_{Z_3}\rangle
\nonumber\\
&\mathop{\Rightarrow}_{2\leftrightarrow 1}
\langle N\mathop{\bullet}^4_0
N-M_0-M_3+k\mathop{\circ}^1_0
N+M_1-M_3+k\mathop{\circ}^2_{-Z_1}
N+M_0-M_3+k\mathop{\bullet}^3_{Z_3}\rangle\nonumber\\
&\mathop{\Rightarrow}_{4\leftrightarrow 1}
\langle N\mathop{\circ}^1_0
N+M_0+M_1+k\mathop{\bullet}^4_0
N+M_1-M_3+k\mathop{\circ}^2_{-Z_1}
N+M_0-M_3+k\mathop{\bullet}^3_{Z_3}
\rangle\nonumber\\
&\mathop{\Rightarrow}_{3\leftrightarrow 1}
\langle N+2M_0+M_1-M_3+3k\mathop{\bullet}^3_{Z_3}
N+M_0+M_1+k\mathop{\bullet}^4_0
N+M_1-M_3+k\mathop{\circ}^2_{-Z_1}
N+M_0-M_3+k\mathop{\circ}^1_{0}\rangle\nonumber\\
&\rightarrow\langle{}\overset{4}{\bullet}
\overset{2}{\circ}
\overset{1}{\circ}
\overset{3}{\bullet}
{}\rangle\Bigr|_{(N,M_0,M_1,M_3)\rightarrow (N',M'_0,M'_1,M'_3)=(N+M_0+M_1+k,M_0+k,M_1+k,M_3)},
\end{align}
where $\underset{a\leftrightarrow b}{\Rightarrow}$ stands for the exchange of 5-branes labeled with $a$ and $b$, which changes the partition function $W$ by the transformation factors $\Omega_\pm$ \eqref{OmegaOmegaprime} or $\Gamma$ \eqref{Gamma}, while $\rightarrow$ stands for the cyclic rotation of the entire brane configuration which does not change the expression of $W$.
Taking into account all the transformation factors, we obtain
\begin{align}
&W_{k,{\bm M}}(N)=
\Gamma^{(-Z_1,0)}\Big[\begin{matrix}N-M_1-M_3+k\\N+M_1-M_3+k\end{matrix}\Big]\nonumber\\
&\quad\times\Omega_+^{(0,0)}\Big[\begin{matrix}N-M_0-M_3+k\\N+M_0+M_1+k\end{matrix}\Big]
\Omega_+^{(Z_3,0)}\Big[\begin{matrix}N\\N+2M_0+M_1-M_3+3k\end{matrix}\Big]
W_{k,{\bm M}'}(N').
\end{align}
Here, again, the equality is obtained under the assumption of replacing the integration by residue sums.
Substituting \eqref{OmegaOmegaprime}, \eqref{Gamma} and converting back to the relation for $Z_{k,{\bm M}}(N)$ through \eqref{convert}, we find
\begin{align}
Z_{k,{\bm M}+(k,k,0)}(N+M_0+M_1+k)
=e^{-2\pi ik(Z_1+Z_3)}(e^{\pi iZ_1}-(-1)^{2M_1}e^{-\pi iZ_1})^{-2M_1}
Z_{k,{\bm M}}(N),
\end{align}
which implies
\begin{align}
\frac{
\Xi_{k,{\bm M}+(k,k,0)}(\kappa)}{\Xi_{k,{\bm M}}(\kappa)}
=e^{-2\pi ik(Z_1+Z_3)}(e^{\pi iZ_1}-(-1)^{2M_1}e^{-\pi iZ_1})^{-2M_1}\kappa^{M_0+M_1+k}.
\end{align}
Interestingly, this precisely reproduces the cascade relation \eqref{22cascades} for $\Delta_{(1,1,0)}$ which is consistent with the bilinear equations.

\begin{table}[!t]
\begin{center}
\begin{tabular}{c|cc}
&5-brane exchanges&
$(N',M'_0,M'_1,M'_3)$\\[0pt]
$\Delta$&\multicolumn{2}{c}{$W_{k,{\bm M}}(N)/W_{k,{\bm M}'}(N')$}\\[0pt]
&$Z_{k,{\bm M}'}(N')/Z_{k,{\bm M}}(N)$&
$\Xi_{k,{\bm M}'}(\kappa)/\Xi_{k,{\bm M}}(\kappa)$
\\[2pt]\hline\hline
&&\\[-14pt]
&$\overset{2}{\circ}\leftrightarrow\overset{1}{\circ}$,
$\overset{4}{{\bullet}}\leftrightarrow\overset{1}{\circ}$,
$\overset{3}{{\bullet}}\leftrightarrow\overset{1}{\circ}$&$(N+M_0+M_1+k,M_0+k,M_1+k,M_3)$\\[6pt]
$\Delta_{(1,1,0)}$
&\multicolumn{2}{c}{$\footnotesize\Gamma^{(-Z_1,0)}\Big[\begin{matrix}N-M_1-M_3+k\\N+M_1-M_3+k\end{matrix}\Big]
\Omega_+^{(0,0)}\Big[\begin{matrix}N-M_0-M_3+k\\N+M_0+M_1+k\end{matrix}\Big]
\Omega_+^{(Z_3,0)}\Big[\begin{matrix}N\\N+2M_0+M_1-M_3+3k\end{matrix}\Big]$}\\[14pt]
&$e^{-2\pi ik(Z_1+Z_3)}(e^{\pi iZ_1}-(-1)^{2M_1}e^{-\pi iZ_1})^{-2M_1}$&$e^{-2\pi ik(Z_1+Z_3)}f_{M_1}(Z_1)\kappa^{M_0+M_1+k}$
\\[2pt]\hline
&&\\[-14pt]
&$\overset{2}{\circ}\leftrightarrow\overset{1}{\circ}$,
$\overset{2}{\circ}\leftrightarrow\overset{3}{{\bullet}}$,
$\overset{2}{\circ}\leftrightarrow\overset{4}{{\bullet}}$&$(N-M_0+M_1+k,M_0-k,M_1+k,M_3)$\\[6pt]
$\Delta_{(-1,1,0)}$
&\multicolumn{2}{c}{$\footnotesize\Gamma^{(-Z_1,0)}\Big[\begin{matrix}N-M_1-M_3+k\\N+M_1-M_3+k\end{matrix}\Big]
\Omega_-^{(-Z_1,Z_3)}\Big[\begin{matrix}N+M_0-M_3+k\\N-M_0+M_1+k\end{matrix}\Big]
\Omega_-^{(-Z_1,0)}\Big[\begin{matrix}N\\N-2M_0+M_1-M_3+3k\end{matrix}\Big]$}\\[14pt]
&$e^{-2\pi ik(Z_1+Z_3)}(e^{\pi iZ_1}-(-1)^{2M_1}e^{-\pi iZ_1})^{-2M_1}$&$e^{-2\pi ik(Z_1+Z_3)}f_{M_1}(Z_1)\kappa^{-M_0+M_1+k}$
\\[2pt]\hline
&&\\[-14pt]
&$\overset{4}{{\bullet}}\leftrightarrow\overset{2}{\circ}$,
$\overset{4}{{\bullet}}\leftrightarrow\overset{1}{\circ}$,
$\overset{4}{{\bullet}}\leftrightarrow\overset{3}{{\bullet}}$&$(N+M_0+M_3+k,M_0+k,M_1,M_3+k)$\\[6pt]
$\Delta_{(1,0,1)}$
&\multicolumn{2}{c}{$\footnotesize\Omega_+^{(0,-Z_1)}\Big[\begin{matrix}N-M_0-M_3+k\\N+M_0-M_1+k\end{matrix}\Big]
\Omega_+^{(0,0)}\Big[\begin{matrix}N-M_1-M_3+k\\N+2M_0+2k\end{matrix}\Big]
\Gamma^{(0,Z_3)}\Big[\begin{matrix}N+M_0-M_3+k\\N+M_0+M_3+k\end{matrix}\Big]$}\\[14pt]
&$(e^{\pi iZ_3}-(-1)^{2M_3}e^{-\pi iZ_3})^{-2M_3}$&$f_{M_3}(Z_3)\kappa^{M_0+M_3+k}$
\\[2pt]\hline
&&\\[-14pt]
&$\overset{1}{\circ}\leftrightarrow\overset{3}{{\bullet}}$,
$\overset{2}{\circ}\leftrightarrow\overset{3}{{\bullet}}$,
$\overset{4}{{\bullet}}\leftrightarrow\overset{3}{{\bullet}}$&$(N-M_0+M_3+k,M_0-k,M_1,M_3+k)$\\[6pt]
$\Delta_{(-1,0,1)}$
&\multicolumn{2}{c}{$\footnotesize\Omega_-^{(0,Z_3)}\Big[\begin{matrix}N+M_0-M_3+k\\N-M_0-M_1+k\end{matrix}\Big]
\Omega_-^{(-Z_1,Z_3)}\Big[\begin{matrix}N-M_1-M_3+k\\N-2M_0+2k\end{matrix}\Big]
\Gamma^{(0,Z_3)}\Big[\begin{matrix}N-M_0-M_3+k\\N-M_0+M_3+k\end{matrix}\Big]$}\\[14pt]
&$(e^{\pi iZ_3}-(-1)^{2M_3}e^{-\pi iZ_3})^{-2M_3}$&
$f_{M_3}(Z_3)\kappa^{-M_0+M_3+k}$\\[-18pt]
\end{tabular}
\end{center}
\caption{Formal derivations of the cascade relations \eqref{22cascades} with the prescription of residue sums.
We display the 5-brane exchanges which the discrete translation consists of, the change of variables $(N,M_0,M_1,M_3)$, $W_{k,{\bm M}}(N)/W_{k,{\bm M}'}(N')$, and the results of $Z_{k,{\bm M}'}(N')/Z_{k,{\bm M}}(N)$ and $\Xi_{k,{\bm M}'}(\kappa)/\Xi_{k,{\bm M}}(\kappa)$.
The function $f_M(Z)$ appearing in $\Xi_{k,{\bm M}'}(\kappa)/\Xi_{k,{\bm M}}(\kappa)$ is defined in \eqref{fM}.
}
\label{der}
\end{table}

Similarly, the same computation works for all of the discrete translations in \eqref{22cascades}.
In table \ref{der} we display necessary information for the derivations, including the 5-brane exchanges which the discrete translation consists of, the change of variables $(N,M_0,M_1,M_3)$, $W_{k,{\bm M}}(N)/W_{k,{\bm M}'}(N')$ computed using \eqref{differentkinds} and \eqref{samekinds}, and the results of $Z_{k,{\bm M}'}(N')/Z_{k,{\bm M}}(N)$ and $\Xi_{k,{\bm M}'}(\kappa)/\Xi_{k,{\bm M}}(\kappa)$ with \eqref{OmegaOmegaprime} and \eqref{Gamma} substituted.
In all cases we find that the cascade relations \eqref{22cascades} are reproduced correctly.

\section{More evidences for cascade relations}
\label{translation}

In this section, we shall provide some more evidences for the cascade relations \eqref{22cascades} with \eqref{fM}.
In the previous section, we apply the prescription of residue sums \eqref{Wblock} with \eqref{formulaIresiduesum} uncritically to derive the cascade relations \eqref{22cascades} in spite of the issue of divergences.
It may be more comfortable if we apply the prescription only in the fundamental domain and try to extend the domain using the $q$-Painlev\'e equations $q\text{P}_\text{VI}$ as when we first reached the cascade relations \eqref{22cascades} in \cite{MN9}.
Also, the structure of affine symmetries should be clearer by presenting the expressions of partition functions explicitly.
For these reasons, in this section, we shall continue the analysis of \cite{MN9} by computing partition functions of higher ranks in the fundamental domain and extending the domain using $q\text{P}_\text{VI}$.
It is important to note that the computation in this section is not completely independent from that in the previous section since we adopt the same prescription of residue sums.
Nevertheless, we stress that the total consistency among the physical argument of duality cascades in section \ref{22}, the formal derivation of the cascade relations \eqref{22cascades} in the previous section and the expressions of partition functions using the prescription only in the fundamental domain in this section is non-trivial.

Let us first recapitulate the computations in \cite{MN9}.
In \cite{MN9} the cascade relations were identified at $N=0,1$ for $k=1$ and at $N=0$ for $k=2$.
Here as in \cite{MN9}, we introduce the short-hand notation for the partition function
\begin{align}
Z_{M_0,M_1,M_3}^{[k](N)}=e^{-i\Theta_{k,{\bm M}}}Z_{k,{\bm M}}(N),
\label{dropphase}
\end{align}
and use the same notation also in appendix \ref{lowestk12}.
Note that the redefinition of the overall phase factor by $e^{i\Theta_{k,{\bm M}}}$ simplifies the expressions of partition functions in tables \ref{k1data}, \ref{k2data1} and \ref{k2data2} in appendix \ref{lowestk12}, although it does not play a role as long as we only consider the ratios $Z_{k,{\bm M}}(1)/Z_{k,{\bm M}}(0)=Z^{[k](1)}_{M_0,M_1,M_3}/Z^{[k](0)}_{M_0,M_1,M_3}$.
Hereafter in this section we shall use the two notations interchangeably.
Also, to explain the computations, we introduce a notation ${\cal D}_N^{[k]}$, denoting the domain where the partition function $Z_{k,{\bm M}}(N)$ is non-vanishing.
For $k=2$, the exact expressions of partition functions $Z_{2,{\bm M}}(0)$ was determined in the fundamental domain ${\cal D}_0^{[2]}$ and the cascade relations \eqref{22cascades} were only confirmed on the boundary.
For $k=1$, besides $Z_{1,{\bm M}}(0)$, $Z_{1,{\bm M}}(1)$ was also determined and the cascade relations \eqref{22cascades} were confirmed more generally.
There, first $Z_{1,{\bm M}}(1)$ in the fundamental domain ${\cal D}_0^{[1]}$ was computed.
Since the grand partition function normalized by $Z_{k,{\bm M}}(0)$ is expressed by the Fredholm determinant of a spectral operator conjectured to be invariant under the Weyl group, the ratio of the partition functions 
$Z_{1,{\bm M}}(1)/Z_{1,{\bm M}}(0)=Z^{[1](1)}_{M_0,M_1,M_3}/Z^{[1](0)}_{M_0,M_1,M_3}$
is expected to be classified by the Weyl group,
\begin{align}
\Delta_0^{[1]}=\frac{Z^{[1](1)}_{0,0,0}}{Z^{[1](0)}_{0,0,0}},\quad
\Delta_1^{[1]}=\frac{Z^{[1](1)}_{\{\pm\frac{1}{2},\pm\frac{1}{2},\pm\frac{1}{2}\}_0}}{Z^{[1](0)}_{\{\pm\frac{1}{2},\pm\frac{1}{2},\pm\frac{1}{2}\}_0}},\quad
\Delta_2^{[1]}=\frac{Z^{[1](1)}_{\{\pm 1,0,0\}}}{Z^{[1](0)}_{\{\pm 1,0,0\}}},\quad
\Delta_3^{[1]}=\frac{Z^{[1](1)}_{\{\pm\frac{1}{2},\pm\frac{1}{2},\pm\frac{1}{2}\}_1}}{Z^{[1](0)}_{\{\pm\frac{1}{2},\pm\frac{1}{2},\pm\frac{1}{2}\}_1}}.
\label{Delta[1]}
\end{align}
Here we have changed notations from \cite{MN9} since the classification is characterized by ${\mathbb Z}_4$.
As in appendix \ref{lowestk12}, the subscripts of $Z$ denote $(M_0,M_1,M_3)$ where the curly brackets indicate that the order can be arbitrary.
Also, the subscripts of the curly brackets $0$ or $1$ imply respectively that the number of minus signs is even or odd.
For example, $\Delta_1^{[1]}$ denotes the four identical quantities
\begin{align}
\Delta_1^{[1]}=\frac{Z^{[1](1)}_{\frac{1}{2},\frac{1}{2},\frac{1}{2}}}{Z^{[1](0)}_{\frac{1}{2},\frac{1}{2},\frac{1}{2}}}
=\frac{Z^{[1](1)}_{\frac{1}{2},-\frac{1}{2},-\frac{1}{2}}}{Z^{[1](0)}_{\frac{1}{2},-\frac{1}{2},-\frac{1}{2}}}
=\frac{Z^{[1](1)}_{-\frac{1}{2},\frac{1}{2},-\frac{1}{2}}}{Z^{[1](0)}_{-\frac{1}{2},\frac{1}{2},-\frac{1}{2}}}
=\frac{Z^{[1](1)}_{-\frac{1}{2},-\frac{1}{2},\frac{1}{2}}}{Z^{[1](0)}_{-\frac{1}{2},-\frac{1}{2},\frac{1}{2}}}.
\end{align}
In \cite{MN9}, it was found that they are given by
\begin{align}
\Delta_0^{[1]}&=-\frac{1}{4\pi}\csc(\pi Z_1)\csc(\pi Z_3)\Big(\pi Z_1Z_3-\pi\cot(\pi Z_1)\cot(\pi Z_3)-\cot(\pi Z_1Z_3)
\nonumber\\
&\qquad+\pi\big(Z_1\cot(\pi Z_1)+Z_3\cot(\pi Z_3)\big)\cot(\pi Z_1Z_3)\Big),\nonumber\\
\Delta_{1,3}^{[1]}&=-\frac{1}{16\pi}\sec(\pi Z_1)\sec(\pi Z_3)\Big(2\pm 4\pi iZ_1Z_3+2\pi\big(Z_1\tan(\pi Z_1)+Z_3\tan(\pi Z_3)\big)\nonumber\\
&\qquad\mp\pi i\tan(\pi Z_1)\tan(\pi Z_3)-e^{\mp 2\pi iZ_1Z_3}\pi\sec(\pi Z_1)\sec(\pi Z_3)\Big),\nonumber\\
\Delta_2^{[1]}&=\frac{1}{4}Z_1Z_3\csc(\pi Z_1)\csc(\pi Z_3).
\label{ratiok1}
\end{align}
Here the upper signs in $\Delta_{1,3}$ are for $\Delta_1$, while the lower ones are for $\Delta_3$.
Then, using the 40 bilinear equations \eqref{D5bilinear} with $Z_{1,{\bm M}}(1)$ in ${\cal D}^{[1]}_0$, we can solve for $Z_{1,{\bm M}}(1)$ outside the fundamental domain and determine all of them in the whole parameter space.
It was found that the domain for non-vanishing $Z_{1,{\bm M}}(1)$ (which is ${\cal D}^{[1]}_1$) extends slightly from the original fundamental domain ${\cal D}^{[1]}_0$.
With the exact expressions $Z_{1,{\bm M}}(1)$ outside the fundamental domain, we still find that the cascade relations \eqref{22cascades} are valid.
Note that since the transformation rules under the translations themselves were identified with the extra phases $e^{i\Theta_{k,{\bm M}}}$ removed \eqref{dropphase}, to obtain \eqref{22cascades} we need to take care of these phases.

Let us add some more evidences for the cascade relations \eqref{22cascades} with \eqref{fM} in this paper.
First, for $k=1$, we computed $Z_{1,{\bm M}}(0)$ and $Z_{1,{\bm M}}(1)$ in the fundamental domain ${\cal D}^{[1]}_0$.
Besides the fact that the ratio $Z_{k,{\bm M}}(1)/Z_{k,{\bm M}}(0)$ is invariant under the Weyl group, from the physical argument of duality cascades it is natural to expect that even outside the fundamental domain the ratio is kept intact as in \eqref{XioverZcascaderule22model}.
Thus we can build $Z_{k,{\bm M}}(2)$ on top of $Z_{k,{\bm M}}(1)$ in the domain ${\cal D}^{[k]}_1\backslash{\cal D}^{[k]}_0$ by assuming $Z_{k,{\bm M}}(2)/Z_{k,{\bm M}}(1)|_{{\cal D}^{[k]}_1\backslash{\cal D}^{[k]}_0}=Z_{k,{\bm M}_\text{FD}}(1)/Z_{k,{\bm M}_\text{FD}}(0)|_{{\cal D}^{[k]}_0}$.
This assumption is reminiscent of the grand partition function of the ABJM theory in figure \ref{ABJMGC}, where the partition functions of higher ranks are built on top of the lowest non-vanishing partition function identically even outside the fundamental domain.
Then, using the 40 bilinear equations \eqref{D5bilinear} with $Z_{k,{\bm M}}(2)$ in ${\cal D}^{[k]}_1$, we can determine $Z_{k,{\bm M}}(2)$ in the whole parameter space except ${\cal D}^{[k]}_0$, ${\mathbb R}^3\backslash{\cal D}^{[k]}_0$.
The results for $k=1$ are given in appendix \ref{k1}.
In table \ref{k1data}, expressions in dark grey, grey and light grey denote respectively $Z^{[1](0)}_{M_0,M_1,M_3}$, $Z^{[1](1)}_{M_0,M_1,M_3}$ and $Z^{[1](2)}_{M_0,M_1,M_3}$ with the phases dropped from $Z_{k,{\bm M}}(N)$ \eqref{dropphase}.
It is difficult, however, to determine $Z_{1,{\bm M}}(2)$ in ${\cal D}^{[1]}_0$ from the bilinear equations, since this time we need to solve intricate difference equations for $Z_{1,{\bm M}}(2)$.

For $k=2$, we need to first determine $Z_{2,{\bm M}}(1)$.
Even with the classification under the Weyl group, it is not easy to compute all of the exact expressions.
Instead, since the depth of the layer from the boundary indicates the complexity of the expression, we note that those on the boundary of the fundamental domain ${\cal D}^{[2]}_0$ are easier to compute and concentrate on them.
Then, using the 40 bilinear equations \eqref{D5bilinear}, we can determine all of $Z_{2,{\bm M}}(1)$ on ${\cal D}^{[2]}_1$.
This time $Z_{2,{\bm M}}(1)$ in the interior of the fundamental domain ${\cal D}^{[2]}_0$ is also obtained since some of the bilinear equations are no more difference equations.
The exact expressions of $Z_{2,{\bm M}}(1)$ are summarized in appendix \ref{higherk2} for those in the fundamental domain ${\cal D}^{[2]}_0$ and in appendix \ref{k2} for those outside, ${\cal D}^{[2]}_1\backslash {\cal D}^{[2]}_0$.
For those inside we disply the ratios $Z_{2,{\bm M}}(1)/Z_{2,{\bm M}}(0)$ classified by the Weyl group, while for those outside we display $Z^{[2](1)}_{M_0,M_1,M_3}$ with the phases dropped.

Using the resulting exact expressions of partition functions at $k=1$ and $k=2$ in appendix \ref{lowestk12}, we can confirm the validity of the cascade relations \eqref{22cascades} for all the possible cases.

\begin{figure}[!t]
\centering\includegraphics[width=\textwidth]{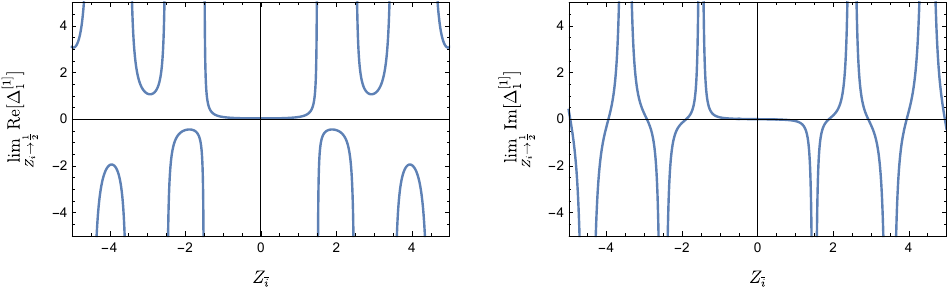}
\caption{Plots of the real part (left) and the imaginary part (right) of the ratio $\Delta_{1}^{[1]}$ of the partition functions in the limit $Z_i\to 1/2$ \eqref{Delta13limit}.
They are not divergent in $|Z_{\bar i}|\le 1/2$.}
\label{Delta13}
\end{figure}

\begin{figure}[!t]
\centering\includegraphics[width=0.5\textwidth]{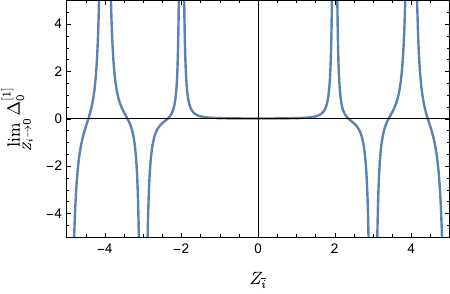}
\caption{
The ratio $\Delta_{0}^{[1]}$ of the partition functions in the limit $Z_i\to 0$ \eqref{Delta2limit} is not divergent in $|Z_{\bar i}|\le 1$.}
\label{Delta2}
\end{figure}

Before closing this subsection, let us comment on an interesting observation for the convergence of the ratios.
Namely, by extending the idea of the fundamental domain with the Weyl group \cite{FMMN}, we note that the resulting ratios for $k=1$ in \eqref{ratiok1} and $k=2$ in table \ref{ratios} are convergent there.
Indeed, instead of the three-dimensional fundamental domain \eqref{22FD} for $(M_0,M_1,M_3)$, we can consider the five-dimensional fundamental domain for $({\widetilde M}_i)_{i=1}^5=(M_0,M_1,M_3,Z_1,Z_3)$
\begin{align}
|{\widetilde M}_i|+|{\widetilde M}_j|\le k,
\label{D5FD}
\end{align}
which is determined from the $D_5$ Weyl group \cite{FMMN}.
Then, we find that the ratios seem to be convergent in the five-dimensional domain.
Let us explain it explicitly with the above example \eqref{ratiok1} with $k=1$ and $N=1$.
For $\Delta_2^{[1]}$, if we substitute the relative ranks $(M_0,M_1,M_3)$ into the inequalities for the fundamental domain \eqref{D5FD}, the FI parameters $(Z_1,Z_3)$ are restricted to the origin.
Since we expect that the ratio is well-defined in the fundamental domain, $\Delta_2^{[1]}$ should be convergent in the limit $(Z_1,Z_3)\to(0,0)$.
Indeed, in the limit we find
\begin{align}
\lim_{(Z_1,Z_3)\to(0,0)}\Delta_2^{[1]}=\frac{1}{4\pi^2}.
\end{align}
Similarly, for $\Delta_1^{[1]}$ and $\Delta_3^{[1]}$, the inequalities for the fundamental domain \eqref{D5FD} are $|Z_1|\le\frac{1}{2}$, $|Z_3|\le\frac{1}{2}$.
We expect the ratios to be convergent in this domain, in spite of the factors of $\sec(\pi Z_1)$ and $\sec(\pi Z_3)$.
Indeed by taking the boundary limit, we find ($\bar 1=3,\bar 3=1$ as in \eqref{prefactor})
\begin{align}
\lim_{Z_i\to\pm\frac{1}{2}}\Delta_1^{[1]}
&=\frac{\sec(\pi Z_{\bar i})}{32\pi}
\big(\pi\mp 8iZ_{\bar i}-4\pi Z_{\bar i}^2\pm\pi i(1-4Z_{\bar i}^2\tan(\pi Z_{\bar i}))\big),\nonumber\\
\lim_{Z_i\to\pm\frac{1}{2}}\Delta_3^{[1]}
&=\frac{\sec(\pi Z_{\bar i})}{32\pi}
\big(\pi\pm 8iZ_{\bar i}-4\pi Z_{\bar i}^2\mp\pi i(1-4Z_{\bar i}^2\tan(\pi Z_{\bar i}))\big),
\label{Delta13limit}
\end{align}
which are convergent for $|Z_{\bar i}|\le\frac{1}{2}$ (see figure \ref{Delta13}).
For $\Delta_0^{[1]}$, the inequalities for the fundamental domain \eqref{D5FD} reduce to $|Z_1|+|Z_3|\le 1$ where the ratio is manifestly finite except on $Z_i=0$.
By taking the limit $Z_i\to 0$, we find
\begin{align}
\lim_{Z_i\to 0}\Delta_0^{[1]}
=\frac{\csc(\pi Z_{\bar i})}{12\pi Z_{\bar i}}\big(1-3Z_{\bar i}^2-\pi Z_{\bar i}\cot(\pi Z_{\bar i})+\pi Z_{\bar i}^3\cot(\pi Z_{\bar i})\big),
\label{Delta2limit}
\end{align}
which converges again (see figure \ref{Delta2}).

\section{Conclusion}

In this paper, we have shown that, once the grand partition function satisfies the bilinear form of the $q$-Painlev\'e equations in the fundamental domain, it also satisfies the same equations outside the fundamental domain.

The extension from the finite fundamental domain into the infinite parameter space is possible since we have clarified the affine symmetries including the translations associated to duality cascades by correctly identifying the transformations of the grand partition function.
The transformation of the lowest power of the fugacity $\kappa$ or the lowest overall rank is identified from the physical argument of duality cascades, while the overall factor appearing in the transformation is proposed by the analysis of the remaining partition function from a few slightly different approaches.
For the ABJM theory the extra factor is absent (i.e.~completely absorbed into our convention for the overall phase of the partition function \eqref{ABJMMM}), while for the four-node $(2,2)$ model it is conjectured in \eqref{22cascades}.

We have made various progresses compared with previous works.
In \cite{BGT3} the grand partition function of the ABJM theory was found to satisfy the bilinear form of $q\text{P}_{\text{III}_3}$, without constructing the extension outside the fundamental domain.
In \cite{HK} the extension of the grand partition function outside the fundamental domain was studied explicitly from the matrix model, though the studies of $q\text{P}_{\text{III}_3}$ was missing.
In \cite{FMMN,FMS} the concept of the fundamental domain of duality cascades was introduced and pointed out to be the affine Weyl chamber, though the computation of the grand partition function was missing.
In \cite{MN9} we made the first step to extend the domain of definition for the grand partition function of the $(2,2)$ model slightly beyond the fundamental domain, where we found the cascade relations \eqref{22cascades} in a few examples.
Here we have extended the analysis further and provided a formal derivation for the cascade relations and show that, with the cascade relations, the 40 $q\text{P}_\text{VI}$ bilinear equations (as well as the $q\text{P}_{\text{III}_3}$ bilinear equation) originally found in the fundamental domain continue to be valid outside.

We shall list in the following several future directions we wish to pursue.

In our studies of the matrix models, the convergence condition is obscure.
Nevertheless, we adopt a prescription with residue sums and study the 5-brane exchanges with it.
Especially, unlike the exchange of 5-branes of different kinds studied in \cite{HK}, an interesting factor \eqref{Gamma} appears in exchanging 5-branes of the same kind.
This suggests that a special care is needed when we discuss parallel 5-branes.
Similarly, in the studies of five-dimensional partition functions with topological vertices \cite{Iqbal:2002we,Aganagic:2003db,Iqbal:2007ii,Taki:2007dh}, it was found that, when there are adjacent parallel 5-branes, we need to subtract a factor originating from extra degrees of freedom of $\text{U}(1)$ \cite{Bao:2013pwa,Hayashi:2013qwa,Taki:2013vka,Taki:2014pba}.
These two situations with parallel 5-branes look similar and it would be interesting to clarify their relations.

Although we have clarified symmetries of the affine Weyl groups for the three-dimensional super Chern-Simons theories, most of questions listed in \cite{MN9} remain unanswered.
Especially, after clarifying the affine symmetries it is interesting to understand the relation to the five-dimensional theories.

So far the examples for the $q$-Painlev\'e equations and the affine Weyl groups at hand are only the ABJM theory and the four-node $(2,2)$ model.
We would like to explore more examples.
One direction would be the $q$-Painlev\'e equations with the affine $E_7$ Weyl group \cite{KMN,FurukawaMSugimoto,Furukawa:2020cjp} and another would be the massive ABJM theory \cite{Nosaka:2020tyv}.
It would be interesting to investigate their bilinear equations.

It is interesting to observe that the pattern of the non-vanishing partition functions in figure \ref{ABJMGC} resembles closely to the representations of the affine $A_1$ Lie algebra (see, for example, figure 14.4 and table 14.4 in \cite{CFTDiFrancesco}).
This is not surprising since, as we have already explained, the partition functions enjoy symmetries of the affine $A_1$ Weyl group.
However, it is not clear whether the pattern is exactly a representation of the affine $A_1$ Lie algebra.
Especially, for representations of affine Lie algebras, it is crucial to understand multiplicities of weights.
The role of multiplicities in the ABJM theory is unclear to us.
This question also applies to the $(2,2)$ model.

\appendix
\section{Exact expressions of partition functions}
\label{lowestk12}

In this appendix we list results on the exact expressions of partition functions for the four-node $(2,2)$ model with short explanations.
The following two subsections are devoted respectively to the cases of $k=1$ and $k=2$.
 
\subsection{$k=1$}
\label{k1}

\begin{table}[!t]
\centering\footnotesize
\begin{minipage}[t]{.3\textwidth}
\centering
\begin{tabular}{c|ccc}
$M_0=-2$&$-1$&$0$&$1$\\\hline
$1$&&\cellcolor[gray]{0.9}$\epsilon^2$&\\
$0$&\cellcolor[gray]{0.9}$-1/s_1^2$&\cellcolor[gray]{0.9}$\epsilon S/(s_1s_3)$&\cellcolor[gray]{0.9}$\epsilon^2$\\
$-1$&&\cellcolor[gray]{0.9}$-1/s_3^2$&
\end{tabular}
\end{minipage}
\quad\quad\quad
\begin{minipage}[t]{.6\textwidth}
\centering
\begin{tabular}{c|cccc}
$M_0=-3/2$&$-3/2$&$-1/2$&$1/2$&$3/2$\\\hline
$3/2$&&\cellcolor[gray]{0.9}$\epsilon^2/(c_1c_3)$&\cellcolor[gray]{0.9}$\epsilon^2/c_3$&\\
$1/2$&\cellcolor[gray]{0.9}$1/c_1^4$&\cellcolor[gray]{0.7}$1/c_1$&\cellcolor[gray]{0.7}$\epsilon^2$&\cellcolor[gray]{0.9}$\epsilon^2/c_1$\\
$-1/2$&\cellcolor[gray]{0.9}$\epsilon^{-2}/(c_1^4c_3)$&\cellcolor[gray]{0.7}$1/(c_1c_3)$&\cellcolor[gray]{0.7}$1/c_3$&\cellcolor[gray]{0.9}$\epsilon^2/(c_1c_3)$\\
$-3/2$&&\cellcolor[gray]{0.9}$\epsilon^{-2}/(c_1c_3^4)$&\cellcolor[gray]{0.9}$1/c_3^4$&
\end{tabular}
\end{minipage}

\begin{tabular}{c|ccccc}
$M_0=-1$&$-2$&$-1$&$0$&$1$&$2$\\\hline
$2$&&&\cellcolor[gray]{0.9}$-\epsilon^2/s_3^2$&&\\
$1$&&\cellcolor[gray]{0.7}$-1/s_1^2$&\cellcolor[gray]{0.7}$\epsilon S/(s_1s_3)$&\cellcolor[gray]{0.7}$\epsilon^2$&\\
$0$&\cellcolor[gray]{0.9}$-\epsilon^{-2}/s_1^6$&\cellcolor[gray]{0.7}$-\epsilon^{-1}S/(s_1^3s_3)$&\cellcolor[gray]{0.5}$1$&\cellcolor[gray]{0.7}$\epsilon S/(s_1s_3)$&\cellcolor[gray]{0.9}$-\epsilon^2/s_1^2$\\
$-1$&&\cellcolor[gray]{0.7}$\epsilon^{-2}/(s_1^2s_3^2)$&\cellcolor[gray]{0.7}$-\epsilon^{-1}S/(s_1s_3^3)$&\cellcolor[gray]{0.7}$-1/s_3^2$&\\
$-2$&&&\cellcolor[gray]{0.9}$-\epsilon^{-2}/s_3^6$&&
\end{tabular}

\begin{tabular}{c|cccc}
$M_0=-1/2$&$-3/2$&$-1/2$&$1/2$&$3/2$\\\hline
$3/2$&\cellcolor[gray]{0.9}$1/(c_1^4c_3)$&\cellcolor[gray]{0.7}$1/(c_1c_3)$&\cellcolor[gray]{0.7}$\epsilon^2/c_3$&\cellcolor[gray]{0.9}$\epsilon^2/(c_1c_3)$\\
$1/2$&\cellcolor[gray]{0.7}$\epsilon^{-2}/c_1^4$&\cellcolor[gray]{0.5}$1/c_1$&\cellcolor[gray]{0.5}$1$&\cellcolor[gray]{0.7}$\epsilon^2/c_1$\\
$-1/2$&\cellcolor[gray]{0.7}$\epsilon^{-2}/(c_1^4c_3)$&\cellcolor[gray]{0.5}$\epsilon^{-2}/(c_1c_3)$&\cellcolor[gray]{0.5}$1/c_3$&\cellcolor[gray]{0.7}$1/(c_1c_3)$\\
$-3/2$&\cellcolor[gray]{0.9}$\epsilon^{-4}/(c_1^4c_3^4)$&\cellcolor[gray]{0.7}$\epsilon^{-2}/(c_1c_3^4)$&\cellcolor[gray]{0.7}$\epsilon^{-2}/c_3^4$&\cellcolor[gray]{0.9}$1/(c_1c_3^4)$
\end{tabular}

\begin{tabular}{c|ccccc}
$M_0=0$&$-2$&$-1$&$0$&$1$&$2$\\\hline
$2$&&\cellcolor[gray]{0.9}$1/(s_1^2s_3^2)$&\cellcolor[gray]{0.9}$-\epsilon S/(s_1s_3^3)$&\cellcolor[gray]{0.9}$-\epsilon^2/s_3^2$&\\
$1$&\cellcolor[gray]{0.9}$-\epsilon^{-2}/s_1^6$&\cellcolor[gray]{0.7}$-\epsilon^{-1}S/(s_1^3s_3)$&\cellcolor[gray]{0.5}$1$&\cellcolor[gray]{0.7}$\epsilon S/(s_1s_3)$&\cellcolor[gray]{0.9}$-\epsilon^2/s_1^2$\\
$0$&\cellcolor[gray]{0.9}$-\epsilon^{-3}S/(s_1^7s_3)$&\cellcolor[gray]{0.5}$-\epsilon^{-2}/s_1^2$&\cellcolor[gray]{0.5}$\epsilon^{-1}S/(s_1s_3)$&\cellcolor[gray]{0.5}$1$&\cellcolor[gray]{0.9}$-\epsilon S/(s_1^3s_3)$\\
$-1$&\cellcolor[gray]{0.9}$\epsilon^{-4}/(s_1^6s_3^2)$&\cellcolor[gray]{0.7}$\epsilon^{-3}S/(s_1^3s_3^3)$&\cellcolor[gray]{0.5}$-\epsilon^{-2}/s_3^2$&\cellcolor[gray]{0.7}$-\epsilon^{-1}S/(s_1s_3^3)$&\cellcolor[gray]{0.9}$1/(s_1^2s_3^2)$\\
$-2$&&\cellcolor[gray]{0.9}$\epsilon^{-4}/(s_1^2s_3^6)$&\cellcolor[gray]{0.9}$-\epsilon^{-3}S/(s_1s_3^7)$&\cellcolor[gray]{0.9}$-\epsilon^{-2}/s_3^6$&
\end{tabular}

\begin{tabular}{c|cccc}
$M_0=1/2$&$-3/2$&$-1/2$&$1/2$&$3/2$\\\hline
$3/2$&\cellcolor[gray]{0.9}$\epsilon^{-2}/(c_1^4c_3)$&\cellcolor[gray]{0.7}$1/(c_1c_3)$&\cellcolor[gray]{0.7}$1/c_3$&\cellcolor[gray]{0.9}$\epsilon^2/(c_1c_3)$\\
$1/2$&\cellcolor[gray]{0.7}$\epsilon^{-2}/c_1^4$&\cellcolor[gray]{0.5}$\epsilon^{-2}/c_1$&\cellcolor[gray]{0.5}$1$&\cellcolor[gray]{0.7}$1/c_1$\\
$-1/2$&\cellcolor[gray]{0.7}$\epsilon^{-4}/(c_1^4c_3)$&\cellcolor[gray]{0.5}$\epsilon^{-2}/(c_1c_3)$&\cellcolor[gray]{0.5}$\epsilon^{-2}/c_3$&\cellcolor[gray]{0.7}$1/(c_1c_3)$\\
$-3/2$&\cellcolor[gray]{0.9}$\epsilon^{-4}/(c_1^4c_3^4)$&\cellcolor[gray]{0.7}$\epsilon^{-4}/(c_1c_3^4)$&\cellcolor[gray]{0.7}$\epsilon^{-2}/c_3^4$&\cellcolor[gray]{0.9}$\epsilon^{-2}/(c_1c_3^4)$
\end{tabular}

\begin{tabular}{c|ccccc}
$M_0=1$&$-2$&$-1$&$0$&$1$&$2$\\\hline
$2$&&&\cellcolor[gray]{0.9}$-1/s_3^2$&&\\
$1$&&\cellcolor[gray]{0.7}$-\epsilon^{-2}/s_1^2$&\cellcolor[gray]{0.7}$\epsilon^{-1}S/(s_1s_3)$&\cellcolor[gray]{0.7}$1$&\\
$0$&\cellcolor[gray]{0.9}$-\epsilon^{-4}/s_1^6$&\cellcolor[gray]{0.7}$-\epsilon^{-3}S/(s_1^3s_3)$&\cellcolor[gray]{0.5}$\epsilon^{-2}$&\cellcolor[gray]{0.7}$\epsilon^{-1}S/(s_1s_3)$&\cellcolor[gray]{0.9}$-1/s_1^2$\\
$-1$&&\cellcolor[gray]{0.7}$\epsilon^{-4}/(s_1^2s_3^2)$&\cellcolor[gray]{0.7}$-\epsilon^{-3}S/(s_1s_3^3)$&\cellcolor[gray]{0.7}$-\epsilon^{-2}/s_3^2$&\\
$-2$&&&\cellcolor[gray]{0.9}$-\epsilon^{-4}/s_3^6$&&
\end{tabular}

\noindent
\begin{minipage}[t]{.55\textwidth}
\centering
\begin{tabular}{c|cccc}
$M_0=3/2$&$-3/2$&$-1/2$&$1/2$&$3/2$\\\hline
$3/2$&&\cellcolor[gray]{0.9}$\epsilon^{-2}/(c_1c_3)$&\cellcolor[gray]{0.9}$1/c_3$&\\
$1/2$&\cellcolor[gray]{0.9}$\epsilon^{-4}/c_1^4$&\cellcolor[gray]{0.7}$\epsilon^{-2}/c_1$&\cellcolor[gray]{0.7}$\epsilon^{-2}$&\cellcolor[gray]{0.9}$1/c_1$\\
$-1/2$&\cellcolor[gray]{0.9}$\epsilon^{-4}/(c_1^4c_3)$&\cellcolor[gray]{0.7}$\epsilon^{-4}/(c_1c_3)$&\cellcolor[gray]{0.7}$\epsilon^{-2}/c_3$&\cellcolor[gray]{0.9}$\epsilon^{-2}/(c_1c_3)$\\
$-3/2$&&\cellcolor[gray]{0.9}$\epsilon^{-4}/(c_1c_3^4)$&\cellcolor[gray]{0.9}$\epsilon^{-4}/c_3^4$&
\end{tabular}
\end{minipage}
\quad\quad
\begin{minipage}[t]{.4\textwidth}
\centering
\begin{tabular}{c|ccc}
$M_0=2$&$-1$&$0$&$1$\\\hline
$1$&&\cellcolor[gray]{0.9}$\epsilon^{-2}$&\\
$0$&\cellcolor[gray]{0.9}$-\epsilon^{-4}/s_1^2$&\cellcolor[gray]{0.9}$\epsilon^{-3}S/(s_1s_3)$&\cellcolor[gray]{0.9}$\epsilon^{-2}$\\
$-1$&&\cellcolor[gray]{0.9}$-\epsilon^{-4}/s_3^2$&
\end{tabular}
\end{minipage}
\caption{The expressions of the lowest non-vanishing partition functions for $k=1$.
We list them horizontally by the value of $M_1$ and vertically by the value of $M_3$.
}
\label{k1data}
\end{table}

Using the strategy explained in section \ref{translation}, we can study the lowest non-vanishing partition functions for the four-node $(2,2)$ model with $k=1$ even outside the fundamental domain.
Namely, since the grand partition function normalized by the lowest non-vanishing partition function is given by the Fredholm determinant of a spectral operator invariant under the Weyl group, the ratio is classified by the Weyl group as in \eqref{Delta[1]}.
Furthermore, from the physical argument of duality cascades, we expect that even outside the fundamental domain, the ratio of the second-lowest non-vanishing partition function to the lowest non-vanishing partition function is identical under the discrete translations of duality cascades, $Z_{1,{\bm M}}(2)/Z_{1,{\bm M}}(1)\big|_{{\cal D}^{[1]}_1\backslash{\cal D}^{[1]}_0}=Z_{1,{\bm M}_\text{FD}}(1)/Z_{1,{\bm M}_\text{FD}}(0)\big|_{{\cal D}^{[1]}_0}$.
Here, as in \eqref{XioverZcascaderule22model}, ${\bm M}_{\text{FD}}$ is the relative rank in ${\cal D}^{[1]}_0$ corresponding to ${\bm M}$ obtained after applying duality cascades \eqref{Deltas}.
Using these facts, we can fix the partition functions $Z_{1,{\bm M}}(2)$ in ${\cal D}^{[1]}_1\backslash{\cal D}^{[1]}_0$ where the lowest non-vanishing partition function is $N=1$.
Then, we can utilize the 40 $q\text{P}_\text{VI}$ bilinear equations to study the lowest partition function outside ${\cal D}_1^{[1]}$.

Here we summarize the exact expressions of partition functions for $k=1$.
We introduce shorthand notations
\begin{align}
&s_1=2\sin\pi Z_1,\quad
s_3=2\sin\pi Z_3,\quad
c_1=2\cos\pi Z_1,\quad
c_3=2\cos\pi Z_3,\nonumber\\
&\epsilon=e^{\pi iZ_1Z_3},\quad
S=2\sin\pi Z_1Z_3.
\label{notationk1}
\end{align}
Then, the exact expressions are given in table \ref{k1data}.
The expressions shaded in dark grey denotes $Z_{M_0,M_1,M_3}^{[1](0)}$ in the fundamental domain ${\cal D}^{[1]}_0$, while those shaded in grey and light grey are respectively $Z_{M_0,M_1,M_3}^{[1](1)}$ in ${\cal D}^{[1]}_1\backslash{\cal D}^{[1]}_0$ and $Z_{M_0,M_1,M_3}^{[1](2)}$ in ${\cal D}^{[1]}_2\backslash{\cal D}^{[1]}_1$.
As in \cite{MN9}, we redefine the partition functions with the phase $\Theta_{k,{\bm M}}$ dropped \eqref{dropphase}.
The results of the expressions $Z_{M_0,M_1,M_3}^{[1](0)}$ and $Z_{M_0,M_1,M_3}^{[1](1)}$ are collected from appendix B.1 and C in \cite{MN9}.
Only the results $Z_{M_0,M_1,M_3}^{[1](2)}$ (in light grey) are new in this paper.

\subsection{$k=2$}
\subsubsection{Higher partition functions}
\label{higherk2}

This appendix is devoted to the ratios invariant under the Weyl group for $k=2$, while the next one is to the lowest partition functions.
Since we assume that the exact expressions of the ratios outside the fundamental domain are repetitions of those inside and those in the fundamental domain enjoy symmetries of the $D_5$ Weyl group after normalized by those of the lowest rank, the exact expressions reduce to the ten ratios
\begin{align}
&\Delta^{[2]}_0=\frac{Z_{0,0,0}^{[2](1)}}{Z_{0,0,0}^{[2](0)}},\quad
\Delta^{[2]}_1=\frac{Z_{\{\pm\frac{1}{2},\pm\frac{1}{2},\pm\frac{1}{2}\}_0}^{[2](1)}}{Z_{\{\pm\frac{1}{2},\pm\frac{1}{2},\pm\frac{1}{2}\}_0}^{[2](0)}},\quad
\Delta^{[2]}_2=\frac{Z_{\{\pm 1,\pm 1,\pm 1\}_0}^{[2](1)}}{Z_{\{\pm 1,\pm 1,\pm 1\}_0}^{[2](0)}},\quad
\Delta^{[2]}_3=\frac{Z_{\{\pm\frac{3}{2},\pm\frac{1}{2},\pm\frac{1}{2}\}_0}^{[2](1)}}{Z_{\{\pm\frac{3}{2},\pm\frac{1}{2},\pm\frac{1}{2}\}_0}^{[2](0)}},\nonumber\\
&\Delta^{[2]}_4=\frac{Z_{\{\pm 2,0,0\}}^{[2](1)}}{Z_{\{\pm 2,0,0\}}^{[2](0)}},\quad
\Delta^{[2]}_5=\frac{Z_{\{\pm\frac{3}{2},\pm\frac{1}{2},\pm\frac{1}{2}\}_1}^{[2](1)}}{Z_{\{\pm\frac{3}{2},\pm\frac{1}{2},\pm\frac{1}{2}\}_1}^{[2](0)}},\quad
\Delta^{[2]}_6=\frac{Z_{\{\pm 1,\pm 1,\pm 1\}_1}^{[2](1)}}{Z_{\{\pm 1,\pm 1,\pm 1\}_1}^{[2](0)}},\quad
\Delta^{[2]}_7=\frac{Z_{\{\pm\frac{1}{2},\pm\frac{1}{2},\pm\frac{1}{2}\}_1}^{[2](1)}}{Z_{\{\pm\frac{1}{2},\pm\frac{1}{2},\pm\frac{1}{2}\}_1}^{[2](0)}},\nonumber\\
&\Gamma^{[2]}_4=\frac{Z_{\{\pm 1,\pm 1,0\}}^{[2](1)}}{Z_{\{\pm 1,\pm 1,0\}}^{[2](0)}},\quad
\Gamma^{[2]}_2=\frac{Z_{\{\pm 1,0,0\}}^{[2](1)}}{Z_{\{\pm 1,0,0\}}^{[2](0)}}.
\label{defratios}
\end{align}
Here, as in section \ref{translation}, the subscripts of $Z$ denote $(M_0,M_1,M_3)$ though we use the curly brackets since the order can be arbitrary.
The subscripts of the curly brackets $0$ or $1$ indicate whether the number of minus signs is even or odd.

\begin{table}[!t]
\centering\footnotesize
\begin{align}
&\Delta_0^{[2]}(Z_1,Z_3)=\frac{1}{8}Z_1Z_3\csc(\pi Z_1)\csc(\pi Z_3)\nonumber\\
&\quad+\frac{1}{2Y}\csc(\pi Z_1)\csc(\pi Z_3)\sin(\pi Z_1Z_3)\big(Z_1\cot(\pi Z_1)+Z_3\cot(\pi Z_3)\big)\nonumber\\
&\quad+\frac{1}{2\pi Y}\csc^2(\pi Z_1)\csc^2(\pi Z_3)\Big({-\sin(\pi Z_1)\sin(\pi Z_3)\sin(\pi Z_1Z_3)}\nonumber\\
&\qquad+4\pi\Big(\sin^4\!\Big(\frac{\pi Z_1}{2}\Big)\sin^4\!\Big(\frac{\pi Z_3}{2}\Big)-\cos(\pi Z_1)\cos(\pi Z_3)
\sin^2\!\Big(\frac{\pi Z_1Z_3}{2}\Big)\Big)\Big),\nonumber\\
&\Delta_{1,7}^{[2]}(Z_1,Z_3)=\pm\frac{i}{8}Z_1Z_3\sec(\pi Z_1)\sec(\pi Z_3)\nonumber\\
&\quad-\frac{1}{16\pi X^\pm}\sec(\pi Z_1)\sec(\pi Z_3)\big(1+\pi (Z_1\tan(\pi Z_1)+Z_3\tan(\pi Z_3))\big)\big(X^\pm+8)\nonumber\\
&\quad+\frac{1}{64X^\pm}\sec^2(\pi Z_1)\sec^2(\pi Z_3)
\Big(4\cos(\pi Z_1)+4\cos(\pi Z_3)\pm 40i\sin(\pi Z_1)\sin(\pi Z_3)-8e^{\mp 2\pi iZ_1Z_3}\nonumber\\
&\quad+e^{\mp\pi iZ_1Z_3}\Big(12e^{\pm\frac{\pi i}{4}}e^{-\frac{\pi i}{2}(Z_1+Z_3)}(1\mp ie^{\pi iZ_1}\mp ie^{\pi iZ_3}+e^{\pi i(Z_1+Z_3)})
\nonumber\\
&\qquad+5e^{\pm\frac{3\pi i}{4}}\Big(e^{-\frac{\pi i}{2}(3Z_1+Z_3)}(1\pm ie^{3\pi iZ_1}\pm ie^{\pi iZ_3}+e^{\pi i(3Z_1+Z_3)})
+(Z_1\leftrightarrow Z_3)\Big)\nonumber\\
&\qquad+2e^{\pm\frac{\pi i}{4}}e^{-\frac{3\pi i}{2}(Z_1+Z_3)}(1\mp ie^{3\pi iZ_1}\mp ie^{3\pi iZ_3}+e^{3\pi i(Z_1+Z_3)})\Big)\Big),\nonumber\\
&\Delta_{2,6}^{[2]}(Z_1,Z_3)=-\frac{1}{8}Z_1Z_3\csc(\pi Z_1)\csc(\pi Z_3)
\mp\frac{i}{8}\csc(\pi Z_1)\csc(\pi Z_3)
\big(Z_1\cot(\pi Z_1)+Z_3\cot(\pi Z_3)\big)\nonumber\\
&\quad\pm\frac{i}{8\pi}\csc(\pi Z_1)\csc(\pi Z_3)
+\frac{1}{8}\cot(\pi Z_1)\cot(\pi Z_3)\csc(\pi Z_1)\csc(\pi Z_3)\nonumber\\
&\quad+\frac{1}{16}e^{\mp\pi iZ_1Z_3}\csc^2(\pi Z_1)\csc^2(\pi Z_3)
\big(1-\cos(\pi Z_1)-\cos(\pi Z_3)-\cos(\pi Z_1)\cos(\pi Z_3)\big),\nonumber\\
&\Delta_{3,5}^{[2]}(Z_1,Z_3)=\mp\frac{i}{8}Z_1Z_3\sec(\pi Z_1)\sec(\pi Z_3)
-\frac{1}{16}\sec(\pi Z_1)\sec(\pi Z_3)
\big(Z_1\tan(\pi Z_1)+Z_3\tan(\pi Z_3)\big)\nonumber\\
&\quad-\frac{1}{64\pi}\sec^2(\pi Z_1)\sec^2(\pi Z_3)
\Big(4\cos(\pi Z_1)\cos(\pi Z_3)
\mp 4\pi i\sin(\pi Z_1)\sin(\pi Z_3)\nonumber\\
&\quad
-2\pi\big(e^{\mp 2\pi iZ_1Z_3}(X^{\mp}+2)
-\cos(\pi Z_1)-\cos(\pi Z_3)\big)\Big),\nonumber\\
&\Delta_4^{[2]}(Z_1,Z_3)=\frac{1}{8}Z_1Z_3\csc(\pi Z_1)\csc(\pi Z_3),\nonumber\\
&\Gamma_4^{[2]}(Z_1,Z_3)=\frac{1}{8}Z_1Z_3\csc(\pi Z_1)\csc(\pi Z_3)+\frac{1}{16}\csc^2(\pi Z_1)\csc^2(\pi Z_3)Y,\nonumber\\
&\Gamma_2^{[2]}(Z_1,Z_3)=-\frac{1}{8}Z_1Z_3\csc(\pi Z_1)\csc(\pi Z_3)
+\frac{1}{8\pi}\csc(\pi Z_1)\csc(\pi Z_3)\cot\!\Big(\frac{\pi Z_1Z_3}{2}\Big)\nonumber\\
&\quad-\frac{1}{8}\csc(\pi Z_1)\csc(\pi Z_3)\cot\!\Big(\frac{\pi Z_1Z_3}{2}\Big)\big(Z_1\cot(\pi Z_1)+Z_3\cot(\pi Z_3)\big)\nonumber\\
&\quad-\frac{1}{16}\csc^2(\pi Z_1)\csc^2(\pi Z_3)\big(1-\cos(\pi Z_1)-\cos(\pi Z_3)-3\cos(\pi Z_1)\cos(\pi Z_3)\big).\nonumber
\end{align}
\caption{
Exact expressions for various ratios $Z_{2,{\bm M}}(1)/Z_{2,{\bm M}}(0)=Z^{[2](1)}_{M_0,M_1,M_3}/Z^{[2](0)}_{M_0,M_1,M_3}$ \eqref{defratios} of the partition functions $Z_{k,{\bm M}}(N)$.
For double signs in $\Delta_{1,7}^{[2]}(Z_1,Z_3)$, $\Delta_{2,6}^{[2]}(Z_1,Z_3)$ and $\Delta_{3,5}^{[2]}(Z_1,Z_3)$, the upper and lower ones denote respectively those for $\Delta_1^{[2]}$, $\Delta_2^{[2]}$, $\Delta_3^{[2]}$ and $\Delta_7^{[2]}$, $\Delta_6^{[2]}$, $\Delta_5^{[2]}$.
}
\label{ratios}
\end{table}

As stressed in \cite{MN9} the rank variable $L$ detects the depth from a boundary of the fundamental domain and the size of the determinant increases with $L$.
Here we concentrate on the ratios on the boundary of the fundamental domain, $\Delta_{2,3,4,5,6}^{[2]}$ and $\Gamma_4^{[2]}$, and compute them with the techniques equipped in \cite{MN9}.
After that using the 40 $q\text{P}_\text{VI}$ bilinear equations \eqref{D5bilinear} we can determine the remaining ratios away from the boundary.
We have found the ratios for $\Delta_{0,1,7}^{[2]}$ and $\Gamma_2^{[2]}$ only from several selected bilinear equations in \eqref{D5bilinear}, which are easier to solve.
Once they are found, we can check numerically that the ratios satisfy all of the 40 bilinear equations to very high accuracy (about 400 digits).
The results are listed as in table \ref{ratios}.
We simplify the expression by introducing
\begin{align}
X^\pm&=e^{\mp\pi iZ_1Z_3}e^{\mp\frac{\pi i}{4}}e^{-\frac{\pi i}{2}(Z_1+Z_3)}(1\pm ie^{\pi iZ_1}\pm ie^{\pi iZ_3}+e^{\pi i(Z_1+Z_3)})-2,\nonumber\\
Y&=1+\cos(\pi Z_1)+\cos(\pi Z_3)-\cos(\pi Z_1)\cos(\pi Z_3)-2\cos(\pi Z_1Z_3),
\end{align}
which enjoy symmetries
\begin{align}
&X^\pm(Z_1,Z_3)=X^\pm(Z_3,Z_1)=X^\pm(-Z_1,-Z_3),
\quad X^-(Z_1,Z_3)=X^+(Z_1,-Z_3),\nonumber\\
&Y(Z_1,Z_3)=Y(Z_3,Z_1)=Y(-Z_1,Z_3).
\end{align}

\subsubsection{Lowest partition functions}\label{k2}

In this subsection we list the exact expressions of partition functions for $k=2$, by substituting the results in table \ref{ratios} into the 40 $q\text{P}_\text{VI}$ bilinear equations.
In addition to \eqref{notationk1} we also need to introduce notations
\begin{align}
s'_1=2\sin\frac{\pi}{2}Z_1,\quad
s'_3=2\sin\frac{\pi}{2}Z_3,\quad
c'_1=2\cos\frac{\pi}{2}Z_1,\quad
c'_3=2\cos\frac{\pi}{2}Z_3,\quad
S'=2\sin\frac{\pi}{2}Z_1Z_3.
\label{notationk2}
\end{align}

\begin{table}[!t]
\centering\footnotesize
\begin{minipage}[t]{.3\textwidth}
\centering
\underline{$M_0=-5/2, (M_1,M_3\in\{-1/2,1/2\})$}
\begin{tabular}{cc}
\cellcolor[gray]{0.7}$1/(\sqrt{2}c_1)$&\cellcolor[gray]{0.7}$\epsilon/\sqrt{2}$\\
\cellcolor[gray]{0.7}$1/(\sqrt{2}c_1c_3)$&\cellcolor[gray]{0.7}$1/(\sqrt{2}c_3)$\\
\end{tabular}
\end{minipage}
\quad\quad\quad
\begin{minipage}[t]{.5\textwidth}
\centering
\underline{$M_0=-2, (M_1,M_3\in\{-1,0,1\})$}
\begin{tabular}{ccc}
\cellcolor[gray]{0.7}$-1/(2s_1^2)$&\cellcolor[gray]{0.7}$\epsilon^{\frac{1}{2}}S'/(2s'_3s_1)$&\cellcolor[gray]{0.7}$\epsilon/2$\\
\cellcolor[gray]{0.7}$-\epsilon^{-\frac{1}{2}}S'/(2s'_1s_1^2s_3)$&\cellcolor[gray]{0.5}$1$&\cellcolor[gray]{0.7}$\epsilon^{\frac{1}{2}}S'/(2s'_1s_3)$\\
\cellcolor[gray]{0.7}$\epsilon^{-1}/(2s_1^2s_3^2)$&\cellcolor[gray]{0.7}$-\epsilon^{-\frac{1}{2}}S'/(2s'_3s_1s_3^2)$&\cellcolor[gray]{0.7}$-1/(2s_3^2)$
\end{tabular}
\end{minipage}

\vspace{0.5cm}
\underline{$M_0=-3/2, (M_1,M_3\in\{-3/2,-1/2,1/2,3/2\})$}\\
\begin{tabular}{cccc}
\cellcolor[gray]{0.7}$1/(\sqrt{2}c_1^3)$&\cellcolor[gray]{0.7}$\epsilon X^+/(2\sqrt{2}c_1^2c_3)$&\cellcolor[gray]{0.7}$X^-/(2\sqrt{2}c_1c_3)$&\cellcolor[gray]{0.7}$\epsilon/\sqrt{2}$\\
\cellcolor[gray]{0.7}$X^+/(2\sqrt{2}c_1^4c_3)$&\cellcolor[gray]{0.5}$1/(\sqrt{2}c_1)$&\cellcolor[gray]{0.5}$1/\sqrt{2}$&\cellcolor[gray]{0.7}$X^-/(2\sqrt{2}c_1c_3)$\\
\cellcolor[gray]{0.7}$\epsilon^{-2}X^-/(2\sqrt{2}c_1^4c_3^2)$&\cellcolor[gray]{0.5}$\epsilon^{-1}/(\sqrt{2}c_1c_3)$&\cellcolor[gray]{0.5}$1/(\sqrt{2}c_3)$&\cellcolor[gray]{0.7}$\epsilon X^+/(2\sqrt{2}c_1c_3^2)$\\
\cellcolor[gray]{0.7}$\epsilon^{-2}/(\sqrt{2}c_1^3c_3^3)$&\cellcolor[gray]{0.7}$\epsilon^{-2}X^-/(2\sqrt{2}c_1^2c_3^4)$&\cellcolor[gray]{0.7}$X^+/(2\sqrt{2}c_1c_3^4)$&\cellcolor[gray]{0.7}$1/(\sqrt{2}c_3^3)$
\end{tabular}

\vspace{0.5cm}
\underline{$M_0=-1, (M_1,M_3\in\{-2,-1,0,1,2\})$}\\
\begin{tabular}{ccccc}
&\cellcolor[gray]{0.7}$-1/(2c'_3s_1^2)$&\cellcolor[gray]{0.7}$\epsilon^{\frac{1}{2}}S'/(2s_1s_3)$&\cellcolor[gray]{0.7}$\epsilon/(2c'_3)$&\\
\cellcolor[gray]{0.7}$\epsilon^{-1}/(2c'_1s_1^4)$&\cellcolor[gray]{0.5}$-1/s_1^2$&\cellcolor[gray]{0.5}$1/(2c'_1)$&\cellcolor[gray]{0.5}$1$&\cellcolor[gray]{0.7}$\epsilon/(2c'_1)$\\
\cellcolor[gray]{0.7}$\epsilon^{-\frac{3}{2}}S'/(2s_1^5s_3)$&\cellcolor[gray]{0.5}$-\epsilon^{-1}/(2c'_3s_1^2)$&\cellcolor[gray]{0.5}$\epsilon^{-\frac{1}{2}}S'/(2s_1s_3)$&\cellcolor[gray]{0.5}$1/(2c'_3)$&\cellcolor[gray]{0.7}$\epsilon^{\frac{1}{2}}S'/(2s_1s_3)$\\
\cellcolor[gray]{0.7}$-\epsilon^{-2}/(2c'_1s_1^4s_3^2)$&\cellcolor[gray]{0.5}$\epsilon^{-2}/(s_1^2s_3^2)$&\cellcolor[gray]{0.5}$-\epsilon^{-1}/(2c'_1s_3^2)$&\cellcolor[gray]{0.5}$-1/s_3^2$&\cellcolor[gray]{0.7}$-1/(2c'_1s_3^2)$\\
&\cellcolor[gray]{0.7}$-\epsilon^{-2}/(2c'_3s_1^2s_3^4)$&\cellcolor[gray]{0.7}$\epsilon^{-\frac{3}{2}}S'/(2s_1s_3^5)$&\cellcolor[gray]{0.7}$\epsilon^{-1}/(2c'_3s_3^4)$&
\end{tabular}

\vspace{0.5cm}
\underline{$M_0=-1/2, (M_1,M_3\in\{-5/2,-3/2,-1/2,1/2,3/2,5/2\})$}
\begin{tabular}{cccccc}
&&\cellcolor[gray]{0.7}$1/(\sqrt{2}c_1c_3)$&\cellcolor[gray]{0.7}$\epsilon/(\sqrt{2}c_3)$&&\\
&\cellcolor[gray]{0.7}$X^+/(2\sqrt{2}c_1^4c_3)$&\cellcolor[gray]{0.5}$1/(\sqrt{2}c_1)$&\cellcolor[gray]{0.5}$1/\sqrt{2}$&\cellcolor[gray]{0.7}$X^-/(2\sqrt{2}c_1c_3)$&\\
\cellcolor[gray]{0.7}$\epsilon^{-2}/(\sqrt{2}c_1^6)$&\cellcolor[gray]{0.5}$\epsilon^{-1}/(\sqrt{2}c_1^3)$&\cellcolor[gray]{0.5}$X^+/(2\sqrt{2}c_1^2c_3)$&\cellcolor[gray]{0.5}$\epsilon^{-1}X^-/(2\sqrt{2}c_1c_3)$&\cellcolor[gray]{0.5}$1/\sqrt{2}$&\cellcolor[gray]{0.7}$\epsilon/(\sqrt{2}c_1)$\\
\cellcolor[gray]{0.7}$\epsilon^{-2}/(\sqrt{2}c_1^6c_3)$&\cellcolor[gray]{0.5}$\epsilon^{-2}/(\sqrt{2}c_1^3c_3)$&\cellcolor[gray]{0.5}$\epsilon^{-2}X^-/(2\sqrt{2}c_1^2c_3^2)$&\cellcolor[gray]{0.5}$X^+/(2\sqrt{2}c_1c_3^2)$&\cellcolor[gray]{0.5}$1/(\sqrt{2}c_3)$&\cellcolor[gray]{0.7}$1/(\sqrt{2}c_1c_3)$\\
&\cellcolor[gray]{0.7}$\epsilon^{-3}X^-/(2\sqrt{2}c_1^4c_3^4)$&\cellcolor[gray]{0.5}$\epsilon^{-2}/(\sqrt{2}c_1c_3^3)$&\cellcolor[gray]{0.5}$\epsilon^{-1}/(\sqrt{2}c_3^3)$&\cellcolor[gray]{0.7}$X^+/(2\sqrt{2}c_1c_3^4)$&\\
&&\cellcolor[gray]{0.7}$\epsilon^{-2}/(\sqrt{2}c_1c_3^6)$&\cellcolor[gray]{0.7}$\epsilon^{-2}/(\sqrt{2}c_3^6)$&&
\end{tabular}

\vspace{0.5cm}
\underline{$M_0=0, (M_1,M_3\in\{-2,-1,0,1,2\})$}
\begin{tabular}{ccccc}
&\cellcolor[gray]{0.7}$-\epsilon^{-\frac{1}{2}}S'/(2s'_1s_1^2s_3)$&\cellcolor[gray]{0.5}$1$&\cellcolor[gray]{0.7}$\epsilon^{\frac{1}{2}}S'/(2s'_1s_3)$&\\
\cellcolor[gray]{0.7}$\epsilon^{-\frac{3}{2}}S'/(2s'_3s_1^5)$&\cellcolor[gray]{0.5}$-\epsilon^{-1}/(2s_1^2)$&\cellcolor[gray]{0.5}$\epsilon^{-\frac{1}{2}}S'/(2s'_3s_1)$&\cellcolor[gray]{0.5}$1/2$&\cellcolor[gray]{0.7}$\epsilon^{\frac{1}{2}}S'/(2s'_3s_1)$\\
\cellcolor[gray]{0.5}$\epsilon^{-2}/s_1^4$&\cellcolor[gray]{0.5}$-\epsilon^{-\frac{3}{2}}S'/(2s'_1s_1^2s_3)$&\cellcolor[gray]{0.5}$-\epsilon^{-1}Y/(s_1^2s_3^2)$&\cellcolor[gray]{0.5}$\epsilon^{-\frac{1}{2}}S'/(2s'_1s_3)$&\cellcolor[gray]{0.5}$1$\\
\cellcolor[gray]{0.7}$-\epsilon^{-\frac{5}{2}}S'/(2s'_3s_1^5s_3^2)$&\cellcolor[gray]{0.5}$\epsilon^{-2}/(2s_1^2s_3^2)$&\cellcolor[gray]{0.5}$-\epsilon^{-\frac{3}{2}}S'/(2s'_3s_1s_3^2)$&\cellcolor[gray]{0.5}$-\epsilon^{-1}/(2s_3^2)$&\cellcolor[gray]{0.7}$-\epsilon^{-\frac{1}{2}}S'/(2s'_3s_1s_3^2)$\\
&\cellcolor[gray]{0.7}$-\epsilon^{-\frac{5}{2}}S'/(2s'_1s_1^2s_3^5)$&\cellcolor[gray]{0.5}$\epsilon^{-2}/s_3^4$&\cellcolor[gray]{0.7}$\epsilon^{-\frac{3}{2}}S'/(2s'_1s_3^5)$&
\end{tabular}
\caption{The lowest non-vanishing partition functions for $k=2$.}
\label{k2data1}
\end{table}

Then, the exact expressions are given in tables \ref{k2data1} and \ref{k2data2}.
This time as in table \ref{k1data}, we list the results horizontally by the value of $M_1$ and vertically by the value of $M_3$.
It is always understood that $M_1$ increases from left to right and $M_3$ increases from bottom to top.
The expressions of partition functions $Z^{[2](0)}_{M_0,M_1,M_3}$ in ${\cal D}^{[2]}_0$ and $Z^{[2](1)}_{M_0,M_1,M_3}$ in ${\cal D}^{[2]}_1\backslash{\cal D}^{[2]}_0$ are shaded respectively in dark grey and grey. 
Only the results $Z^{[2](1)}_{M_0,M_1,M_3}$ (in grey) are new.

\begin{table}[!t]
\centering\footnotesize
\underline{$M_0=1/2, (M_1,M_3\in\{-5/2,-3/2,-1/2,1/2,3/2,5/2\})$}
\begin{tabular}{cccccc}
&&\cellcolor[gray]{0.7}$1/(\sqrt{2}c_1c_3)$&\cellcolor[gray]{0.7}$1/(\sqrt{2}c_3)$&&\\
&\cellcolor[gray]{0.7}$\epsilon^{-2}X^-/(2\sqrt{2}c_1^4c_3)$&\cellcolor[gray]{0.5}$\epsilon^{-1}/(\sqrt{2}c_1)$&\cellcolor[gray]{0.5}$1/\sqrt{2}$&\cellcolor[gray]{0.7}$\epsilon X^+/(2\sqrt{2}c_1c_3)$&\\
\cellcolor[gray]{0.7}$\epsilon^{-2}/(\sqrt{2}c_1^6)$&\cellcolor[gray]{0.5}$\epsilon^{-2}/(\sqrt{2}c_1^3)$&\cellcolor[gray]{0.5}$\epsilon^{-2}X^-/(2\sqrt{2}c_1^2c_3)$&\cellcolor[gray]{0.5}$X^+/(2\sqrt{2}c_1c_3)$&\cellcolor[gray]{0.5}$1/\sqrt{2}$&\cellcolor[gray]{0.7}$1/(\sqrt{2}c_1)$\\
\cellcolor[gray]{0.7}$\epsilon^{-3}/(\sqrt{2}c_1^6c_3)$&\cellcolor[gray]{0.5}$\epsilon^{-2}/(\sqrt{2}c_1^3c_3)$&\cellcolor[gray]{0.5}$\epsilon^{-1}X^+/(2\sqrt{2}c_1^2c_3^2)$&\cellcolor[gray]{0.5}$\epsilon^{-2}X^-/(2\sqrt{2}c_1c_3^2)$&\cellcolor[gray]{0.5}$\epsilon^{-1}/(\sqrt{2}c_3)$&\cellcolor[gray]{0.7}$1/(\sqrt{2}c_1c_3)$\\
&\cellcolor[gray]{0.7}$\epsilon^{-2}X^+/(2\sqrt{2}c_1^4c_3^4)$&\cellcolor[gray]{0.5}$\epsilon^{-2}/(\sqrt{2}c_1c_3^3)$&\cellcolor[gray]{0.5}$\epsilon^{-2}/(\sqrt{2}c_3^3)$&\cellcolor[gray]{0.7}$\epsilon^{-2}X^-/(2\sqrt{2}c_1c_3^4)$&\\
&&\cellcolor[gray]{0.7}$\epsilon^{-3}/(\sqrt{2}c_1c_3^6)$&\cellcolor[gray]{0.7}$\epsilon^{-2}/(\sqrt{2}c_3^6)$&&
\end{tabular}

\vspace{0.5cm}
\underline{$M_0=1, (M_1,M_3\in\{-2,-1,0,1,2\})$}\\
\begin{tabular}{ccccc}
&\cellcolor[gray]{0.7}$-\epsilon^{-1}/(2c'_3s_1^2)$&\cellcolor[gray]{0.7}$\epsilon^{-\frac{1}{2}}S'/(2s_1s_3)$&\cellcolor[gray]{0.7}$1/(2c'_3)$&\\
\cellcolor[gray]{0.7}$\epsilon^{-2}/(2c'_1s_1^4)$&\cellcolor[gray]{0.5}$-\epsilon^{-2}/s_1^2$&\cellcolor[gray]{0.5}$\epsilon^{-1}/(2c'_1)$&\cellcolor[gray]{0.5}$1$&\cellcolor[gray]{0.7}$1/(2c'_1)$\\
\cellcolor[gray]{0.7}$\epsilon^{-\frac{5}{2}}S'/(2s_1^5s_3)$&\cellcolor[gray]{0.5}$-\epsilon^{-2}/(2c'_3s_1^2)$&\cellcolor[gray]{0.5}$\epsilon^{-\frac{3}{2}}S'/(2s_1s_3)$&\cellcolor[gray]{0.5}$\epsilon^{-1}/(2c'_3)$&\cellcolor[gray]{0.7}$\epsilon^{-\frac{1}{2}}S'/(2s_1s_3)$\\
\cellcolor[gray]{0.7}$-\epsilon^{-3}/(2c'_1s_1^4s_3^2)$&\cellcolor[gray]{0.5}$\epsilon^{-2}/(s_1^2s_3^2)$&\cellcolor[gray]{0.5}$-\epsilon^{-2}/(2c'_1s_3^2)$&\cellcolor[gray]{0.5}$-\epsilon^{-2}/s_3^2$&\cellcolor[gray]{0.7}$-\epsilon^{-1}/(2c'_1s_3^2)$\\
&\cellcolor[gray]{0.7}$-\epsilon^{-3}/(2c'_3s_1^2s_3^4)$&\cellcolor[gray]{0.7}$\epsilon^{-\frac{5}{2}}S'/(2s_1s_3^5)$&\cellcolor[gray]{0.7}$\epsilon^{-2}/(2c'_3s_3^4)$&
\end{tabular}

\vspace{0.5cm}
\underline{$M_0=3/2, (M_1,M_3\in\{-3/2,-1/2,1/2,3/2\})$}\\
\begin{tabular}{cccc}
\cellcolor[gray]{0.7}$\epsilon^{-2}/(\sqrt{2}c_1^3)$&\cellcolor[gray]{0.7}$\epsilon^{-2}X^-/(2\sqrt{2}c_1^2c_3)$&\cellcolor[gray]{0.7}$X^+/(2\sqrt{2}c_1c_3)$&\cellcolor[gray]{0.7}$1/\sqrt{2}$\\
\cellcolor[gray]{0.7}$\epsilon^{-3}X^-/(2\sqrt{2}c_1^4c_3)$&\cellcolor[gray]{0.5}$\epsilon^{-2}/(\sqrt{2}c_1)$&\cellcolor[gray]{0.5}$\epsilon^{-1}/\sqrt{2}$&\cellcolor[gray]{0.7}$X^+/(2\sqrt{2}c_1c_3)$\\
\cellcolor[gray]{0.7}$\epsilon^{-2}X^+/(2\sqrt{2}c_1^4c_3^2)$&\cellcolor[gray]{0.5}$\epsilon^{-2}/(\sqrt{2}c_1c_3)$&\cellcolor[gray]{0.5}$\epsilon^{-2}/(\sqrt{2}c_3)$&\cellcolor[gray]{0.7}$\epsilon^{-2}X^-/(2\sqrt{2}c_1c_3^2)$\\
\cellcolor[gray]{0.7}$\epsilon^{-3}/(\sqrt{2}c_1^3c_3^3)$&\cellcolor[gray]{0.7}$\epsilon^{-2}X^+/(2\sqrt{2}c_1^2c_3^4)$&\cellcolor[gray]{0.7}$\epsilon^{-3}X^-/(2\sqrt{2}c_1c_3^4)$&\cellcolor[gray]{0.7}$\epsilon^{-2}/(\sqrt{2}c_3^3)$
\end{tabular}

\vspace{0.5cm}
\begin{minipage}[t]{.5\textwidth}
\centering
\underline{$M_0=2, (M_1,M_3\in\{-1,0,1\})$}
\begin{tabular}{ccc}
\cellcolor[gray]{0.7}$-\epsilon^{-2}/(2s_1^2)$&\cellcolor[gray]{0.7}$\epsilon^{-\frac{3}{2}}S'/(2s'_3s_1)$&\cellcolor[gray]{0.7}$\epsilon^{-1}/2$\\
\cellcolor[gray]{0.7}$-\epsilon^{-\frac{5}{2}}S'/(2s'_1s_1^2s_3)$&\cellcolor[gray]{0.5}$\epsilon^{-2}$&\cellcolor[gray]{0.7}$\epsilon^{-\frac{3}{2}}S'/(2s'_1s_3)$\\
\cellcolor[gray]{0.7}$\epsilon^{-3}/(2s_1^2s_3^2)$&\cellcolor[gray]{0.7}$-\epsilon^{-\frac{5}{2}}S'/(2s'_3s_1s_3^2)$&\cellcolor[gray]{0.7}$-\epsilon^{-2}/(2s_3^2)$
\end{tabular}
\end{minipage}
\quad\quad\quad
\begin{minipage}[t]{.3\textwidth}
\centering
\underline{$M_0=5/2, (M_1,M_3\in\{-1/2,1/2\})$}
\begin{tabular}{cc}
\cellcolor[gray]{0.7}$\epsilon^{-2}/(\sqrt{2}c_1)$&\cellcolor[gray]{0.7}$\epsilon^{-2}/\sqrt{2}$\\
\cellcolor[gray]{0.7}$\epsilon^{-3}/(\sqrt{2}c_1c_3)$&\cellcolor[gray]{0.7}$\epsilon^{-2}/(\sqrt{2}c_3)$\\
\end{tabular}
\end{minipage}
\caption{The lowest non-vanishing partition functions for $k=2$ (continued).}
\label{k2data2}
\end{table}

\section*{Acknowledgements}
We are grateful to Chuan-Tsung Chan, Heng-Yu Chen, Nick Dorey, Hirotaka Hayashi, Chiung Hwang, Yosuke Imamura, Naotaka Kubo, Sangmin Lee, Tomoki Nakanishi, Tadashi Okazaki, Shou Tanigawa, Yasuhiko Yamada for valuable discussions and comments.
The work of S.~M.~is supported by JSPS Grant-in-Aid for Scientific Research (C) \#19K03829 and \#22K03598.
S.~M.~would like to thank Yukawa Institute for Theoretical Physics at Kyoto University for warm hospitality.
Preliminary results of this paper were presented in an international conference ``14th Taiwan String Workshop'' at Taipei and Kaohsiung, Taiwan.
We are grateful to the organizers and also thank the participants for various valuable discussions.

\bibliography{references_for_draft.bib}

\providecommand{\href}[2]{#2}\begingroup\raggedright\begin{thebibliography}{10}

\bibitem{MP}
M.~Marino and P.~Putrov, ``{ABJM theory as a Fermi gas},''
  \href{http://dx.doi.org/10.1088/1742-5468/2012/03/P03001}{{\em J. Stat.
  Mech.} {\bfseries 1203} (2012) P03001},
  \href{http://arxiv.org/abs/1110.4066}{{\ttfamily arXiv:1110.4066 [hep-th]}}.

\bibitem{ABJM}
O.~Aharony, O.~Bergman, D.~L. Jafferis, and J.~Maldacena, ``{N=6 superconformal
  Chern-Simons-matter theories, M2-branes and their gravity duals},''
  \href{http://dx.doi.org/10.1088/1126-6708/2008/10/091}{{\em JHEP} {\bfseries
  10} (2008) 091}, \href{http://arxiv.org/abs/0806.1218}{{\ttfamily
  arXiv:0806.1218 [hep-th]}}.

\bibitem{BGT3}
G.~Bonelli, A.~Grassi, and A.~Tanzini, ``{Quantum curves and $q$-deformed
  Painlev\'e equations},''
  \href{http://dx.doi.org/10.1007/s11005-019-01174-y}{{\em Lett. Math. Phys.}
  {\bfseries 109} no.~9, (2019) 1961--2001},
  \href{http://arxiv.org/abs/1710.11603}{{\ttfamily arXiv:1710.11603
  [hep-th]}}.

\bibitem{MN3}
S.~Moriyama and T.~Nosaka, ``{Exact Instanton Expansion of Superconformal
  Chern-Simons Theories from Topological Strings},''
  \href{http://dx.doi.org/10.1007/JHEP05(2015)022}{{\em JHEP} {\bfseries 05}
  (2015) 022}, \href{http://arxiv.org/abs/1412.6243}{{\ttfamily arXiv:1412.6243
  [hep-th]}}.

\bibitem{KMN}
N.~Kubo, S.~Moriyama, and T.~Nosaka, ``{Symmetry Breaking in Quantum Curves and
  Super Chern-Simons Matrix Models},''
  \href{http://dx.doi.org/10.1007/JHEP01(2019)210}{{\em JHEP} {\bfseries 01}
  (2019) 210}, \href{http://arxiv.org/abs/1811.06048}{{\ttfamily
  arXiv:1811.06048 [hep-th]}}.

\bibitem{BGKNT}
G.~Bonelli, F.~Globlek, N.~Kubo, T.~Nosaka, and A.~Tanzini, ``{M2-branes and
  ${\mathfrak {q}}$-Painlev\'e equations},''
  \href{http://dx.doi.org/10.1007/s11005-022-01597-0}{{\em Lett. Math. Phys.}
  {\bfseries 112} no.~6, (2022) 109},
  \href{http://arxiv.org/abs/2202.10654}{{\ttfamily arXiv:2202.10654
  [hep-th]}}.

\bibitem{MN9}
S.~Moriyama and T.~Nosaka, ``{40 bilinear relations of q-Painlev\'e VI from $
  \mathcal{N} $ = 4 super Chern-Simons theory},''
  \href{http://dx.doi.org/10.1007/JHEP08(2023)191}{{\em JHEP} {\bfseries 08}
  (2023) 191}, \href{http://arxiv.org/abs/2305.03978}{{\ttfamily
  arXiv:2305.03978 [hep-th]}}.

\bibitem{HLLLP2}
K.~Hosomichi, K.-M. Lee, S.~Lee, S.~Lee, and J.~Park, ``{N=5,6 Superconformal
  Chern-Simons Theories and M2-branes on Orbifolds},''
  \href{http://dx.doi.org/10.1088/1126-6708/2008/09/002}{{\em JHEP} {\bfseries
  09} (2008) 002}, \href{http://arxiv.org/abs/0806.4977}{{\ttfamily
  arXiv:0806.4977 [hep-th]}}.

\bibitem{ABJ}
O.~Aharony, O.~Bergman, and D.~L. Jafferis, ``{Fractional M2-branes},''
  \href{http://dx.doi.org/10.1088/1126-6708/2008/11/043}{{\em JHEP} {\bfseries
  11} (2008) 043}, \href{http://arxiv.org/abs/0807.4924}{{\ttfamily
  arXiv:0807.4924 [hep-th]}}.

\bibitem{Kapustin:2009kz}
A.~Kapustin, B.~Willett, and I.~Yaakov, ``{Exact Results for Wilson Loops in
  Superconformal Chern-Simons Theories with Matter},''
  \href{http://dx.doi.org/10.1007/JHEP03(2010)089}{{\em JHEP} {\bfseries 03}
  (2010) 089}, \href{http://arxiv.org/abs/0909.4559}{{\ttfamily arXiv:0909.4559
  [hep-th]}}.

\bibitem{DMP1}
N.~Drukker, M.~Marino, and P.~Putrov, ``{From weak to strong coupling in ABJM
  theory},'' \href{http://dx.doi.org/10.1007/s00220-011-1253-6}{{\em Commun.
  Math. Phys.} {\bfseries 306} (2011) 511--563},
  \href{http://arxiv.org/abs/1007.3837}{{\ttfamily arXiv:1007.3837 [hep-th]}}.

\bibitem{FHM}
H.~Fuji, S.~Hirano, and S.~Moriyama, ``{Summing Up All Genus Free Energy of
  ABJM Matrix Model},'' \href{http://dx.doi.org/10.1007/JHEP08(2011)001}{{\em
  JHEP} {\bfseries 08} (2011) 001},
  \href{http://arxiv.org/abs/1106.4631}{{\ttfamily arXiv:1106.4631 [hep-th]}}.

\bibitem{HMO2}
Y.~Hatsuda, S.~Moriyama, and K.~Okuyama, ``{Instanton Effects in ABJM Theory
  from Fermi Gas Approach},''
  \href{http://dx.doi.org/10.1007/JHEP01(2013)158}{{\em JHEP} {\bfseries 01}
  (2013) 158}, \href{http://arxiv.org/abs/1211.1251}{{\ttfamily arXiv:1211.1251
  [hep-th]}}.

\bibitem{CM}
F.~Calvo and M.~Marino, ``{Membrane instantons from a semiclassical TBA},''
  \href{http://dx.doi.org/10.1007/JHEP05(2013)006}{{\em JHEP} {\bfseries 05}
  (2013) 006}, \href{http://arxiv.org/abs/1212.5118}{{\ttfamily arXiv:1212.5118
  [hep-th]}}.

\bibitem{HMO3}
Y.~Hatsuda, S.~Moriyama, and K.~Okuyama, ``{Instanton Bound States in ABJM
  Theory},'' \href{http://dx.doi.org/10.1007/JHEP05(2013)054}{{\em JHEP}
  {\bfseries 05} (2013) 054}, \href{http://arxiv.org/abs/1301.5184}{{\ttfamily
  arXiv:1301.5184 [hep-th]}}.

\bibitem{HMMO}
Y.~Hatsuda, M.~Marino, S.~Moriyama, and K.~Okuyama, ``{Non-perturbative effects
  and the refined topological string},''
  \href{http://dx.doi.org/10.1007/JHEP09(2014)168}{{\em JHEP} {\bfseries 09}
  (2014) 168}, \href{http://arxiv.org/abs/1306.1734}{{\ttfamily arXiv:1306.1734
  [hep-th]}}.

\bibitem{MM}
S.~Matsumoto and S.~Moriyama, ``{ABJ Fractional Brane from ABJM Wilson Loop},''
  \href{http://dx.doi.org/10.1007/JHEP03(2014)079}{{\em JHEP} {\bfseries 03}
  (2014) 079}, \href{http://arxiv.org/abs/1310.8051}{{\ttfamily arXiv:1310.8051
  [hep-th]}}.

\bibitem{G}
S.~Matsuno and S.~Moriyama, ``{Giambelli Identity in Super Chern-Simons Matrix
  Model},'' \href{http://dx.doi.org/10.1063/1.4978229}{{\em J. Math. Phys.}
  {\bfseries 58} no.~3, (2017) 032301},
  \href{http://arxiv.org/abs/1603.04124}{{\ttfamily arXiv:1603.04124
  [hep-th]}}.

\bibitem{JT}
T.~Furukawa and S.~Moriyama, ``{Jacobi-Trudi Identity in Super Chern-Simons
  Matrix Model},'' \href{http://dx.doi.org/10.3842/SIGMA.2018.049}{{\em SIGMA}
  {\bfseries 14} (2018) 049}, \href{http://arxiv.org/abs/1711.04893}{{\ttfamily
  arXiv:1711.04893 [hep-th]}}.

\bibitem{2DTL}
T.~Furukawa and S.~Moriyama, ``{ABJM Matrix Model and 2D Toda Lattice
  Hierarchy},'' \href{http://dx.doi.org/10.1007/JHEP03(2019)197}{{\em JHEP}
  {\bfseries 03} (2019) 197}, \href{http://arxiv.org/abs/1901.00541}{{\ttfamily
  arXiv:1901.00541 [hep-th]}}.

\bibitem{MS2}
S.~Moriyama and T.~Suyama, ``{Orthosymplectic Chern-Simons Matrix Model and
  Chirality Projection},''
  \href{http://dx.doi.org/10.1007/JHEP04(2016)132}{{\em JHEP} {\bfseries 04}
  (2016) 132}, \href{http://arxiv.org/abs/1601.03846}{{\ttfamily
  arXiv:1601.03846 [hep-th]}}.

\bibitem{MN5}
S.~Moriyama and T.~Nosaka, ``{Orientifold ABJM Matrix Model: Chiral Projections
  and Worldsheet Instantons},''
  \href{http://dx.doi.org/10.1007/JHEP06(2016)068}{{\em JHEP} {\bfseries 06}
  (2016) 068}, \href{http://arxiv.org/abs/1603.00615}{{\ttfamily
  arXiv:1603.00615 [hep-th]}}.

\bibitem{closed}
K.~Kiyoshige and S.~Moriyama, ``{Dualities in ABJM Matrix Model from Closed
  String Viewpoint},'' \href{http://dx.doi.org/10.1007/JHEP11(2016)096}{{\em
  JHEP} {\bfseries 11} (2016) 096},
  \href{http://arxiv.org/abs/1607.06414}{{\ttfamily arXiv:1607.06414
  [hep-th]}}.

\bibitem{KMZ}
R.~Kashaev, M.~Marino, and S.~Zakany, ``{Matrix Models from Operators and
  Topological Strings, 2},''
  \href{http://dx.doi.org/10.1007/s00023-016-0471-z}{{\em Annales Henri
  Poincare} {\bfseries 17} no.~10, (2016) 2741--2781},
  \href{http://arxiv.org/abs/1505.02243}{{\ttfamily arXiv:1505.02243
  [hep-th]}}.

\bibitem{MNN}
S.~Moriyama, S.~Nakayama, and T.~Nosaka, ``{Instanton Effects in Rank Deformed
  Superconformal Chern-Simons Theories from Topological Strings},''
  \href{http://dx.doi.org/10.1007/JHEP08(2017)003}{{\em JHEP} {\bfseries 08}
  (2017) 003}, \href{http://arxiv.org/abs/1704.04358}{{\ttfamily
  arXiv:1704.04358 [hep-th]}}.

\bibitem{MNY}
S.~Moriyama, T.~Nosaka, and K.~Yano, ``{Superconformal Chern-Simons Theories
  from del Pezzo Geometries},''
  \href{http://dx.doi.org/10.1007/JHEP11(2017)089}{{\em JHEP} {\bfseries 11}
  (2017) 089}, \href{http://arxiv.org/abs/1707.02420}{{\ttfamily
  arXiv:1707.02420 [hep-th]}}.

\bibitem{KM}
N.~Kubo and S.~Moriyama, ``{Hanany-Witten Transition in Quantum Curves},''
  \href{http://dx.doi.org/10.1007/JHEP12(2019)101}{{\em JHEP} {\bfseries 12}
  (2019) 101}, \href{http://arxiv.org/abs/1907.04971}{{\ttfamily
  arXiv:1907.04971 [hep-th]}}.

\bibitem{MN1}
S.~Moriyama and T.~Nosaka, ``{Partition Functions of Superconformal
  Chern-Simons Theories from Fermi Gas Approach},''
  \href{http://dx.doi.org/10.1007/JHEP11(2014)164}{{\em JHEP} {\bfseries 11}
  (2014) 164}, \href{http://arxiv.org/abs/1407.4268}{{\ttfamily arXiv:1407.4268
  [hep-th]}}.

\bibitem{MN2}
S.~Moriyama and T.~Nosaka, ``{ABJM membrane instanton from a pole cancellation
  mechanism},'' \href{http://dx.doi.org/10.1103/PhysRevD.92.026003}{{\em Phys.
  Rev. D} {\bfseries 92} no.~2, (2015) 026003},
  \href{http://arxiv.org/abs/1410.4918}{{\ttfamily arXiv:1410.4918 [hep-th]}}.

\bibitem{Hatsuda:2015lpa}
Y.~Hatsuda, M.~Honda, and K.~Okuyama, ``{Large N non-perturbative effects in
  $\mathcal{N}=4$ superconformal Chern-Simons theories},''
  \href{http://dx.doi.org/10.1007/JHEP09(2015)046}{{\em JHEP} {\bfseries 09}
  (2015) 046}, \href{http://arxiv.org/abs/1505.07120}{{\ttfamily
  arXiv:1505.07120 [hep-th]}}.

\bibitem{Sakai}
H.~Sakai, ``{Rational Surfaces Associated with Affine Root Systems and Geometry
  of the Painlev\'e Equations},''
  \href{http://dx.doi.org/10.1007/s002200100446}{{\em Commun. Math. Phys.}
  {\bfseries 220} (2001) 165--229}.

\bibitem{ME8}
S.~Moriyama, ``{Spectral Theories and Topological Strings on del Pezzo
  Geometries},'' \href{http://dx.doi.org/10.1007/JHEP10(2020)154}{{\em JHEP}
  {\bfseries 10} (2020) 154}, \href{http://arxiv.org/abs/2007.05148}{{\ttfamily
  arXiv:2007.05148 [hep-th]}}.

\bibitem{MY}
S.~Moriyama and Y.~Yamada, ``{Quantum Representation of Affine Weyl Groups and
  Associated Quantum Curves},''
  \href{http://dx.doi.org/10.3842/SIGMA.2021.076}{{\em SIGMA} {\bfseries 17}
  (2021) 076}, \href{http://arxiv.org/abs/2104.06661}{{\ttfamily
  arXiv:2104.06661 [math.QA]}}.

\bibitem{Elitzur:1997fh}
S.~Elitzur, A.~Giveon, and D.~Kutasov, ``{Branes and N=1 duality in string
  theory},'' \href{http://dx.doi.org/10.1016/S0370-2693(97)00375-4}{{\em Phys.
  Lett. B} {\bfseries 400} (1997) 269--274},
  \href{http://arxiv.org/abs/hep-th/9702014}{{\ttfamily arXiv:hep-th/9702014}}.

\bibitem{Klebanov:2000hb}
I.~R. Klebanov and M.~J. Strassler, ``{Supergravity and a confining gauge
  theory: Duality cascades and chi SB resolution of naked singularities},''
  \href{http://dx.doi.org/10.1088/1126-6708/2000/08/052}{{\em JHEP} {\bfseries
  08} (2000) 052}, \href{http://arxiv.org/abs/hep-th/0007191}{{\ttfamily
  arXiv:hep-th/0007191}}.

\bibitem{Aharony:2009fc}
O.~Aharony, A.~Hashimoto, S.~Hirano, and P.~Ouyang, ``{D-brane Charges in
  Gravitational Duals of 2+1 Dimensional Gauge Theories and Duality
  Cascades},'' \href{http://dx.doi.org/10.1007/JHEP01(2010)072}{{\em JHEP}
  {\bfseries 01} (2010) 072}, \href{http://arxiv.org/abs/0906.2390}{{\ttfamily
  arXiv:0906.2390 [hep-th]}}.

\bibitem{Evslin:2009pk}
J.~Evslin and S.~Kuperstein, ``{ABJ(M) and Fractional M2's with Fractional M2
  Charge},'' \href{http://dx.doi.org/10.1088/1126-6708/2009/12/016}{{\em JHEP}
  {\bfseries 12} (2009) 016}, \href{http://arxiv.org/abs/0906.2703}{{\ttfamily
  arXiv:0906.2703 [hep-th]}}.

\bibitem{Hanany:1996ie}
A.~Hanany and E.~Witten, ``{Type IIB superstrings, BPS monopoles, and
  three-dimensional gauge dynamics},''
  \href{http://dx.doi.org/10.1016/S0550-3213(97)00157-0}{{\em Nucl. Phys. B}
  {\bfseries 492} (1997) 152--190},
  \href{http://arxiv.org/abs/hep-th/9611230}{{\ttfamily arXiv:hep-th/9611230}}.

\bibitem{HK}
M.~Honda and N.~Kubo, ``{Non-perturbative tests of duality cascades in three
  dimensional supersymmetric gauge theories},''
  \href{http://dx.doi.org/10.1007/JHEP07(2021)012}{{\em JHEP} {\bfseries 07}
  (2021) 012}, \href{http://arxiv.org/abs/2010.15656}{{\ttfamily
  arXiv:2010.15656 [hep-th]}}.

\bibitem{FMMN}
T.~Furukawa, K.~Matsumura, S.~Moriyama, and T.~Nakanishi, ``{Duality cascades
  and affine Weyl groups},''
  \href{http://dx.doi.org/10.1007/JHEP05(2022)132}{{\em JHEP} {\bfseries 05}
  (2022) 132}, \href{http://arxiv.org/abs/2112.13616}{{\ttfamily
  arXiv:2112.13616 [hep-th]}}.

\bibitem{FMS}
T.~Furukawa, S.~Moriyama, and H.~Sasaki, ``{Duality cascades and
  parallelotopes},'' \href{http://dx.doi.org/10.1088/1751-8121/acc2fb}{{\em J.
  Phys. A} {\bfseries 56} no.~16, (2023) 165401},
  \href{http://arxiv.org/abs/2205.08039}{{\ttfamily arXiv:2205.08039
  [hep-th]}}.

\bibitem{Yaakov:2013fza}
I.~Yaakov, ``{Redeeming Bad Theories},''
  \href{http://dx.doi.org/10.1007/JHEP11(2013)189}{{\em JHEP} {\bfseries 11}
  (2013) 189}, \href{http://arxiv.org/abs/1303.2769}{{\ttfamily arXiv:1303.2769
  [hep-th]}}.

\bibitem{Nosaka:2017ohr}
T.~Nosaka and S.~Yokoyama, ``{Complete factorization in minimal $ \mathcal{N}=4
  $ Chern-Simons-matter theory},''
  \href{http://dx.doi.org/10.1007/JHEP01(2018)001}{{\em JHEP} {\bfseries 01}
  (2018) 001}, \href{http://arxiv.org/abs/1706.07234}{{\ttfamily
  arXiv:1706.07234 [hep-th]}}.

\bibitem{Gaiotto:2008ak}
D.~Gaiotto and E.~Witten, ``{S-Duality of Boundary Conditions In N=4 Super
  Yang-Mills Theory},''
  \href{http://dx.doi.org/10.4310/ATMP.2009.v13.n3.a5}{{\em Adv. Theor. Math.
  Phys.} {\bfseries 13} no.~3, (2009) 721--896},
  \href{http://arxiv.org/abs/0807.3720}{{\ttfamily arXiv:0807.3720 [hep-th]}}.

\bibitem{Honda:2014npa}
M.~Honda and K.~Okuyama, ``{Exact results on ABJ theory and the refined
  topological string},'' \href{http://dx.doi.org/10.1007/JHEP08(2014)148}{{\em
  JHEP} {\bfseries 08} (2014) 148},
  \href{http://arxiv.org/abs/1405.3653}{{\ttfamily arXiv:1405.3653 [hep-th]}}.

\bibitem{Hatsuda:2012hm}
Y.~Hatsuda, S.~Moriyama, and K.~Okuyama, ``{Exact Results on the ABJM Fermi
  Gas},'' \href{http://dx.doi.org/10.1007/JHEP10(2012)020}{{\em JHEP}
  {\bfseries 10} (2012) 020}, \href{http://arxiv.org/abs/1207.4283}{{\ttfamily
  arXiv:1207.4283 [hep-th]}}.

\bibitem{Gomis:2008vc}
J.~Gomis, D.~Rodriguez-Gomez, M.~Van~Raamsdonk, and H.~Verlinde, ``{A Massive
  Study of M2-brane Proposals},''
  \href{http://dx.doi.org/10.1088/1126-6708/2008/09/113}{{\em JHEP} {\bfseries
  09} (2008) 113}, \href{http://arxiv.org/abs/0807.1074}{{\ttfamily
  arXiv:0807.1074 [hep-th]}}.

\bibitem{Nosaka:2020tyv}
T.~Nosaka, ``{SU(N) q-Toda equations from mass deformed ABJM theory},''
  \href{http://dx.doi.org/10.1007/JHEP06(2021)060}{{\em JHEP} {\bfseries 06}
  (2021) 060}, \href{http://arxiv.org/abs/2012.07211}{{\ttfamily
  arXiv:2012.07211 [hep-th]}}.

\bibitem{Nosaka:2024gle}
T.~Nosaka, ``{Large N expansion of mass deformed ABJM matrix model:
  M2-instanton condensation and beyond},''
  \href{http://dx.doi.org/10.1007/JHEP03(2024)087}{{\em JHEP} {\bfseries 03}
  (2024) 087}, \href{http://arxiv.org/abs/2401.11484}{{\ttfamily
  arXiv:2401.11484 [hep-th]}}.

\bibitem{Kim:2012uz}
H.-C. Kim, J.~Kim, S.~Kim, and K.~Lee, ``{Vortices and 3 dimensional
  dualities},'' \href{http://arxiv.org/abs/1204.3895}{{\ttfamily
  arXiv:1204.3895 [hep-th]}}.

\bibitem{Giacomelli:2023zkk}
S.~Giacomelli, C.~Hwang, F.~Marino, S.~Pasquetti, and M.~Sacchi, ``{Probing bad
  theories with the dualization algorithm. Part I},''
  \href{http://dx.doi.org/10.1007/JHEP04(2024)008}{{\em JHEP} {\bfseries 04}
  (2024) 008}, \href{http://arxiv.org/abs/2309.05326}{{\ttfamily
  arXiv:2309.05326 [hep-th]}}.

\bibitem{Iqbal:2002we}
A.~Iqbal, ``{All genus topological string amplitudes and five-brane webs as
  Feynman diagrams},'' \href{http://arxiv.org/abs/hep-th/0207114}{{\ttfamily
  arXiv:hep-th/0207114}}.

\bibitem{Aganagic:2003db}
M.~Aganagic, A.~Klemm, M.~Marino, and C.~Vafa, ``{The Topological vertex},''
  \href{http://dx.doi.org/10.1007/s00220-004-1162-z}{{\em Commun. Math. Phys.}
  {\bfseries 254} (2005) 425--478},
  \href{http://arxiv.org/abs/hep-th/0305132}{{\ttfamily arXiv:hep-th/0305132}}.

\bibitem{Iqbal:2007ii}
A.~Iqbal, C.~Kozcaz, and C.~Vafa, ``{The Refined topological vertex},''
  \href{http://dx.doi.org/10.1088/1126-6708/2009/10/069}{{\em JHEP} {\bfseries
  10} (2009) 069}, \href{http://arxiv.org/abs/hep-th/0701156}{{\ttfamily
  arXiv:hep-th/0701156}}.

\bibitem{Taki:2007dh}
M.~Taki, ``{Refined Topological Vertex and Instanton Counting},''
  \href{http://dx.doi.org/10.1088/1126-6708/2008/03/048}{{\em JHEP} {\bfseries
  03} (2008) 048}, \href{http://arxiv.org/abs/0710.1776}{{\ttfamily
  arXiv:0710.1776 [hep-th]}}.

\bibitem{Bao:2013pwa}
L.~Bao, V.~Mitev, E.~Pomoni, M.~Taki, and F.~Yagi, ``{Non-Lagrangian Theories
  from Brane Junctions},''
  \href{http://dx.doi.org/10.1007/JHEP01(2014)175}{{\em JHEP} {\bfseries 01}
  (2014) 175}, \href{http://arxiv.org/abs/1310.3841}{{\ttfamily arXiv:1310.3841
  [hep-th]}}.

\bibitem{Hayashi:2013qwa}
H.~Hayashi, H.-C. Kim, and T.~Nishinaka, ``{Topological strings and 5d $T_N$
  partition functions},'' \href{http://dx.doi.org/10.1007/JHEP06(2014)014}{{\em
  JHEP} {\bfseries 06} (2014) 014},
  \href{http://arxiv.org/abs/1310.3854}{{\ttfamily arXiv:1310.3854 [hep-th]}}.

\bibitem{Taki:2013vka}
M.~Taki, ``{Notes on Enhancement of Flavor Symmetry and 5d Superconformal
  Index},'' \href{http://arxiv.org/abs/1310.7509}{{\ttfamily arXiv:1310.7509
  [hep-th]}}.

\bibitem{Taki:2014pba}
M.~Taki, ``{Seiberg Duality, 5d SCFTs and Nekrasov Partition Functions},''
  \href{http://arxiv.org/abs/1401.7200}{{\ttfamily arXiv:1401.7200 [hep-th]}}.

\bibitem{FurukawaMSugimoto}
T.~Furukawa, S.~Moriyama, and Y.~Sugimoto, ``{Quantum Mirror Map for Del Pezzo
  Geometries},'' \href{http://dx.doi.org/10.1088/1751-8121/ab93fe}{{\em J.
  Phys. A} {\bfseries 53} no.~38, (2020) 38},
  \href{http://arxiv.org/abs/1908.11396}{{\ttfamily arXiv:1908.11396
  [hep-th]}}.

\bibitem{Furukawa:2020cjp}
T.~Furukawa, S.~Moriyama, and T.~Nakanishi, ``{Brane transitions from
  exceptional groups},''
  \href{http://dx.doi.org/10.1016/j.nuclphysb.2021.115477}{{\em Nucl. Phys. B}
  {\bfseries 969} (2021) 115477},
  \href{http://arxiv.org/abs/2010.15402}{{\ttfamily arXiv:2010.15402
  [hep-th]}}.

\bibitem{CFTDiFrancesco}
P.~Di~Francesco, P.~Mathieu, and D.~S\'en\'echal, {\em {Conformal Field
  Theory}}.
\newblock Springer, 1997.

\end{thebibliography}\endgroup

\end{document}